\documentclass[aps,a4paper,superscriptaddress,twocolumn]{revtex4}
\usepackage{tensor}
\usepackage{graphicx}
\usepackage{amsmath}
\usepackage{amssymb}
\usepackage{enumerate}
\usepackage{subfigure}
\usepackage{tabularx}
\usepackage[colorlinks=true, pdfstartview=FitV, linkcolor=blue, citecolor=red, urlcolor=black, breaklinks=true]{hyperref}
\newcommand{\be}{\begin{equation}}
\newcommand{\ee}{\end{equation}}
\newcommand{\ben}{\begin{eqnarray}}
\newcommand{\een}{\end{eqnarray}}
\newcommand{\bes}{\begin{subequations}}
\newcommand{\ees}{\end{subequations}}
\def\bal#1\eal{\begin{align}#1\end{align}}

\newcommand{\sech}{{\rm sech}}
\newcommand{\LL}{{\mathcal L}}

\begin{document}
\title{Electrically charged multi-field configurations}
\author{D. Bazeia}%\email{dbazeia@fisica.ufpb.br}
\affiliation{Departamento de F\'\i sica, Universidade Federal da Para\'\i ba, 58051-970 Jo\~ao Pessoa, PB, Brazil}
\author{M.A. Marques}%\email{marques@cbiotec.ufpb.br}
\affiliation{Departamento de Biotecnologia, Universidade Federal da Para\'\i ba, 58051-900 Jo\~ao Pessoa, PB, Brazil}
\author{M. Paganelly}%\email{matheuspaganelly@gmail.com}
\affiliation{Departamento de F\'\i sica, Universidade Federal da Para\'\i ba, 58051-970 Jo\~ao Pessoa, PB, Brazil}
\begin{abstract}
In this work we investigate the presence of electrically charged structures that are localized in two and three spatial dimensions. We use the Maxwell-scalar Lagrangian to describe several systems with distinct interactions for the scalar fields. The procedure relies on finding first order differential equations that solve the equations of motion and ensure stability of the corresponding minimum energy solutions. We illustrate the many possibilities in two and in three spatial dimensions, examining different examples of electrically charged solutions that engender internal structure. 
\end{abstract}
\date{\today}
\maketitle 

\section{Introduction}

In high energy physics, in the decade of 1970 several important works have driven the physicists attention towards the study of localized magnetic structures such as vortices in the plane \cite{NO} and magnetic monopoles \cite{tH,Po} in three spatial dimensions; see also \cite{PS,Bo,MS} for other interesting investigations. Together with these theoretical studies of magnetic monopoles, one can also construct dyons, which are spatially localized structures that carry both electric and magnetic charges \cite{JZ}. Similar configurations can also appear in the plane, as shown in Refs. \cite{HK,JW}, where the study of vortices with Chern-Simons dynamics ensured the presence of magnetic vortices engendering electric charge.

We can also consider electrically charged localized structures with trivial magnetic behavior. Long ago,  phenomenological models inspired by the Landau–Ginzburg–Devonshire (LGD) procedure \cite{La,Gi,De} have been considered to investigate physical properties of ferroelectric materials; see, e.g., the recent reviews \cite{FE1,FE2} and references therein for other related investigations. In high energy physics, in the decade of 1970, in Ref. \cite{Le1,Le2,Le3,Review,Review2} the authors considered modification of the electric properties of the medium to account for the possibility to describe bag-like models for solitons. Moreover, in the decade of 1980, other studies examined models capable of supporting spherically symmetric, charged and spatially localized configurations referred to as q-balls in Refs. \cite{QB1,QB2,QB3}. In the recent years, some of us have investigated q-balls in distinct scenarios in Refs. \cite{QB4,QB5,QB6,QB7}. In particular, in \cite{QB5,QB6,QB7} we considered the possibility of constructing compact \cite{QB5}, split \cite{QB6} and quasi-compact \cite{QB7} objects of the q-ball type, with the different profiles of the solutions directly related to the presence of distinct self-interactions in the potential that controls the complex scalar field. 

More recently, in Ref. \cite{electric} the study considered the construction of models that support structures engendering electric charge in the absence of magnetic properties in another context, different from the case of q-balls. Moreover, in \cite{ED} we used electric dipole in a medium with electric behavior controlled by a real scalar field to induce new effects of current interest. These recent results have inspired us to go further and explore other possibilities in the present work. The main line of investigation is implemented considering multi-field scenarios, in which two or three real scalar fields are added to account for distinct physical properties of the solutions and control their internal structures. The presence of several fields leads to more complicate scenario, since we are dealing with coupled nonlinear differential equations. However, under some specific conditions, we have been able to develop an interesting procedure, which help us to simplify the mathematical calculations and find stable analytical configurations of current interest in high energy physics, with potential application to other areas of nonlinear science.

In order to describe the construction of localized and electrically charged structures, we organize the present work as follows. In Sec. \ref{sec2}, we start reviewing some basic properties of the standard classical electrodynamics in the presence of a external charge, and then including a single real scalar field which couples to the electromagnetic field via an unusual non-minimal coupling. This gives rise to the Maxwell-scalar model \cite{electric,ED} which we further study in this work. Before doing this, however, in Sec. \ref{sec2} we also review the construction of domain wall that mimics Bloch wall, as investigated before in Refs. \cite{b1,B2} in the presence of two real scalar fields. We then introduce one interesting model in Sec. \ref{sec3}, where we consider two real scalar field non-minimally coupled to the electromagnetic field to describe charged field configurations. In Sec. \ref{sec4} we examine another model, in the presence of three scalar fields, the third one used to control the internal structure of the charged configurations. We then end the work in Sec. \ref{sec5}, adding some comments and distinct possibilities of continuation of the present study.

\section{General Considerations}\label{sec2}

In this work, we deal with the construction of spatially localized configurations in Minkowski spacetime with $(D,1)$ dimensions, with $D$ specifying the number of spatial coordinates. Our main interest here is to describe electrically charged configurations, so we first review the case of the electromagnetic field in the presence of an external charge, the Maxwell-scalar model and Bloch wall solutions.

\subsection{Single point charge}
To better understand the present investigation, let us first review the standard electromagnetic model, described by the action $S=\int dt\, d^Dx\, \LL$, with the Lagrangian density
\be\label{elestd}
\begin{aligned}
   \mathcal{L}&=-\frac{1}{4}F_{\mu\nu}F^{\mu\nu} -A_{\mu}j^{\mu}.
\end{aligned}
\ee
Here, $A^\mu$ stands for the electromagnetic field, $F_{\mu\nu}=\partial_\mu A_\nu-\partial_\nu A_\mu$ represents the  electromagnetic strength tensor and $j^\mu$ is a (conserved) source current. As it is well known, the equation of motion associated to this model is
\be\label{mstd}
\partial_\mu F^{\mu\nu} = j^\nu.
\ee
We now focus on electrostatics, on the particular case of a single point charge $e$, taking
\be\label{current}
j^0=e\Omega(D)\delta(\textbf{r}) \quad\text{and}\quad j^i=0,
\ee
where $\Omega(D)=2\pi^{D/2}/\Gamma(D/2)$ denotes the $D$--dimensional solid angle, introduced in this expression for convenience. Considering that $\vec{E} = \left(F^{10},F^{20},\ldots,F^{D0}\right)$, one can show that Eq.~\eqref{mstd} leads to
\be\label{estd}
{\bf E} = \frac{e}{r^{D-1}}\,\hat{r}.
\ee
This field is divergent at the charge location, $r=0$. It was found in Ref.~\cite{electric} that one can regularize this behavior by introducing a scalar field to control the electric permittivity of the system, with the Lagrangian density \eqref{elestd} changed to
\be\label{elemod}
\begin{aligned}
   \mathcal{L}&=-\frac{1}{4}P(\phi)F_{\mu\nu}F^{\mu\nu}+\frac12\partial_{\mu}\phi\partial^{\mu}\phi -A_{\mu}j^{\mu}.
\end{aligned}
\ee
The coupling between the scalar and gauge field is unusual, non-minimal, and the system is referred to as the Maxwell-scalar model; see, e.g., Refs. \cite{electric,ED}. In this situation, with the same charge considered in Eq. \eqref{current}, the electric field in Eq.~\eqref{estd} is modified to
\be\label{estd}
{\bf E} = \frac{e}{r^{D-1}P(\phi)}\,\hat{r}.
\ee
The above equation allows one to show that the scalar field is independent from the gauge field. If the permittivity is $P(\phi)=e^2/W_\phi^2$, the scalar field $\phi=\phi(r)$ is governed by the first order equation
\be
\phi^\prime = \frac{W_\phi}{r^{D-1}},
\ee
where the prime represents derivative with respect to the radial coordinate $r$, and $W_\phi=dW/d\phi$. So, for instance, one can consider a planar system, with $D=2$, and take $W=\phi-\phi^3/3$; in this case one gets $\phi = (r^2-r_{0}^{2})/(r^2+r_{0}^{2})$ and ${\bf E} =\hat{r}\, [2rr_{0}/(r^2+r_{0}^{2})]^{4}e^{-1}r^{-1}$. If, instead, one considers $D\geq3$, it is possible to find the analytical expressions $\phi=\tanh(-1/(D-2)r^{D-2})$ and ${\bf E} = \hat{r}\,\sech^4(-1/(D-2)r^{D-2})e^{-1}r^{1-D}$.

In this paper, we investigate the behavior of a single point charge in a medium with the electric permittivity controlled by two scalar fields that mimic domain wall of current interest, which we briefly review below.

\subsection{Domain walls}
It was shown in Refs.~\cite{b1,B2,Juan} that a model consisted of two scalar fields can be used to mimic the behavior of domain wall with internal structure. The Lagrangian density of interest is
\be\label{blochmodel}
{\cal L}=\frac12\partial_\mu\phi\partial^\mu\phi+\frac12 \partial_\mu\chi\partial^\mu\chi-V(\phi,\chi),
\ee
where the potential depends on both $\phi$ and $\chi$. In this subsection we consider a single spatial dimension. Minimum energy configurations, which lead to first order equations, may be obtained if the potential has the form
\be
V=\frac12W_\phi^2+\frac12 W_\chi^2,
\ee
where the subscripts represent partial derivatives with respect to the scalar fields. The model of interest arises with the auxiliary function $W=W(\phi,\chi)$ in the form
\be\label{bnrt}
W=\phi-\frac13 \phi^3- k \phi\chi^2,
\ee
in which $k>0$ is a real parameter. It describes interaction between the two scalar fields and induces interesting new behavior in the localized configurations which we explore in the present work. Remember that we are using natural units, and also, dimensionless fields and coordinates.

With $W$ given by Eq. \eqref{bnrt}, the potential becomes
\be
V(\phi,\chi) = \frac12\left(1-\phi^2-k\chi^2\right)^2 + 2k^2\phi^2\chi^2,
\ee
and static fields $\phi=\phi(x)$ and $\chi=\chi(x)$ ($D=1$) are governed by the following equations
\bes
\bal
&\frac{d\phi}{dx}=W_{\phi}=1-\phi^2- k \chi^2,\\
&\frac{d\chi}{dx}=W_{\chi}=-2  k \phi\chi.
\eal
\ees
For $0<k<1/2$, one gets the analytical solutions
\bes
\bal
&\phi=\tanh(2kx),\\
&\chi_\pm=\pm\sqrt{\frac1k-2}\, \sech(2kx).
\eal
\ees
These solutions connects the minima $(-1,0)$ and $(1,0)$, representing $(\phi,\chi)$. They are found with the use of the trial orbit method \cite{Raja,bflrorbit}, which allows us to decouple the first order equations. The solutions engender energy $E=4/3$.

\section{Charged configurations}\label{sec3}
We now investigate the following model in $(D,1)$ flat spacetime dimensions
\be\label{lelebloch}
\begin{aligned}
   \mathcal{L}&=-\frac{1}{4}P(\phi,\chi)F_{\mu\nu}F^{\mu\nu}+\frac12\partial_{\mu}\phi\partial^{\mu}\phi + \frac12\partial_{\mu}\chi\partial^{\mu}\chi -A_{\mu}j^{\mu}.
\end{aligned}
\ee
This Lagrangian density is inspired by the ones in Eqs.~\eqref{elemod} and \eqref{blochmodel}. Notice that, although this model has two scalar fields, $\phi$ and $\chi$, it is quite different from the one in Ref.~\cite{electric}, since the scalar fields are coupled exclusively by the electric permittivity here. The equations of motion are given by
\bes\begin{align}
    &\label{eqm15}\partial_{\mu}\partial^{\mu}\phi+\frac{P_{\phi}}{4}F_{\mu\nu}F^{\mu\nu}=0,\\
    &\label{eqm16}\partial_{\mu}\partial^{\mu}\chi+\frac{P_{\chi}}{4}F_{\mu\nu}F^{\mu\nu}=0,\\
   &\label{18eq} \partial_{\mu}(PF^{\mu\nu})-j^{\nu}=0.
\end{align}\ees
Since we are interested in studying a single point charge fixed at the origin, $\textbf{r}=0$, we take the $j^\mu$ given by Eq.~\eqref{current}. Considering static fields and solving the Gauss' law which arises from $\nu=0$ in Eq.~\eqref{18eq}, we get the electric field
\be
\label{Efield}
{\bf E} = \frac{e}{r^{D-1}P(\phi,\chi)}\hat{r}.
\ee
This expression allows us to write the equations of motion in Eqs.~\eqref{eqm15} and \eqref{eqm16} for static fields as
%%%
\bes
\bal
   & \label{eqm19}\frac{1}{r^{D-1}}(r^{D-1}\phi ')'-\frac{e^2}{2r^{2D-2}}\frac{\partial }{\partial\phi}\bigg(\frac{1}{P}\bigg)=0,\\
   & \label{eqm20}\frac{1}{r^{D-1}}(r^{D-1}\chi ')'-\frac{e^2}{2r^{2D-2}}\frac{\partial }{\partial\chi}\bigg(\frac{1}{P}\bigg)=0.
\eal
\ees
They are of second order and usually have nonlinearities and couplings between the fields due to the presence of the generalized permittivity. In order to get first order equations, we follow the lines of Ref.~\cite{Bo}. The energy density for this model is given by
\be\label{1density}
\begin{aligned}
\rho &=\frac{1}{2}{\phi'}^{2}+\frac{1}{2}{\chi'}^{2}+\frac{P}{2}{|\textbf{E}|}^{2}\\
		 &= \frac{1}{2}{\phi'}^{2}+\frac{1}{2}{\chi'}^{2}+\frac{1}{2P}\frac{e^2}{r^{2D-2}}.
\end{aligned}
\ee
To obtain the latter expression, we have used the electric field in Eq.~\eqref{Efield}. One can introduce the auxiliary function $W(\phi,\chi)$ to write the above expression in the form
\be\begin{aligned}   
   \label{edensity2}
    \rho &=\frac{1}{2}\bigg(\phi'\mp\frac{W_{\phi}}{r^{D-1}}\bigg)^{2}+\frac{1}{2}\bigg(\chi'\mp\frac{W_{\chi}}{r^{D-1}}\bigg)^{2}\\
    &+\frac{1}{2r^{2D-2}}\left(\frac{e^2}{P}-\left(W_{\phi}^2+W_{\chi}^2\right)\right) \pm \frac{1}{r^{D-1}} W'.
\end{aligned}
\ee
%%%
If the permittivity has the form
\begin{equation}
\label{permit}
    P(\phi,\chi)=\frac{e^{2}}{W^{2}_{\phi}+W^{2}_{\chi}},
\end{equation}
one gets that the energy has a minimum value, that is,
\be\label{ebogoelectric1}
E\geq E_B = \Omega(D)\left|\Delta W\right|,
\ee
where $\Delta W = W(\phi(\infty),\chi(\infty))-W(\phi(0),\chi(0))$. The minimum energy, $E=E_B$, is attained when the following first order equations are satisfied 
\bes
\bal
\label{phisolutionD11}
&\phi'=\pm\frac{W_{\phi}}{r^{D-1}},\\
\label{chisolutionD11}
&\chi'=\pm\frac{W_{\chi}}{r^{D-1}}.
\eal
\ees
These equations allow us to rewrite the energy density \eqref{1density} as
\begin{align}\label{energyd3}
    \rho(r)=\phi'^2+\chi'^2\;.
\end{align}
We then consider the function in Eq.~\eqref{bnrt}, with the inclusion of a parameter, $\sigma$, in the form
%%%%
\begin{align}
\label{newW}
    W(\phi,\chi)=\sigma\phi-\frac{1}{3}\sigma\phi^{3}-k\phi\chi^{2}.
\end{align}
%%%%
  Here, $\sigma$ is real and positive parameter. It is  not needed when working in a single spatial dimensions (see \cite{b1,B2,Juan}). Nevertheless, as we shall see below, it is important here to avoid divergences in the energy density. Let us first deal with the first order equations \eqref{phisolutionD11} and \eqref{chisolutionD11}, which now read
%%%%
\bes
\bal
\label{phisolutionDn1}
&\phi'=\pm\frac{\sigma-\sigma\phi^{2}-k\chi^{2}}{r^{D-1}},\\
\label{chisolutionDn1}
&\chi'=\mp\frac{2k\phi\chi}{r^{D-1}}.
\eal
\ees
Using the trial orbit method \cite{bflrorbit}, one can show that these equations support the elliptic orbit
\begin{equation}
\label{orbit1}
    \phi^2+\frac{k}{\sigma-2k}\chi^{2}=1,
\end{equation}
with $0<k<\sigma/2$. It connects the minima $v_{1}=(-1,0)$ and $v_{2}=(1,0)$, for the pair of fields $(\phi,\chi)$. Using the above expressions, one can decouple the first order equation \eqref{phisolutionDn1}:
\be\label{fophiorbitD1}
\phi'=\pm\frac{2k(1-\phi^{2})}{r^{D-1}}.
\ee
It is of interest to notice that the explicit dependence on the radial coordinate $r$ in the right hand side of the latter equations is similar to the case studied in Ref.~\cite{prlbmm}. For simplicity, from now on, we consider the above equation with the upper sign. Next, we investigate the solutions for $D=2$ and $D=3$.

%%%%%%%%%%%%%%%%%%%%
\subsection{Two spatial dimensions}\label{twospatial}
\begin{figure}[t!]
		\centering
		\includegraphics[width=6.2cm,trim={0cm 0cm 0 0},clip]{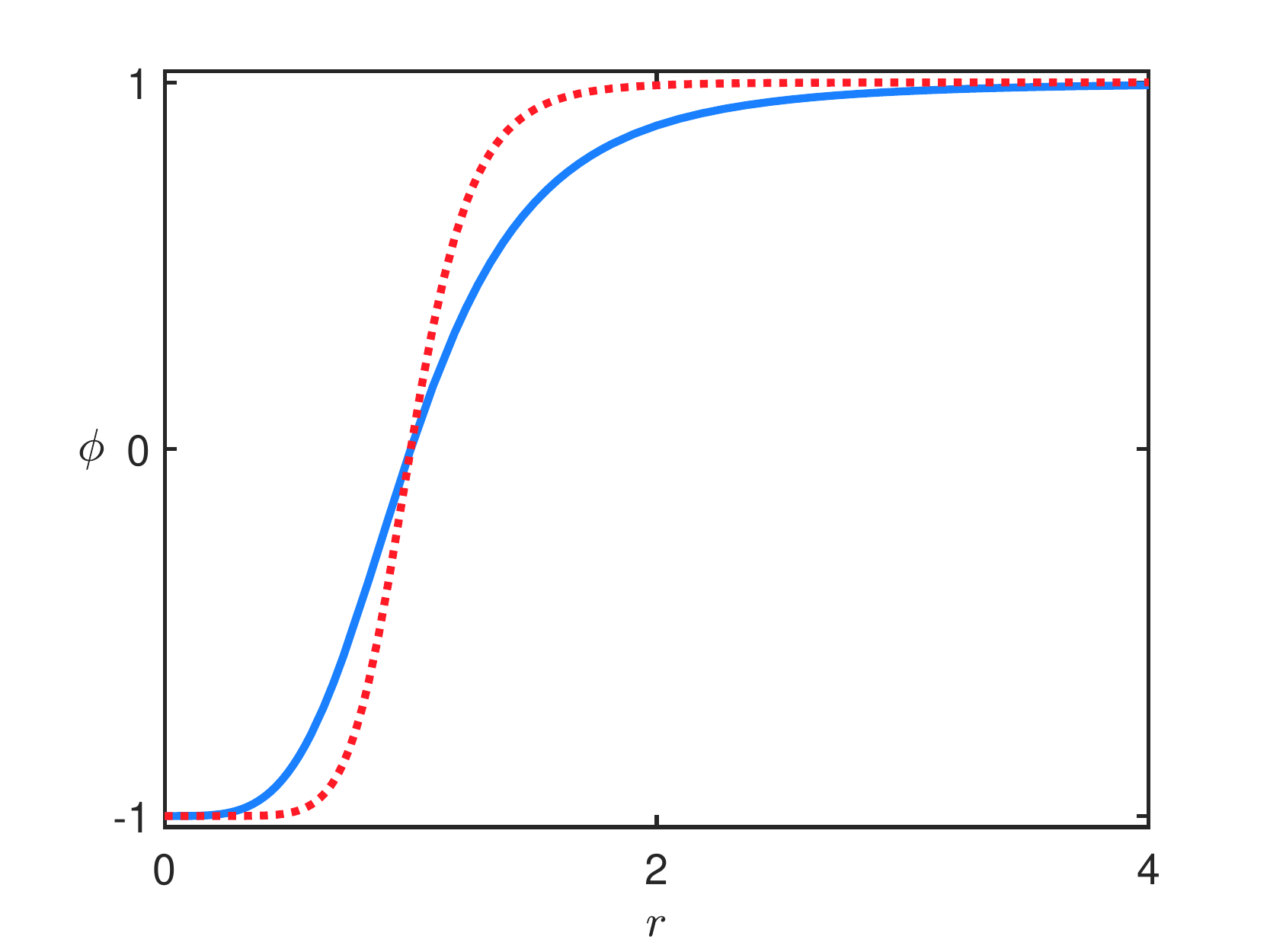}
		\includegraphics[width=6.2cm,trim={0cm 0cm 0 0},clip]{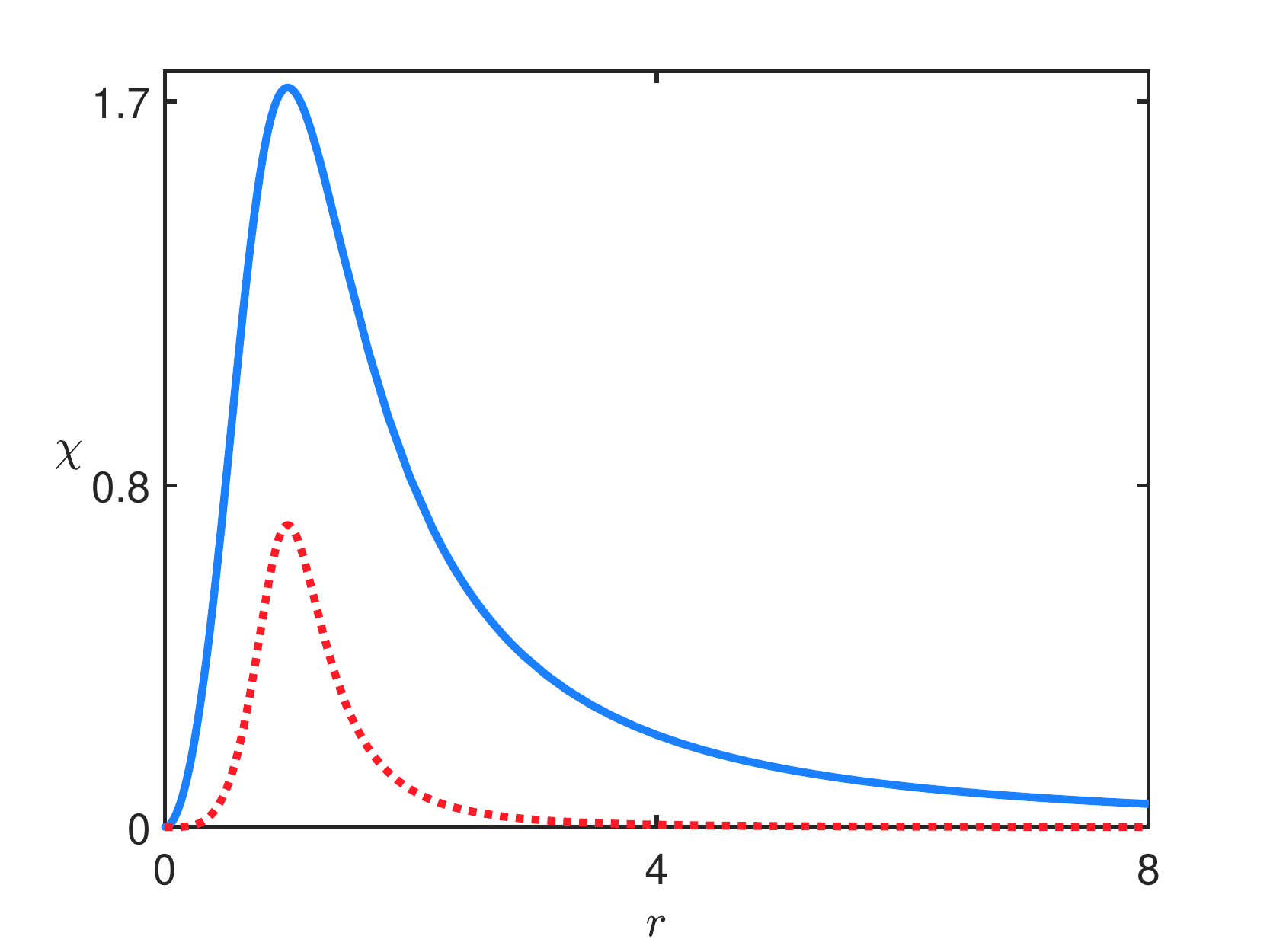}
		\caption{The solutions $\phi(r)$ (top) and $\chi(r)$ (bottom) associated to the model in Sec.~\ref{twospatial} for $\sigma=5$, with $k=1$ (solid, blue line) and $2$ (dotted, red line).}
		\label{figex}
		\end{figure}

\begin{figure}[t!]
		\centering
		\includegraphics[width=4.27cm,trim={0cm 0cm 0 0},clip]{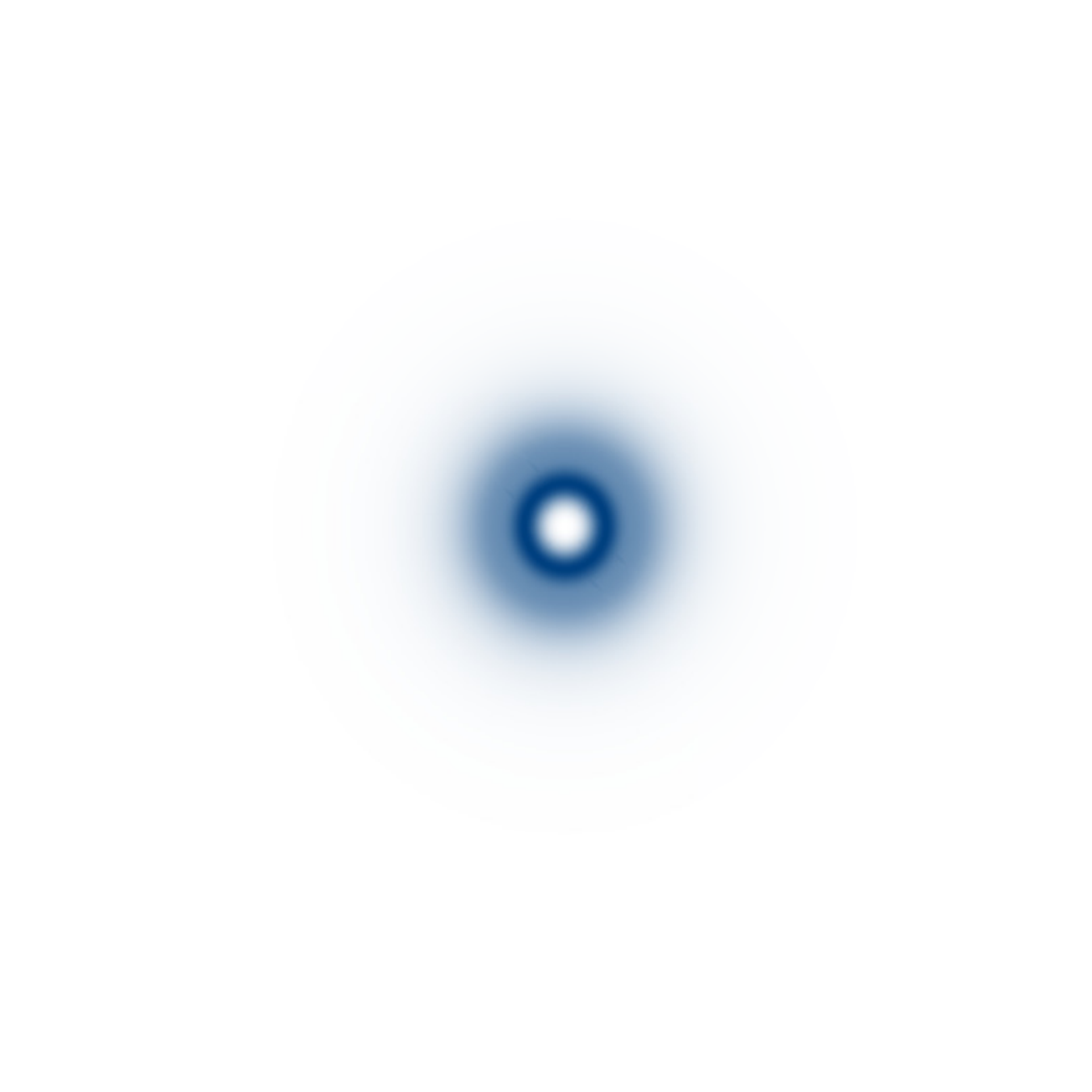}
		\includegraphics[width=4.27cm,trim={0cm 0cm 0 0},clip]{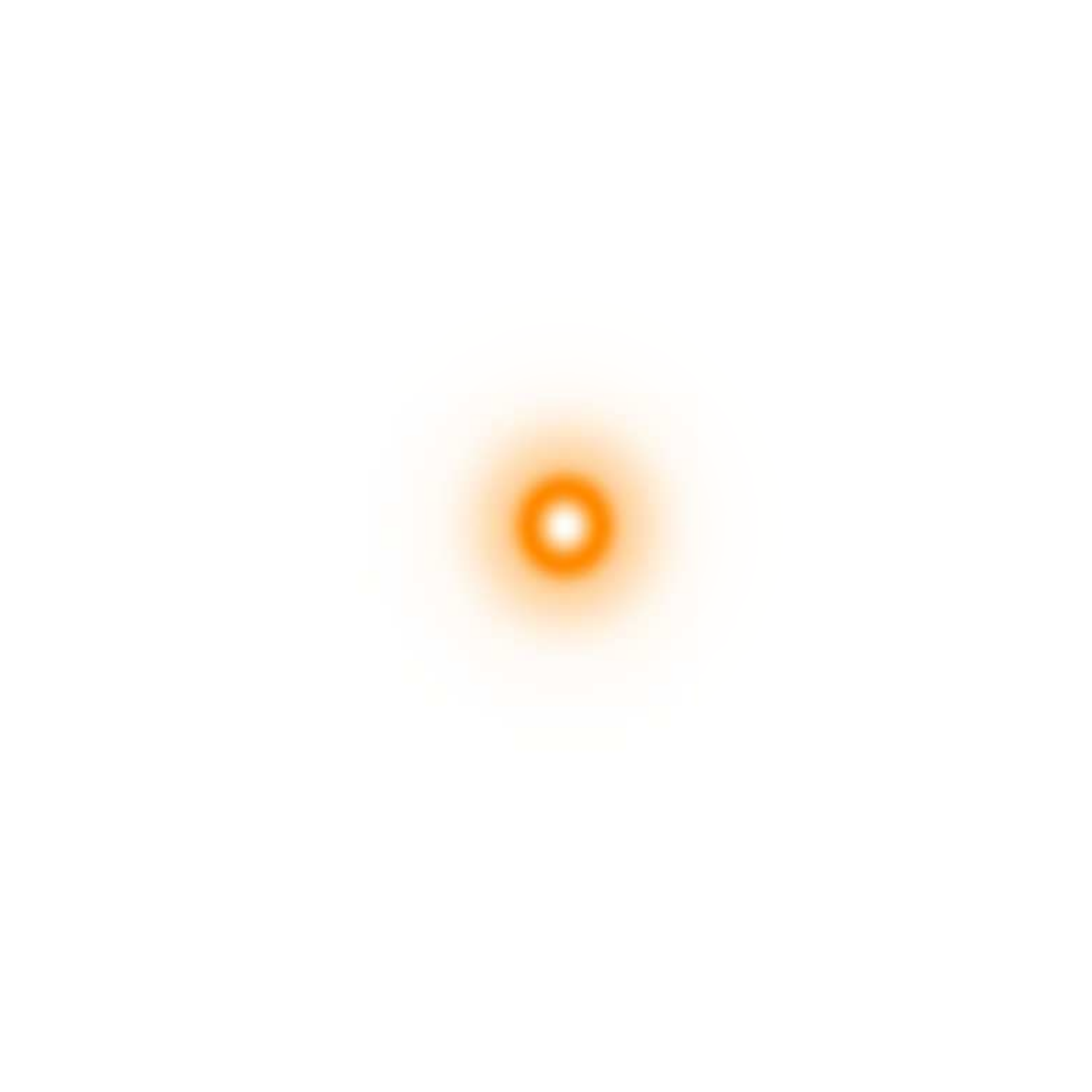}
		\includegraphics[width=4.27cm,trim={0cm 0cm 0 0},clip]{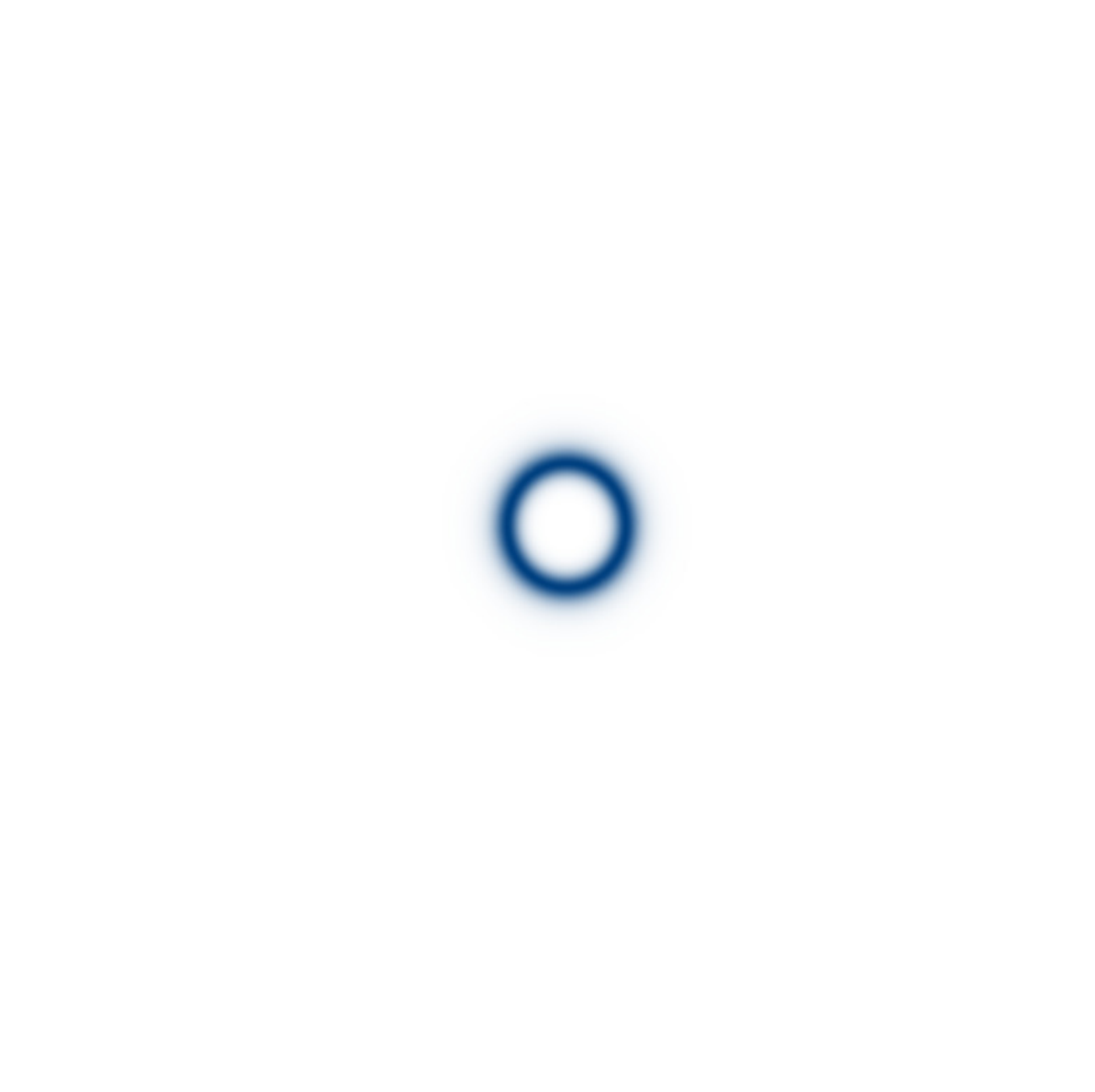}
		\includegraphics[width=4.27cm,trim={0cm 0cm 0 0},clip]{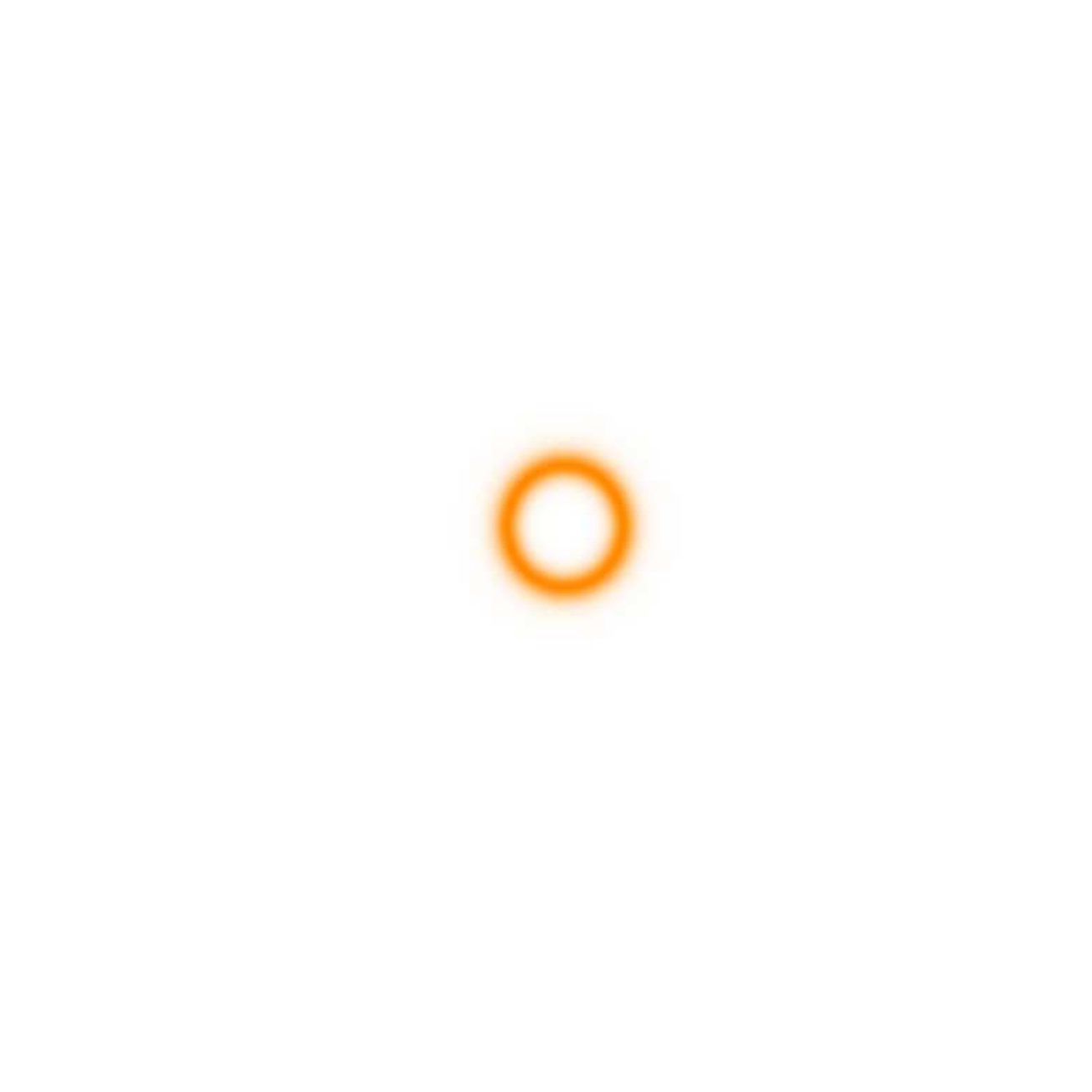}
		\caption{The radial component of the electric field (left, blue) with $e=1$ and the energy density (right, orange) associated to the model in Sec.~\ref{twospatial} depicted in the plane for $\sigma=5$, with $k=1$ (top), and $k=2$ (bottom). The intensity of the blue and orange colors increases with the increasing of the electric field and energy density, respectively.}
		\label{figex5}
		\end{figure}

In two spatial dimensions, the first order equation \eqref{fophiorbitD1} becomes
\be\label{fophiorbitD2}
\phi'=\frac{2k(1-\phi^{2})}{r}.
\ee
It supports the solution
%%%%%%%%%%%%%%%%%%%%
\begin{align}
    \phi(r)=\frac{r^{4k}-1}{r^{4k}+1}.
\end{align}
To find $\chi$, we use the orbit in Eq.~\ref{orbit1}. It leads us to
\begin{align}
\label{chi32}
    \chi(r)=2\sqrt{\frac{\sigma}{k}-2}\;\frac{r^{2k}}{r^{4k}+1}.
\end{align}
%%%%%%%%%%%%%%%%%%%
In the above expression, one must take $\sigma/k> 2$ to ensure that $\chi$ is real and non-vanishing, matching with the condition imposed below Eq.~\eqref{orbit1}. The solutions $\phi(r)$ and $\chi(r)$ are plotted in Fig.~\ref{figex}. Using the energy density in Eq.~\eqref{energyd3}, we get
%%%%%%%%%%%%%%%%%%%
%%%%%%%%%%%%%%%%%%%
%%%%%%%%%%%%%%%%%%%
\begin{figure}[t!]
		\centering
		\includegraphics[width=6.2cm,trim={0cm 0cm 0 0},clip]{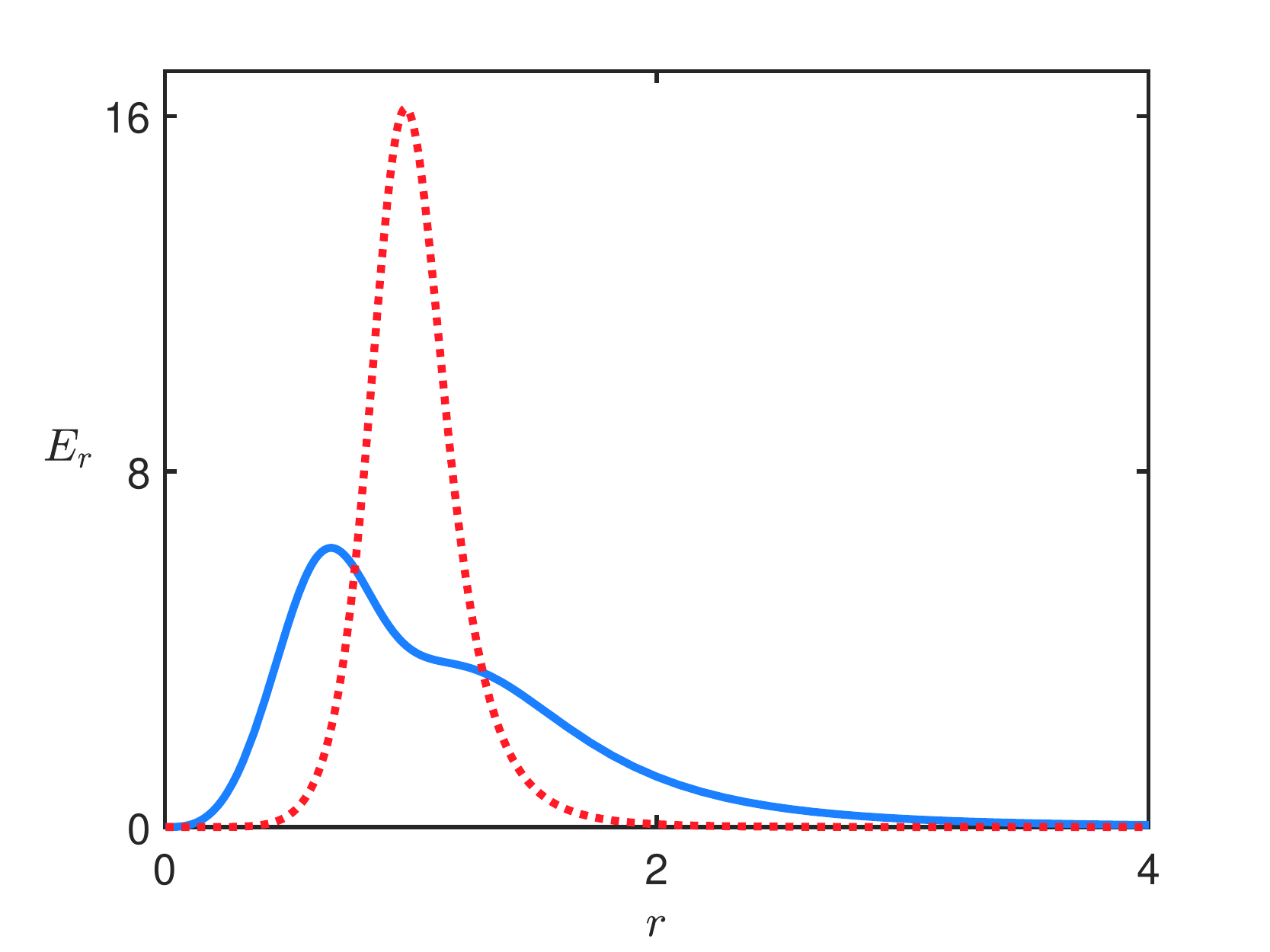}
		\includegraphics[width=6.2cm,trim={0cm 0cm 0 0},clip]{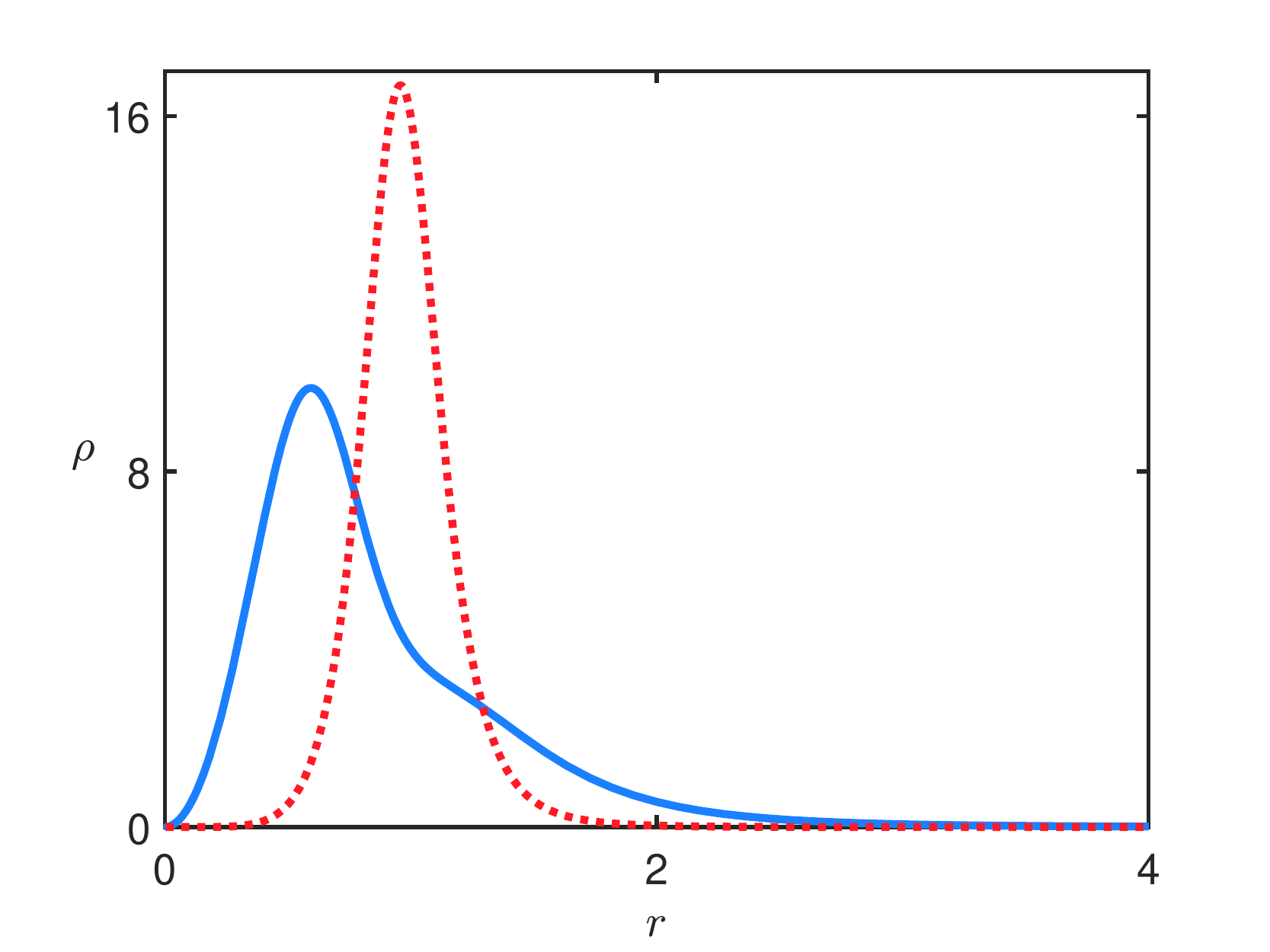}
		\caption{The radial component of the electric field (top) for $e=1$ and the energy density (bottom) associated to the model in Sec.~\ref{twospatial} for $\sigma=5$, with $k=1$ (solid, blue line) and $2$ (dotted, red line).}
		\label{figex3}
		\end{figure}
%%%%%%%%%%%%%%%%%%%
\be\label{energydensity35}
\begin{aligned}
    \rho(r)&=\frac{4k^2}{r^2}\bigg[\bigg(\frac{2r^{2k}}{r^{4k}+1}\bigg)^{4}\\
    &+\bigg(\frac{2r^{2k}}{r^{4k}+1}\bigg)^{2}\bigg(\frac{r^{4k}-1}{r^{4k}+1}\bigg)^{2}\bigg(\frac{\sigma}{k}-2\bigg)\bigg].
\end{aligned}
\ee
The electric field in Eq.~\eqref{Efield} with the permittivity $P(\phi,\chi)$ in Eq.~\eqref{permit} take the form
\be\label{1EFD2}\begin{aligned}
   E_r&=\frac{4k^2}{er}\bigg[\bigg(\frac{2r^{2k}}{r^{4k}+1}\bigg)^{4}\\
    &+\bigg(\frac{2r^{2k}}{r^{4k}+1}\bigg)^{2}\bigg(\frac{r^{4k}-1}{r^{4k}+1}\bigg)^{2}\bigg(\frac{\sigma}{k}-2\bigg)\bigg],
\end{aligned}
\ee
where we have taken $\textbf{E}(r)=E_{r}\hat{r}$. Note that, asymptotically, both the energy density and the electric field goes to zero. On the other hand, near the origin, the first and second terms of the energy density are proportional to $r^{8k-2}$ and $r^{4k-2}$, respectively. In order to avoid divergences in the energy density, we take $k>1/2$. Since $\sigma/k>2$ must be satisfied, this condition imposes that $\sigma>1$. We have checked that these conditions are also sufficient to preserve the finiteness and the single valued character of the electric field. In Fig.~\ref{figex3} we depict the electric field \eqref{1EFD2} and the energy density \eqref{energydensity35}. The electric field has a distinct behavior when compared to the standard expression in Eq.~\eqref{estd}, as it engenders a valley near the origin and vanishes asymptotically. Thus, the scalar fields that models the Bloch wall regularize the electric field (and the energy density) engendered by a single point charge, localizing it in a ringlike structure. This feature can be seen in Fig.~\ref{figex5}, which displays the electric field and the energy density in the plane. The size of hole at $r=0$ is controlled by $k$, which gives the intensity of the coupling between the fields; see Eq.~\eqref{newW}. As expected from Eq.~\eqref{ebogoelectric1}, the total energy of this structure is $E=8\pi\sigma/3$. 

%%%%%%%%%%%%%%%%%%%
%%%%%%%%%%%%%%%%%%%%
\subsection{Three spatial dimensions}\label{sec2mb}
\begin{figure}
		\centering
		\includegraphics[width=6.2cm,trim={0cm 0cm 0 0},clip]{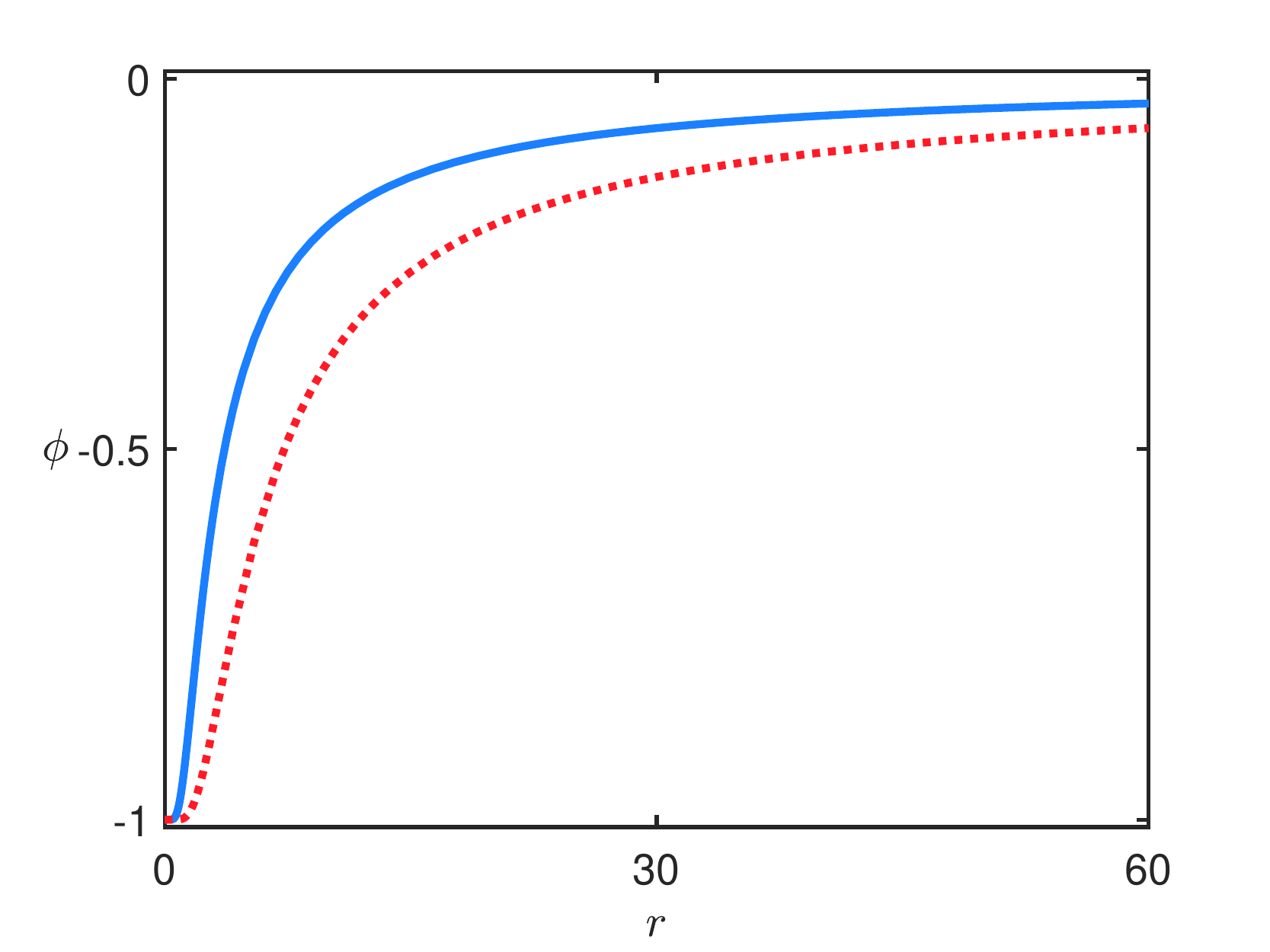}
		\includegraphics[width=6.2cm,trim={0cm 0cm 0 0},clip]{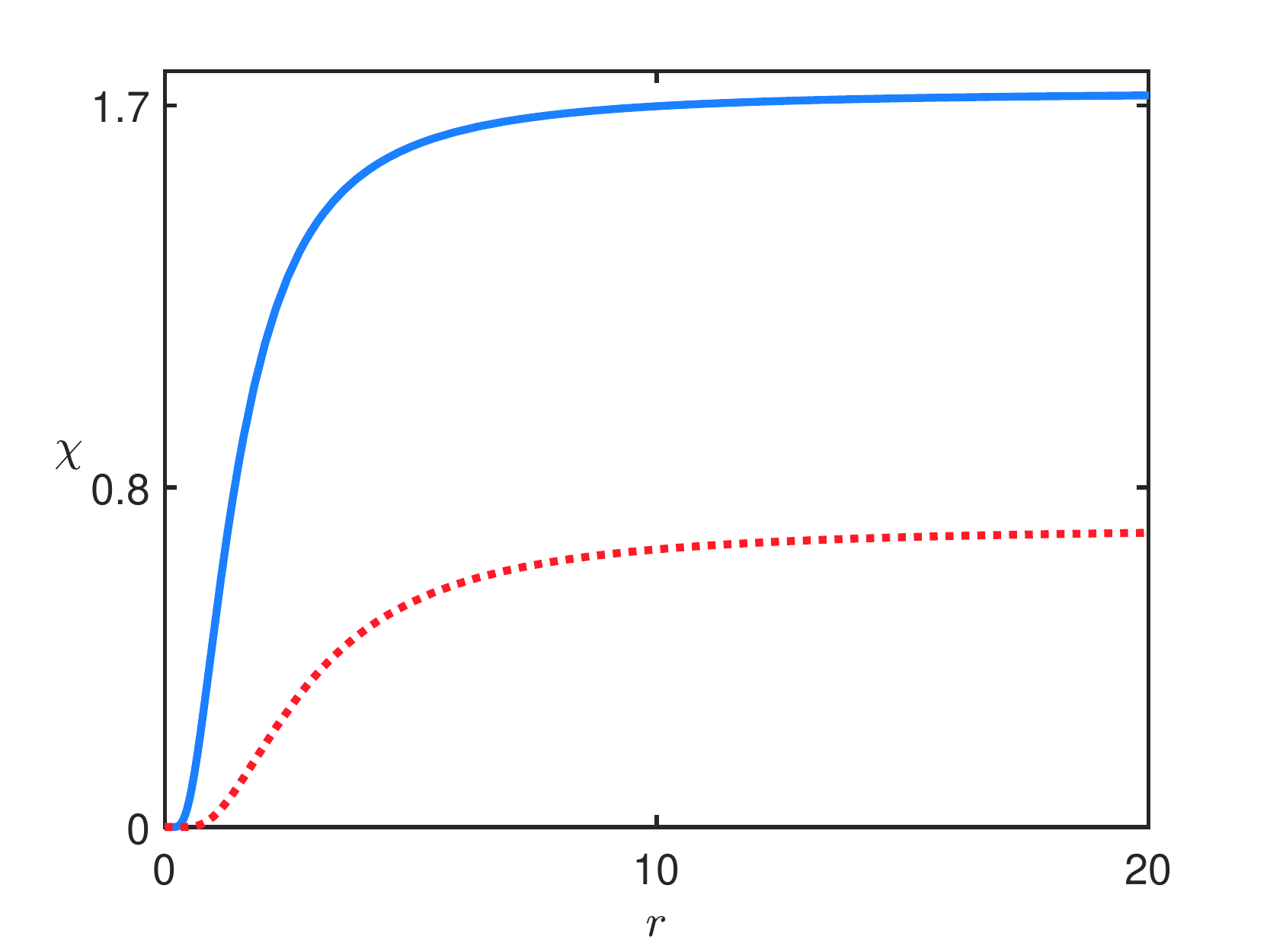}
		\caption{The solutions $\phi(r)$ (top) and $\chi(r)$ (bottom) associated to the model in Sec.~\ref{sec2mb} for $\sigma=5$, with $k=1$ (solid, blue line) and $2$ (dotted, red line).}
		\label{figex2}
		\end{figure}

We now investigate the structure with $D=3$. Following similar steps of the planar case ($D=2$), one gets the solutions
\bes
\bal
\label{1phi}
    &\phi(r)=\tanh(\xi(r))
    \\
    \label{1chi}
    &\chi(r)=\sqrt{\frac{\sigma}{k}-2}\;\sech(\xi(r)),
\eal
\ees
where $\xi(r)$ plays the role of a geometrical coordinate, with
\begin{align}
    \xi(r)=-\frac{2k}{r}.
\end{align}
We display the scalar fields in Fig.~\ref{figex2}. Notice that $\phi$ goes from $\phi=-1$ to $\phi=0$. Interestingly, $\chi$ becomes a monotonically increasing function in this scenario (differently from the planar case), connecting $\chi=0$ to $\chi=\sqrt{\sigma/k-2}$.
%%%%%%%%%%%%%%%%%%%
%%%%%%%%%%%%%%%%%%%
The energy density of this structure can be written in terms of the coordinate $\xi(r)$, as 
\begin{align}
\label{rho38}
    \rho(r)&\!=\!\frac{4k^2}{r^4}\bigg[\sech(\xi)^{4}+\tanh(\xi)^{2}\sech(\xi)^{2}\bigg(\frac{\sigma}{k}\!-2\bigg)\bigg],
\end{align}
%
%%%%%%%%%%%%
and the electric field is given by
\begin{align}
\label{electricf1}
   \! E_{r}(r)&\!=\!\frac{4k^2}{er^2}\!\bigg[\sech(\xi)^{4}\!+\tanh(\xi)^{2}\sech(\xi)^{2}\!\bigg(\frac{\sigma}{k}\!-2\bigg)\bigg]\! .
\end{align}
%%%%
%%%%%%%%%%%%%%%%%%%%%%
	\begin{figure}
		\centering
		\includegraphics[width=6.2cm,trim={0cm 0cm 0 0},clip]{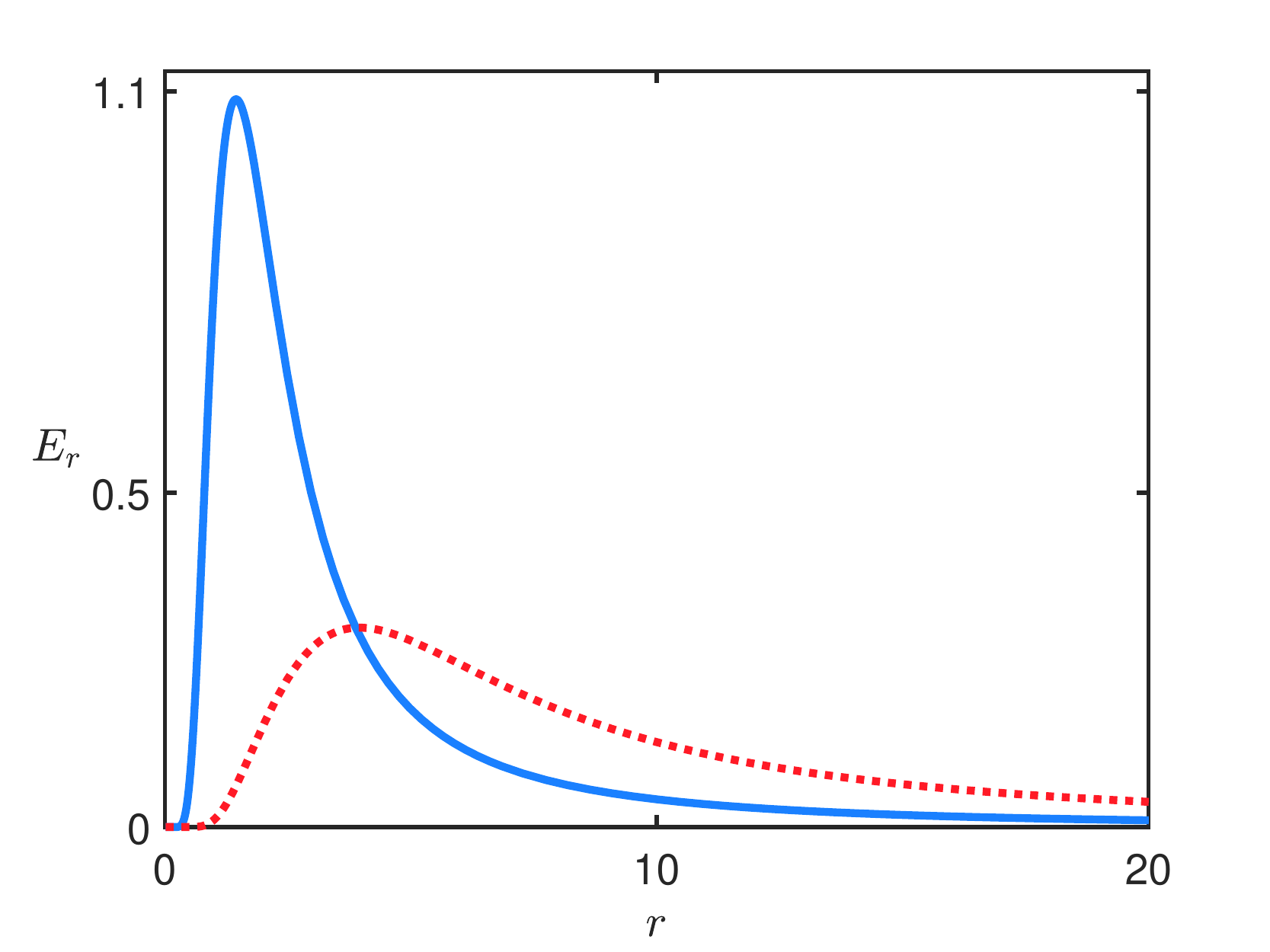}
		\includegraphics[width=6.2cm,trim={0cm 0cm 0 0},clip]{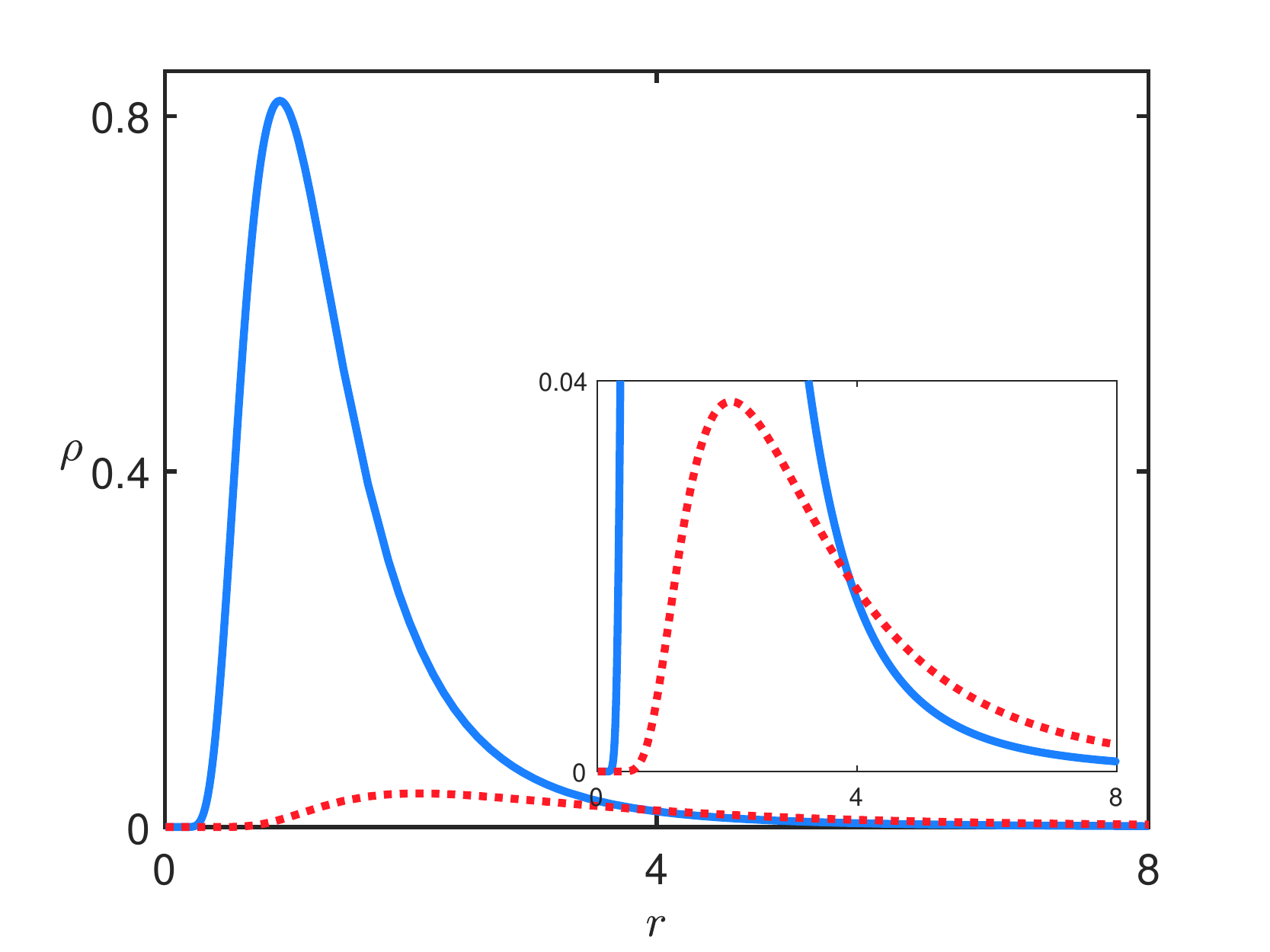}
		\caption{The radial component of the electric field (top) with $e=1$ and the energy density (bottom) associated to the model in Sec.~\ref{sec2mb} for $\sigma=5$, with $k=1$ (solid, blue line) and $2$ (dotted, red line). The inset in the bottom figure shows the interval $\rho\in[0,0.04]$.}
		\label{figex4}
		\end{figure}

\begin{figure}[t!]
		\centering
		\includegraphics[width=4.27cm,trim={0cm 0cm 0 0},clip]{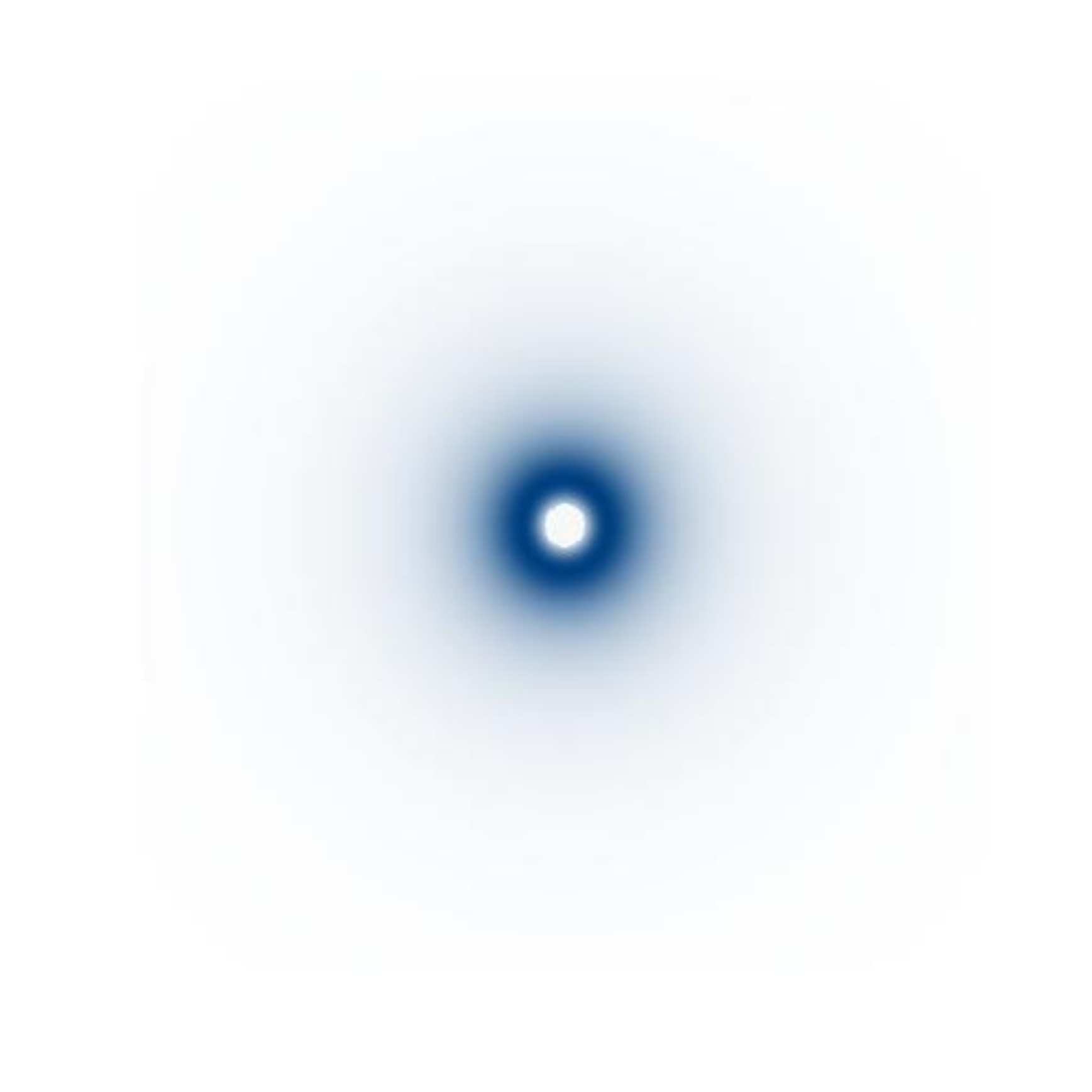}
		\includegraphics[width=4.27cm,trim={0cm 0cm 0 0},clip]{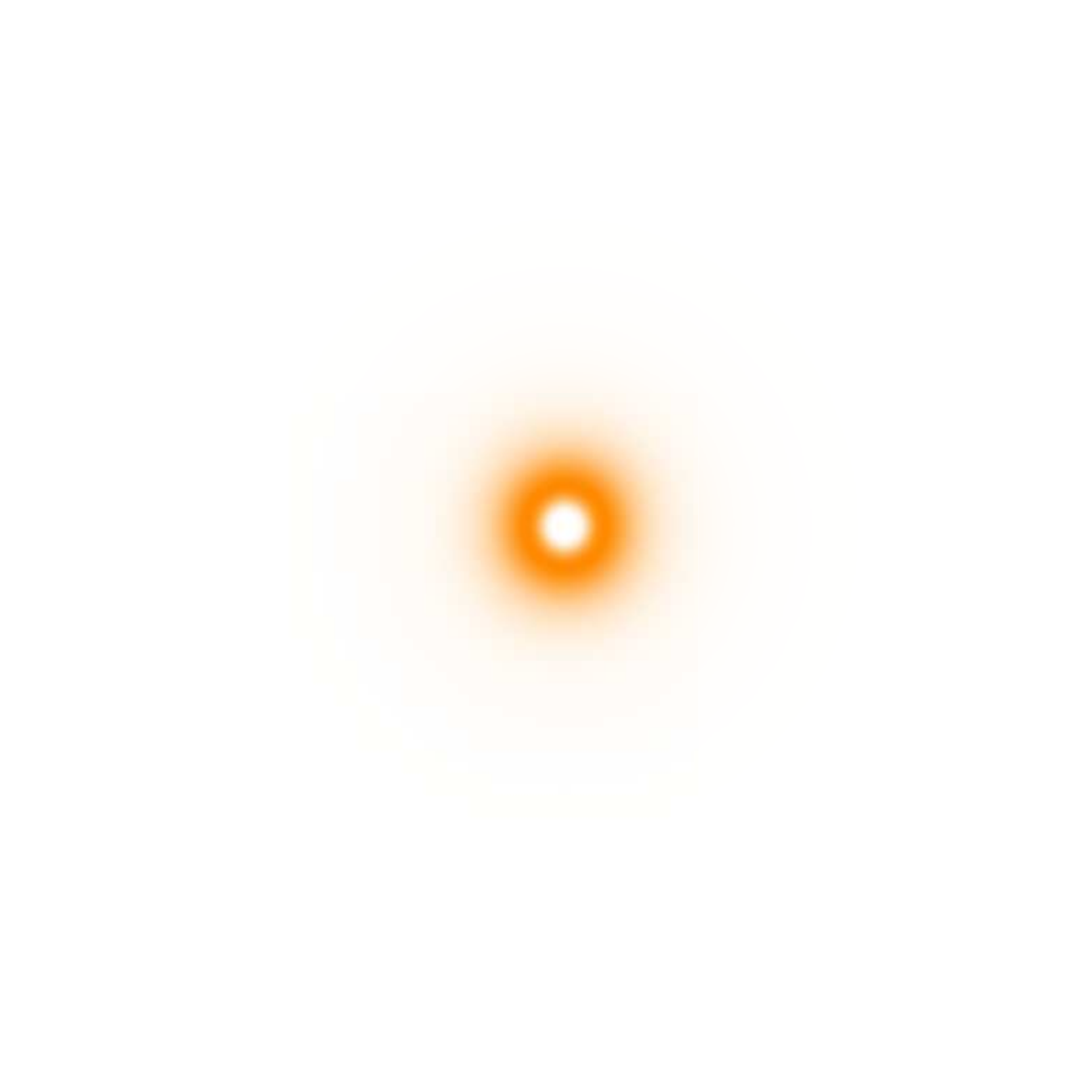}
		\caption{The electric field in the plane (left, blue) with $e=1$ and the energy density (right, orange) associated to the model in Sec.~\ref{sec2mb} for $\sigma=5$ and $k=1$. The intensity of the blue and orange colors  increases with the increasing of the electric field and energy density, respectively.}
		\label{Efielddensi}
		\end{figure}

In Fig.~\ref{figex4} we display the electric field \eqref{electricf1} and the energy density \eqref{rho38} for some values of $k$ and in Fig.~\ref{Efielddensi} we illustrate their behavior in the plane passing through the center of the structure. Notice that, contrary to the $D=2$ case, these physical quantities are finite for any value in the allowed range of $k$. This means that $\sigma$ is not required here to avoid divergences and to ensure that the electric field is single valued. The total energy of this structure is $E=8\pi\sigma/3$.

		%%%%%%
%%%%%%%%%%%%%%%%%%%%

\section{Internal structure of charged configurations}
\label{sec4}
 In the previous section, we have shown that the permittivity controlled by scalar fields can regularize the behavior of the electric field generated by a single point charge, localizing it in a ringlike structure. Modifications on the profile of kinklike structures can be made through a function that changes the kinematic term \cite{multikink}. Inspired by this result, we investigate another model with three scalar fields described by the Lagrange density 
\be\label{lele}\begin{aligned}
   \mathcal{L}&=-\frac{1}{4}P(\phi, \chi, \psi)F_{\mu\nu}F^{\mu\nu}+\frac{1}{2}f(\psi)\partial_{\mu}\phi\partial^{\mu}\phi\\
    &+\frac{1}{2}f(\psi)\partial_{\mu}\chi\partial^{\mu}\chi+\frac{1}{2}\partial_{\mu}\psi\partial^{\mu}\psi -A_{\mu}j^{\mu},
\end{aligned}
\ee
where $f(\psi)$ is a real and positive function. Notice that the field $\psi$ is dynamical and was introduced to modify the kinematics of the fields $\phi$ and $\chi$ associated to the Bloch wall. The scalar fields are also coupled in the electric permittivity, $P(\phi, \chi, \psi)$. We now have to solve the corresponding three equations of motion:
\bes\begin{align}
   &\label{eqm33}\partial_{\mu}(f\partial^{\mu}\phi)+\frac{P_{\phi}}{4}F_{\mu\nu}F^{\mu\nu}=0,\\
    &\label{eqm34}\partial_{\mu}(f\partial^{\mu}\chi)+\frac{P_{\chi}}{4}F_{\mu\nu}F^{\mu\nu}=0,\\
   &\label{eqm35}\partial_{\mu}\partial^{\mu}\psi+\frac{P_{\psi}}{4}F_{\mu\nu}F^{\mu\nu}-\frac{f_{\psi}}{2}(\partial_{\mu}\phi\partial^{\mu}\phi+\partial_{\mu}\chi\partial^{\mu}\chi)=0.
\end{align}\ees
As before, let us consider static fields with the current given as in Eq.~\eqref{current}. By solving Gauss' law, we get the electric field in Eq.~\eqref{Efield} with the permittivity modified to $P(\phi,\chi,\psi)$. The set of equations for static fields is written as 
\bes\begin{align}
   & \label{eqm33}\frac{1}{r^{D-1}}(r^{D-1}f\phi ')'-\frac{e^2}{2r^{2D-2}}\frac{\partial }{\partial\phi}\bigg(\frac{1}{P}\bigg)=0,\\
   & \label{eqm34}\frac{1}{r^{D-1}}(r^{D-1}f\chi ')'-\frac{e^2}{2r^{2D-2}}\frac{\partial }{\partial\chi}\bigg(\frac{1}{P}\bigg)=0,\\
   & \label{eqm35}\frac{1}{r^{D-1}}(r^{D-1}\psi ')'-\frac{e^2}{2r^{2D-2}}\frac{\partial }{\partial\psi}\bigg(\frac{1}{P}\bigg)-\frac{f_{\psi}}{2}(\chi'^{2}  +\phi'^{2})=0.
\end{align}\ees
These equations are independent from the gauge field. However, they present couplings between the scalar fields. The inclusion of $\psi$ in the Lagrangian density makes the problem harder than before, as one now has to solve three nonlinear differential equations of second order. We then take advantage of the procedure in Ref.~\cite{Bo} to find first order equations to describe our model. The energy density associated to the field configuration has the form
\be
\label{energyd22}
\begin{split}
	%\rho &= \frac{f}{2}({\phi'}^{2}+{\chi'}^{2})+\frac{1}{2}{\psi'}^{2}+\frac{P}{2}{|\textbf{E}|}^{2}\\
		 \rho&= \frac{f}{2}({\phi'}^{2}+{\chi'}^{2})+\frac{1}{2}{\psi'}^{2}+\frac{1}{2P}\frac{e^2}{r^{2D-2}}.
\end{split}
\ee
If one consider the function $W=W(\phi,\chi,\psi)$ and the permittivity is written as 
\begin{equation}
\label{permit2}
    P(\phi,\chi,\psi)=e^{2}\bigg(\frac{W^{2}_{\phi}}{f(\psi)}+\frac{W^{2}_{\chi}}{f(\psi)}+W^{2}_{\psi}\bigg)^{-1},
\end{equation}
%Thus the energy density \eqref{energyd22} can be written as
%\be\begin{aligned}   
%   \label{edensity2}
 %   \rho &=\frac{f}{2}\bigg(\phi'\mp\frac{W_{\phi}}{fr^{D-1}}\bigg)^{2}+\frac{f}{2}\bigg(\chi'\mp\frac{W_{\chi}}{fr^{D-1}}\bigg)^{2}\\
 %		 & +\frac{1}{2}\bigg(\psi'\mp\frac{W_{\psi}}{r^{D-1}}\bigg)^{2}\pm \frac{1}{r^{D-1}} W'.
%\end{aligned}
%\ee
%By integrating this expression, one gets that the energy is bounded, that is,
%\be\label{ebogoelectric}
%E\geq E_B = \Omega(D)\left|\Delta W\right|,
%\ee
%for $\Delta W = W(\phi(\infty),\chi(\infty),\psi(\infty))-W(\phi(0),\chi(0),\psi(0))$. If the first order equations
the equations of motion are compatible with the first order equations
\bes\label{solutionsD}
\bal
\label{phisolutionD}
&\phi'=\pm\frac{W_{\phi}}{fr^{D-1}},\\
\label{chisolutionD}
&\chi'=\pm\frac{W_{\chi}}{fr^{D-1}},\\
\label{psisolutionD}
&\psi'=\pm \frac{W_{\psi}}{r^{D-1}}.
\eal
\ees
In this situation, the energy is minimized to $E_B=\Omega(D)\left|\Delta W\right|$, where $\Delta W = W(\phi(\infty),\chi(\infty),\psi(\infty))-W(\phi(0),\chi(0),\psi(0))$. By looking at Eqs.~\eqref{phisolutionD} and \eqref{chisolutionD}, we see that they contain the function $f(\psi)$. On the other hand, Eq.~\eqref{psisolutionD} is only driven by $W_\psi$, admitting a special case, for $W(\phi,\chi,\psi)=W_{1}(\phi,\chi)+W_{2}(\psi)$. This specific form of $W$ makes the first order equation \eqref{psisolutionD} become independent of $\phi$ and $\chi$, having the very same form of the first order equations for kinklike solutions investigated in Ref.~\cite{prlbmm}. By knowing the form of $\psi$, one can use it to feed the function $f(\psi)$, which can be seen as a source of geometrical constrictions \cite{multikink} in the Bloch wall, modeled by the solutions $\phi$ and $\chi$. The independence of $\psi$ can also be seen in the energy density \eqref{energyd22}, which can be written as the sum of two contributions, $\rho = \rho_1+\rho_2$, with
\bes
\bal\label{rho1ele}
\rho_1&= f(\psi)({\phi'}^{2}+{\chi'}^{2}),\\ \label{rho2ele}
\rho_2&={\psi'}^{2}.
\eal
\ees
%%%%%%%
	\begin{figure}
		\centering
		\includegraphics[width=6.2cm,trim={0cm 0cm 0 0},clip]{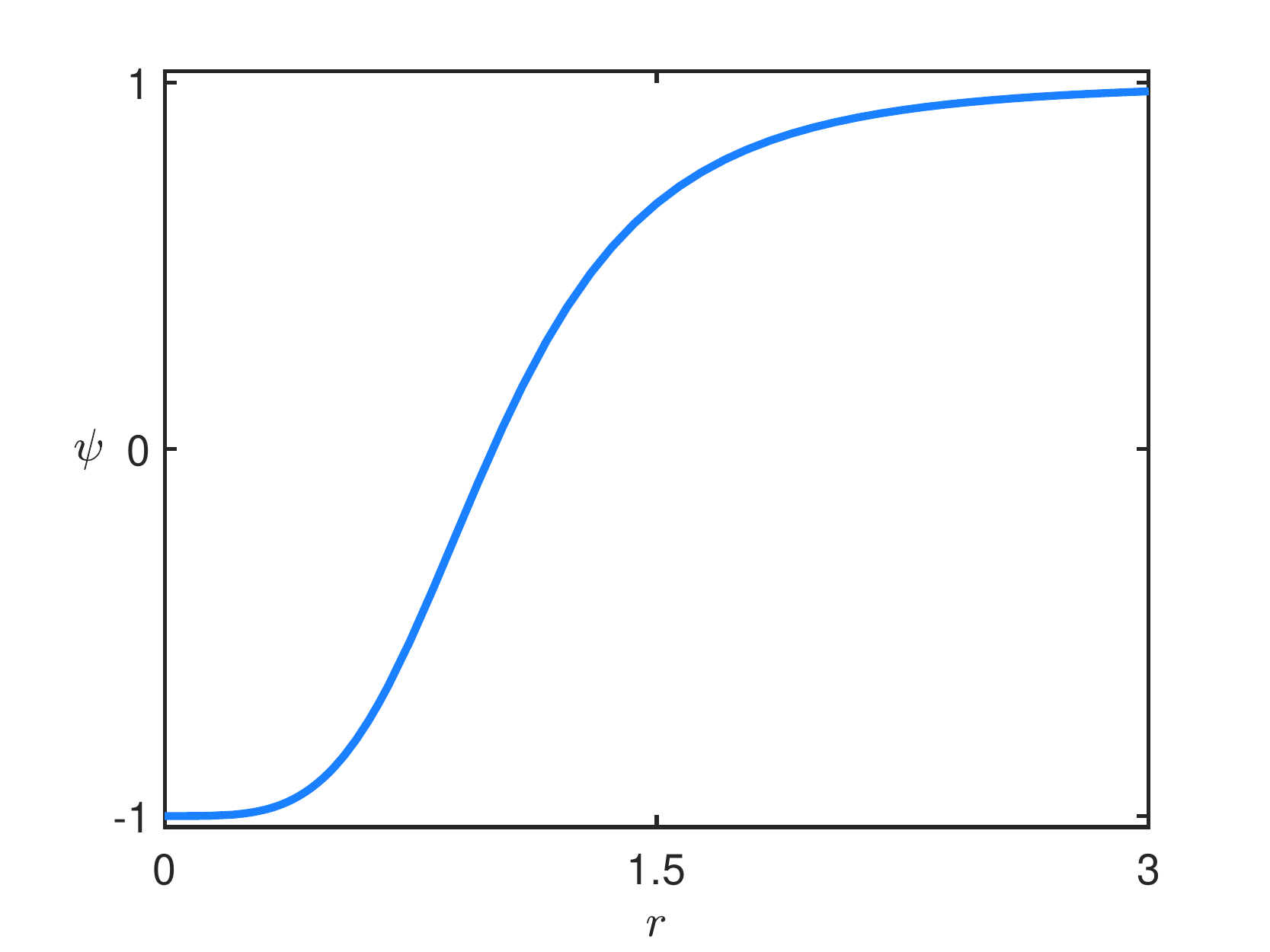}
		\includegraphics[width=6.2cm,trim={0cm 0cm 0 0},clip]{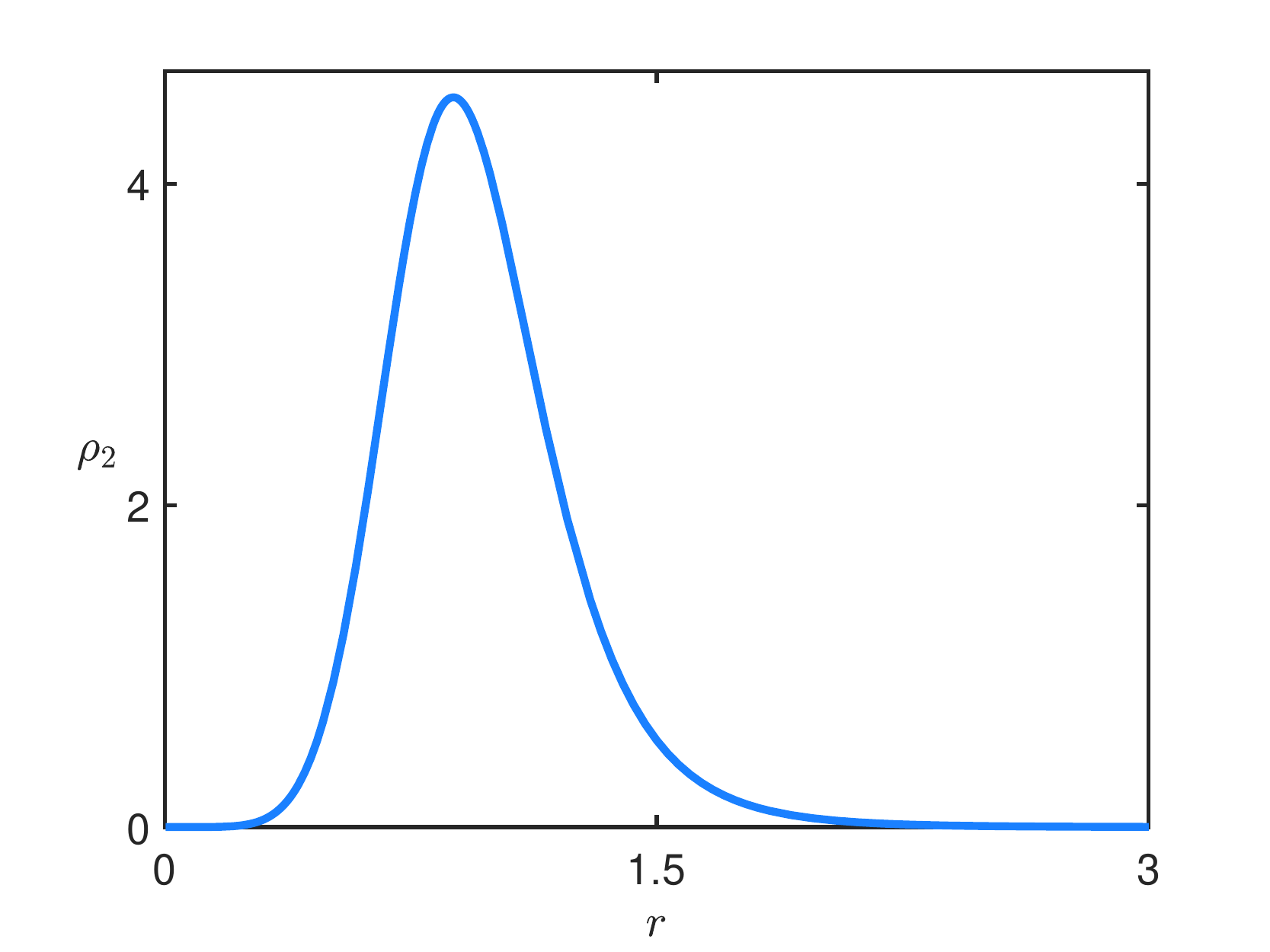}
		\caption{The solutions $\psi(r)$ in Eq.~\eqref{psid2fpsi2} (top), and the energy density in Eq.~\eqref{energydensityd2fpsi22} (bottom) for $\alpha=2$.}
		\label{psid2densityd2}
		\end{figure}
		%%%%%%
%%%%%%%%%%%%%%%%%%%%%%%%%
		\begin{figure}[t!]
		\centering
		\includegraphics[width=6.2cm,trim={0cm 0cm 0 0},clip]{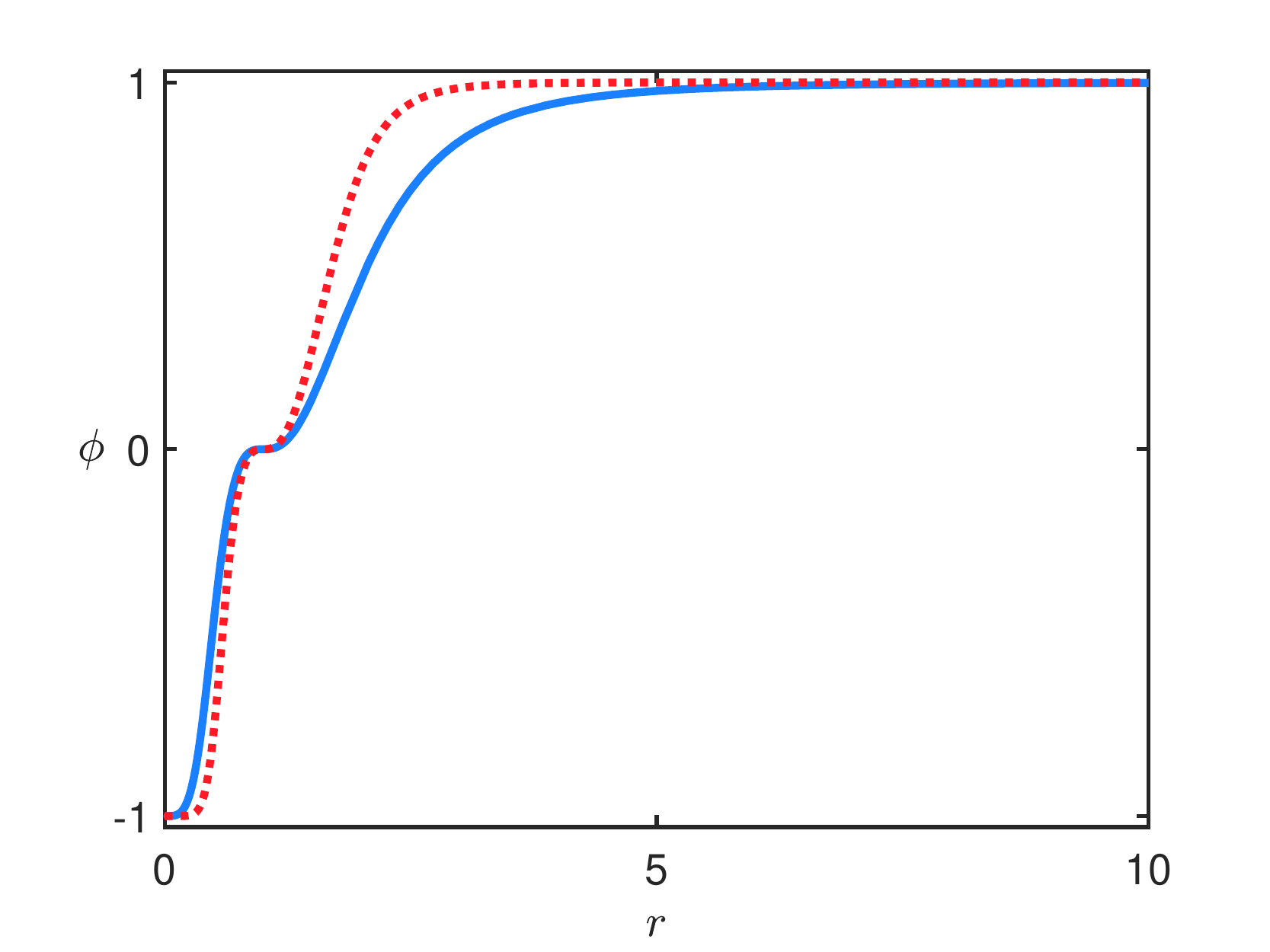}
		\includegraphics[width=6.2cm,trim={0cm 0cm 0 0},clip]{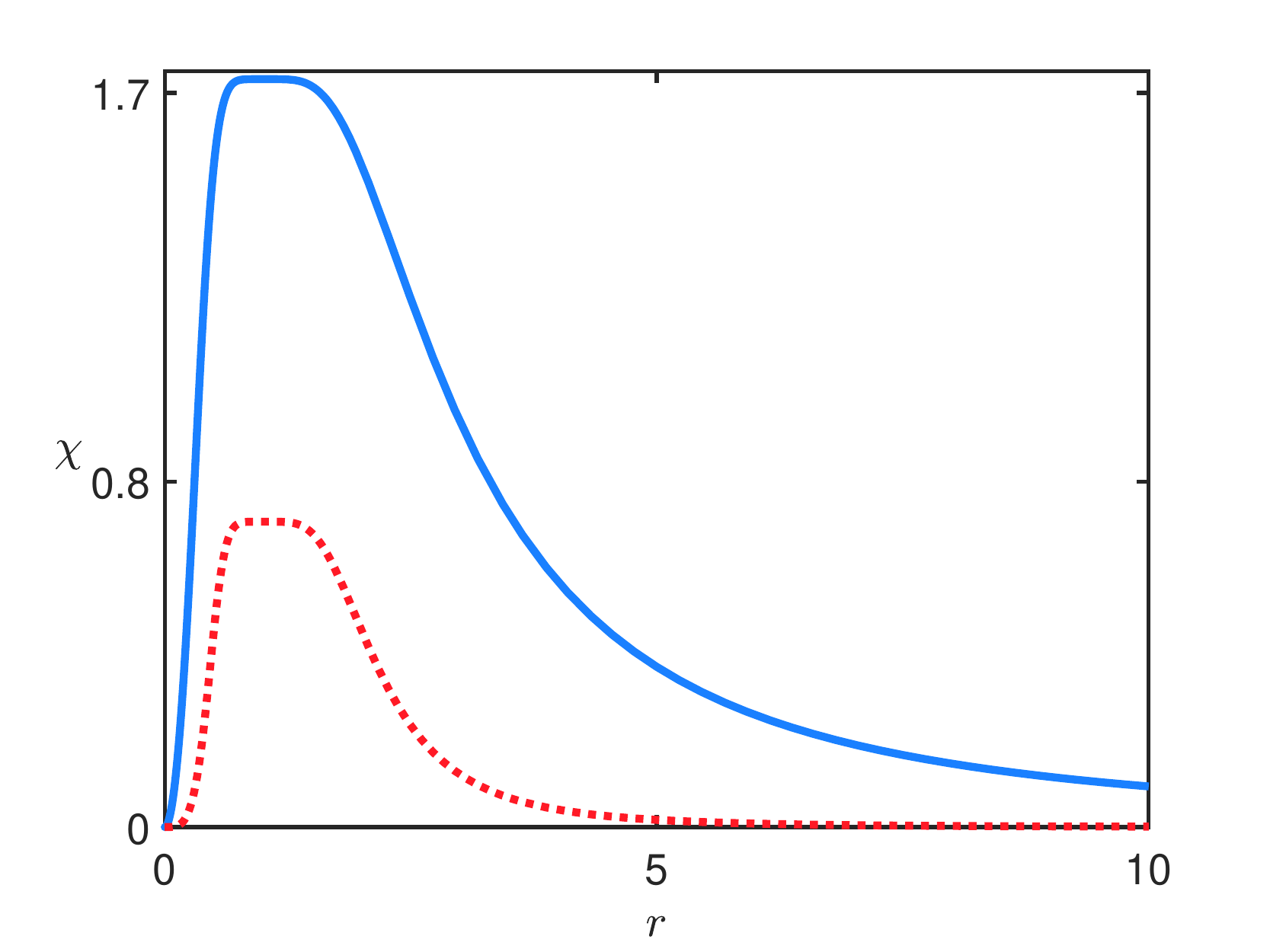}
		\caption{The solutions $\phi(r)$ (top) and $\chi(r)$ (bottom) associated to the model in Sec.~\ref{model2da} for $\sigma=5$ and $\alpha=2$, with $k=1$ (solid, blue line) and $2$ (dotted, red line).}
		\label{figdensityn2}
		\end{figure}
%%%%%%%%%%%%%%%%%%%%%%%%%%%
To investigate how the inclusion of $\psi$ modifies the charged structure, we take $W_1(\phi,\chi)$ as the auxiliary function shown in Eq.~\eqref{newW} and
%%%%%%
\bes
\bal
\label{W1b}
W_{2}=\alpha\psi-\frac{1}{3}\alpha\psi^{3},
\eal
\ees
%%%%
where $\alpha$ is a real and positive parameter. The first order equations \eqref{phisolutionD}-\eqref{psisolutionD} in this scenario are given by 
%%%%
\bes
\bal
\label{phisolutionDn}
&\phi'=\pm\frac{\sigma-\sigma\phi^{2}-k\chi^{2}}{fr^{D-1}},\\
\label{chisolutionDn}
&\chi'=\pm\frac{2k\phi\chi}{fr^{D-1}},\\
\label{psisolutionDn}
&\psi'=\pm \frac{\alpha(1-\psi^{2})}{r^{D-1}}.
\eal
\ees
%%%%
One can show that the first order equations \eqref{phisolutionDn} and \eqref{chisolutionDn} support the same orbit in Eq.~\eqref{orbit1}, which leads us 
\be\label{fophiorbitD}
\phi'=\pm\frac{2k(1-\phi^{2})}{f(\psi)r^{D-1}}.
\ee
Notice the presence of $f(\psi)$ in this expression. As shown in Ref.~\cite{multikink}, it can be associated to geometrical constrictions in the solutions. 

Let us now illustrate the above features, exploring distinct models in two and three spatial dimensions.

%%%%%%%%
\subsection{Two spatial dimensions}\label{secftwo}
We first study the electrically charged structure in the special case with two spatial dimensions, $D=2$. The solution $\psi(r)$ that feeds the function $f(r)$ can be found from Eq.~\eqref{psisolutionDn}, which leads us to
\begin{align}
    \label{psid2fpsi2}
\psi(r)=\frac{r^{2\alpha}-1}{r^{2\alpha}+1},
\end{align}
where we have taken the condition $\psi(1)=0$, for simplicity. Its associated energy density \eqref{rho2ele} has the form
\begin{align}
     \label{energydensityd2fpsi22}
   \rho_{2}&=\frac{16\alpha^{2}r^{4\alpha-2}}{(r^{2\alpha}+1)^4}.
\end{align}
Notice that, the parameter $\alpha$ plays an important hole here as it may lead to a divergent behavior at the origin. To avoid divergences at $r=0$, we take $\alpha\geq0.5$. In Fig.~\ref{psid2densityd2} we plot the solution \eqref{psid2fpsi2} and the energy density \eqref{energydensityd2fpsi22} for $\alpha=2$. We then use this solution to feed the function $f(\psi)$ in the models which we present below.

%%%%%%%%%%%%%%%%%%%%%%%%%%%%%%%%%%%
\subsubsection{First model}\label{model2da}
%%%%%%%%%%%%%%%%%%%%%%%%%%%%%%%%%%%
The first model of interest arises with $f=1/\psi^2$. In this case, the fields associated to the Bloch wall are similar to Eqs.~\eqref{1phi} and \eqref{1chi}, with the geometrical coordinate $\xi(r)$ now given by
%
%\bes
%\bal
%\label{phi50a}
%    \phi(r)&=\tanh(\xi(r)) \\
%    \label{chi50a}
%    \chi(r)&=\sqrt{\frac{\sigma}{k}-2}\sech(\xi(r)),
%\eal
%\ees
%
\begin{equation}
    \xi(r)=2k\bigg(\ln(r)-\frac{1}{\alpha}\frac{r^{2\alpha}-1}{r^{2\alpha}+1}\bigg).
\end{equation}

%%%%%%%%%%%%%
\begin{figure}[t!]
		\centering
		\includegraphics[width=6.2cm,trim={0cm 0cm 0 0},clip]{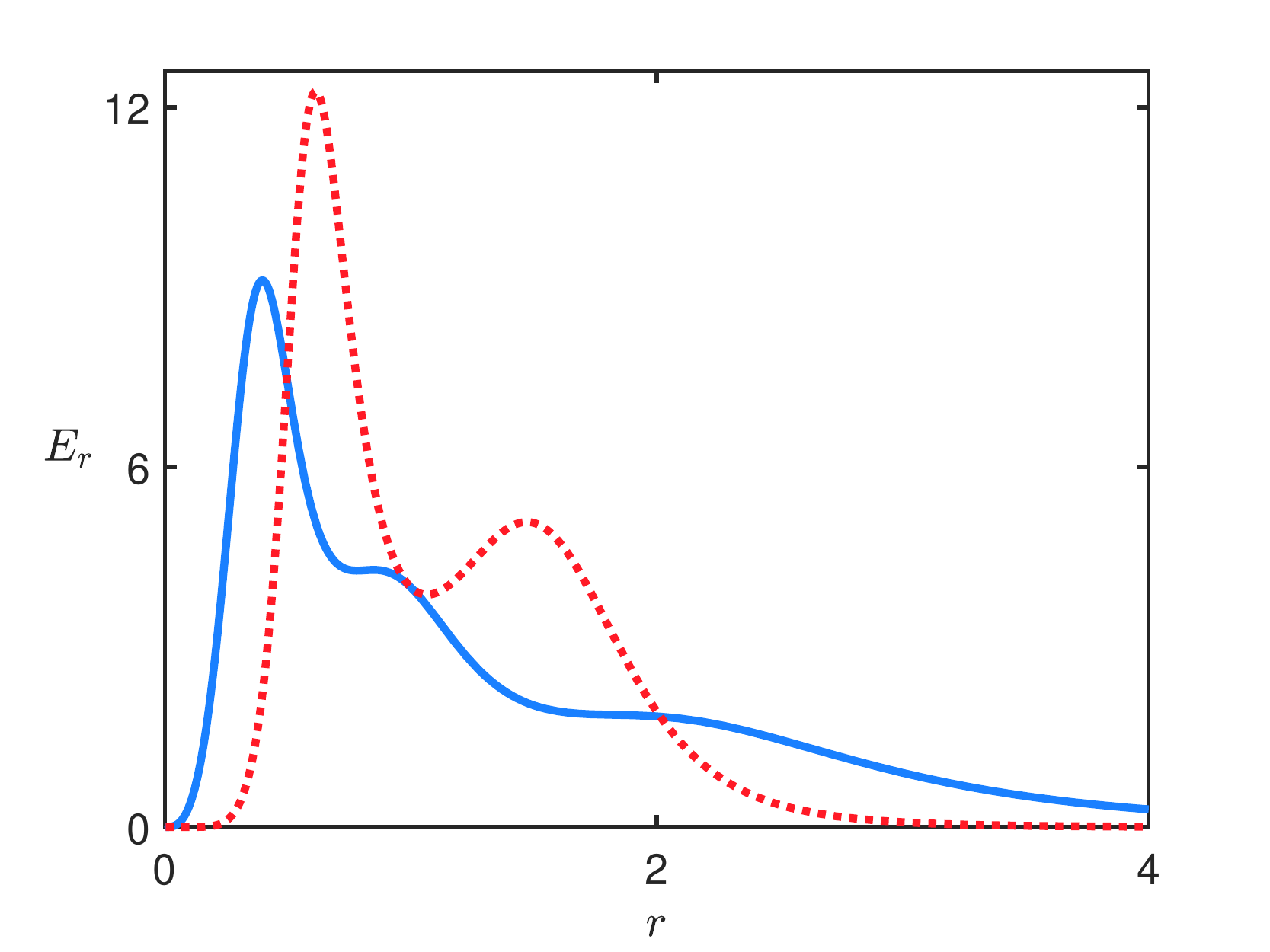}
		\includegraphics[width=6.2cm,trim={0cm 0cm 0 0},clip]{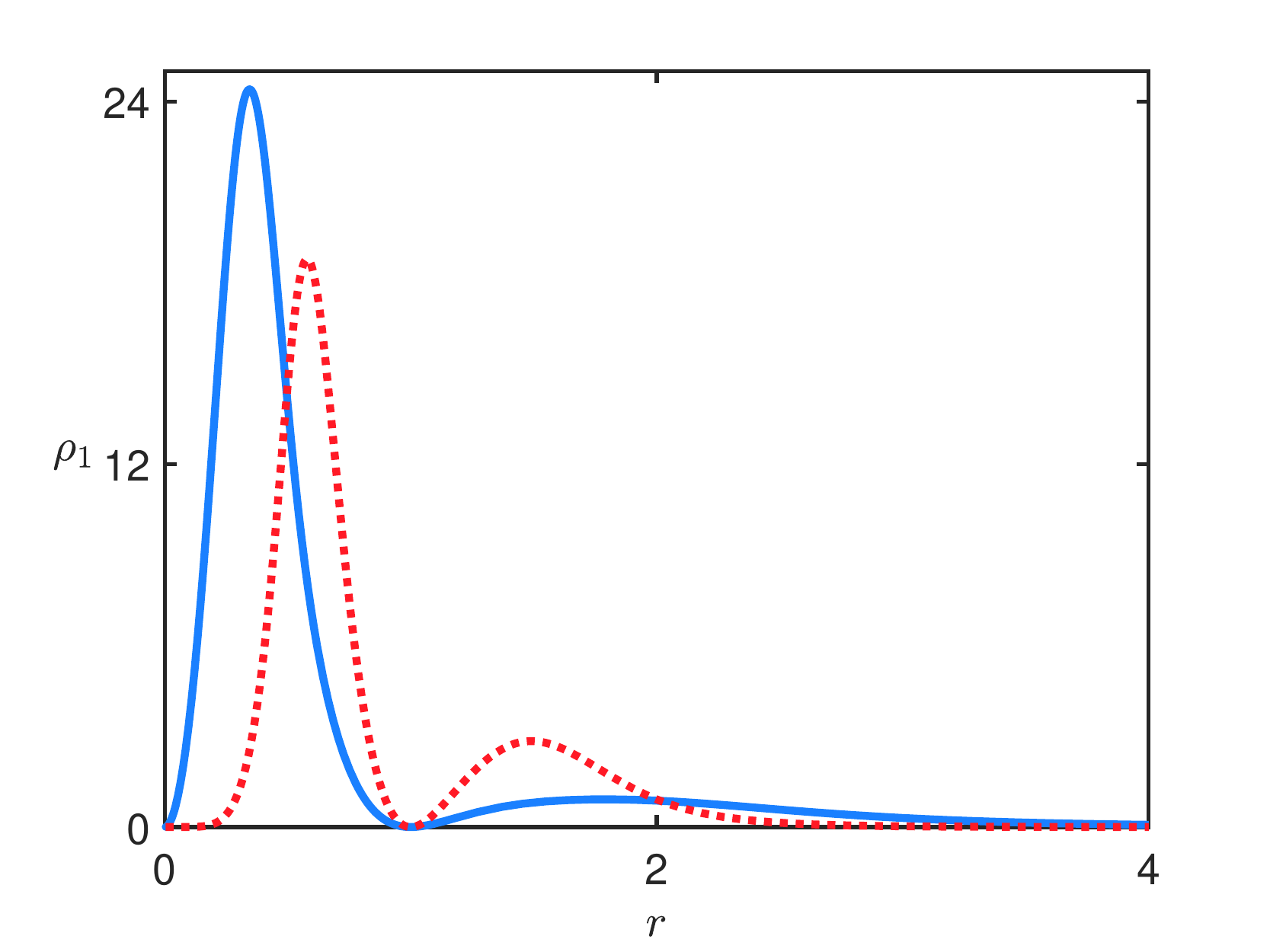}
		\caption{The radial component of the electric field (top) with $e=1$ and the energy density (bottom) associated to the model in Sec.~\ref{model2da} for $\sigma=5$, $\alpha=2$, with $k=1$ (solid, blue line) and $2$ (dotted, red line).}
		\label{figEFfpsi222}
		\end{figure}
%%%%%%%%%%%%

\begin{figure}[t!]
		\centering
		\includegraphics[width=4.27cm,trim={0cm 0cm 0 0},clip]{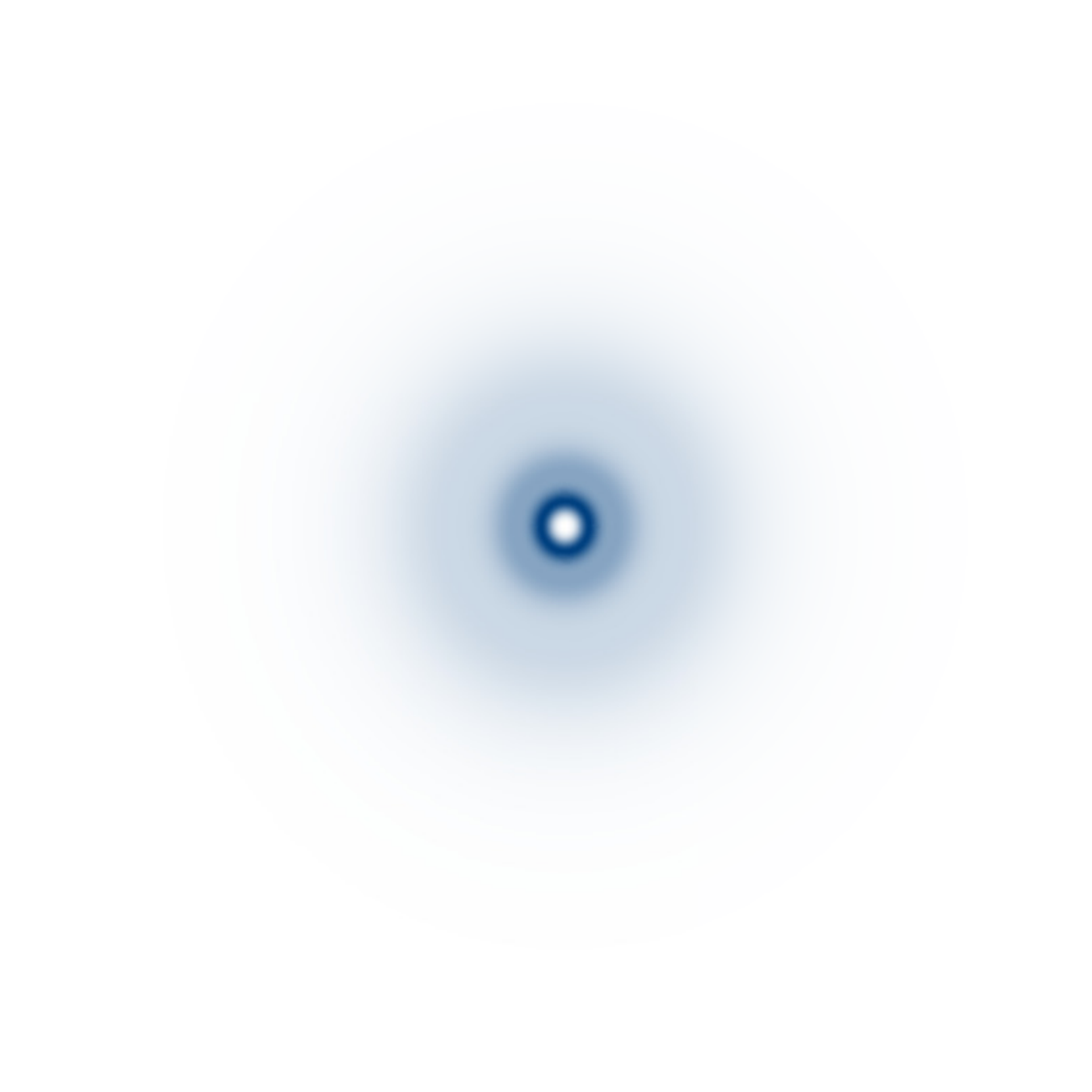}
		\includegraphics[width=4.27cm,trim={0cm 0cm 0 0},clip]{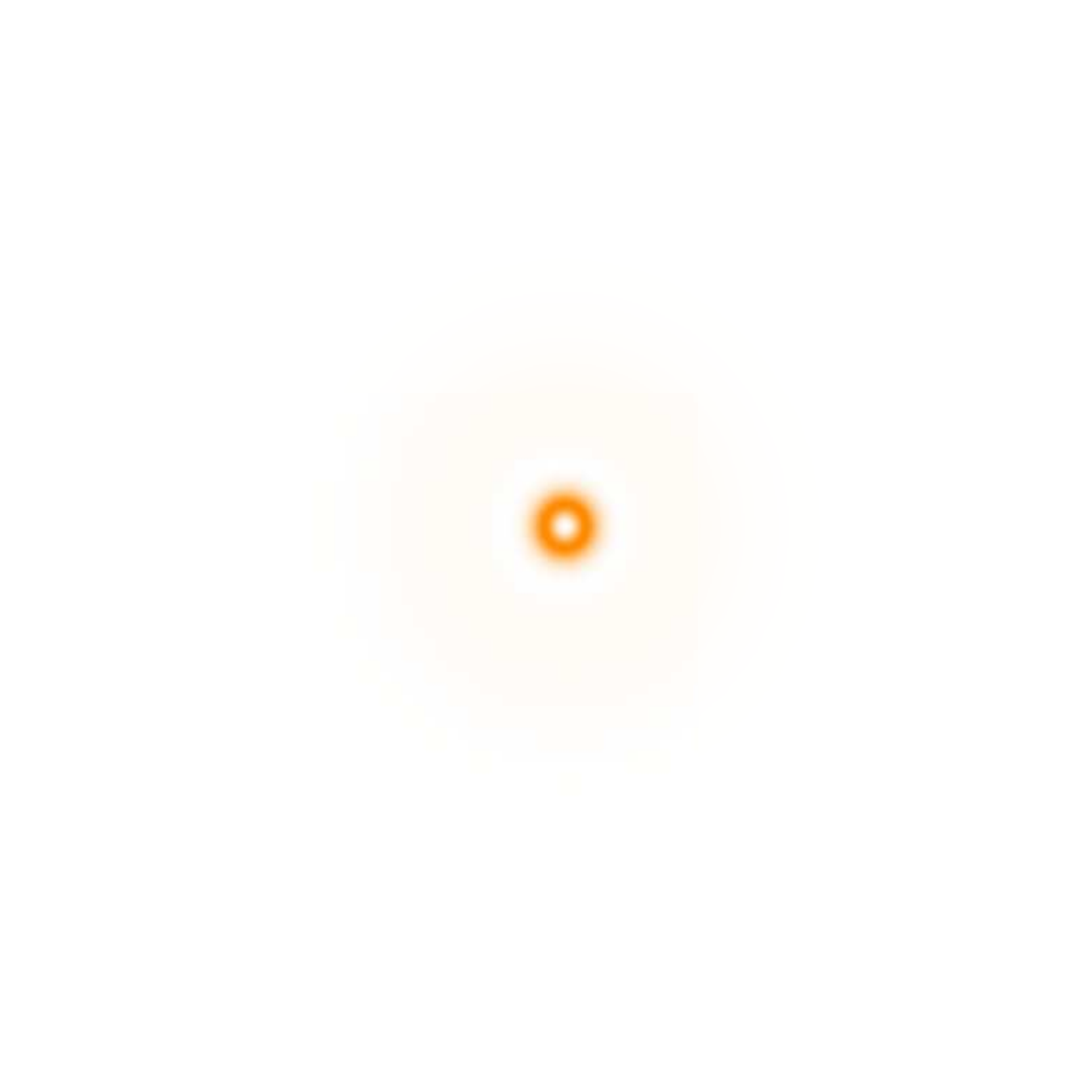}
		\includegraphics[width=4.27cm,trim={0cm 0cm 0 0},clip]{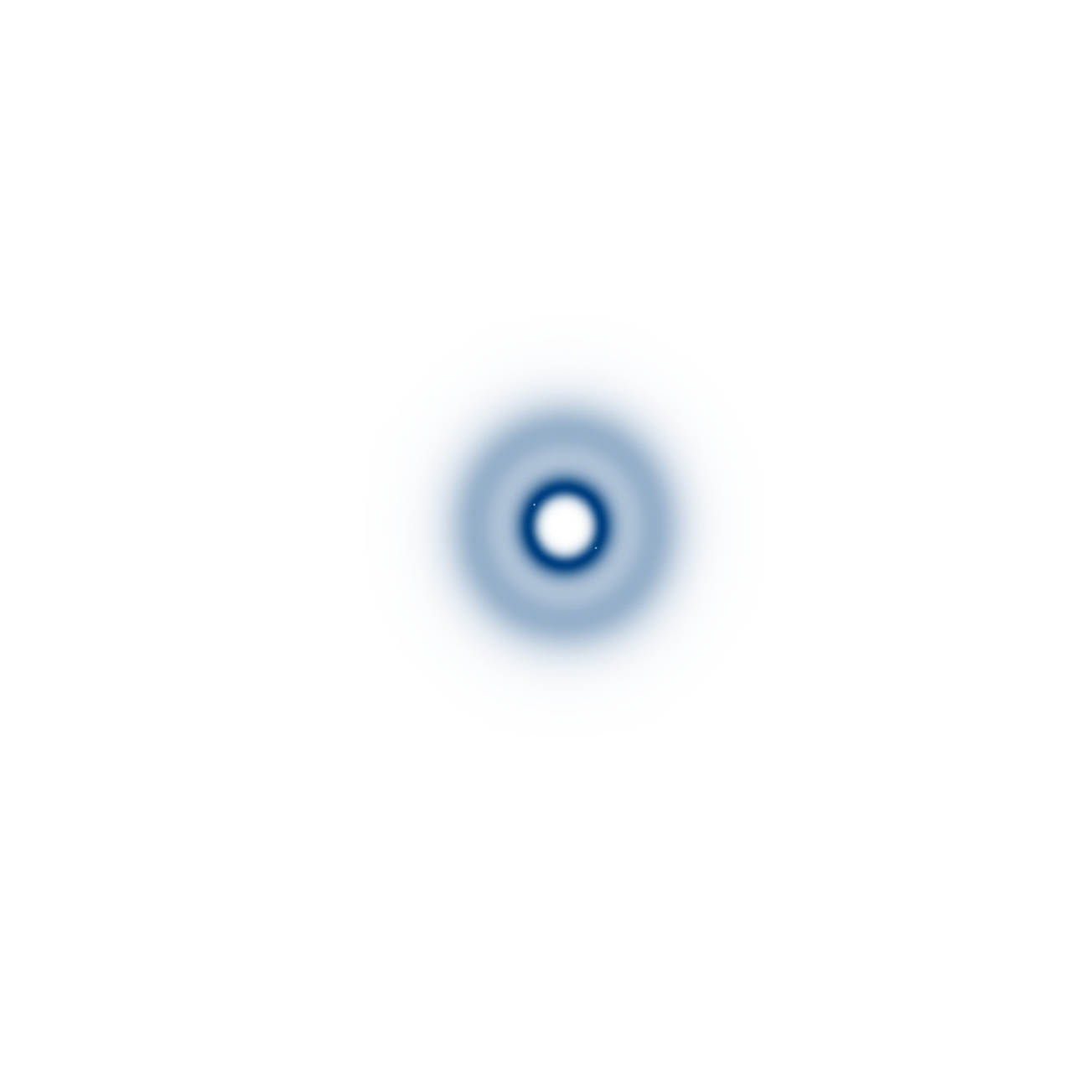}
		\includegraphics[width=4.27cm,trim={0cm 0cm 0 0},clip]{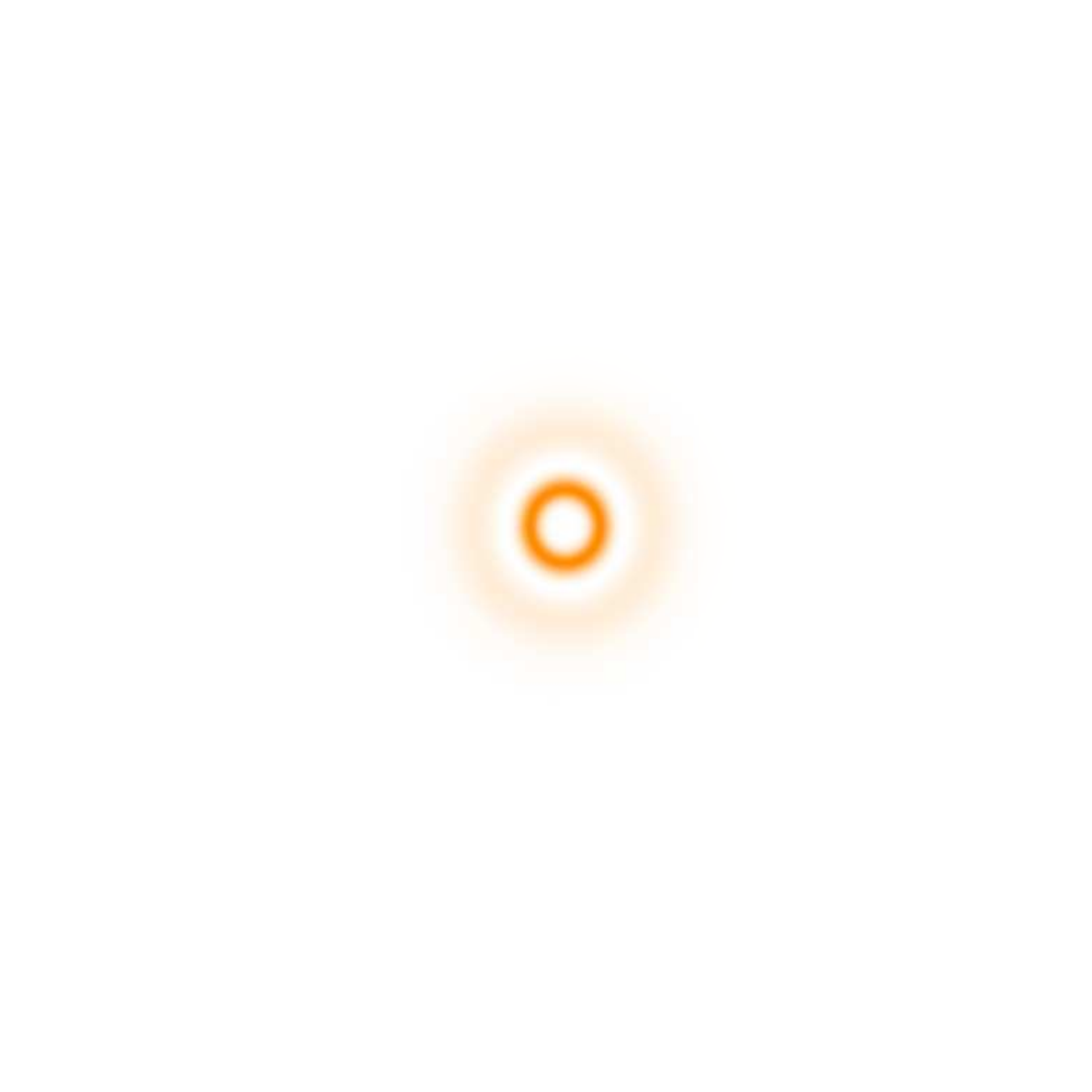}
		\caption{The radial component of the electric field (left, blue) with $e=1$ and the energy density (right, orange) associated to the model in Sec.~\ref{model2da} depicted in the plane for $\sigma=5$ and $\alpha=2$, with $k=1$ (top), and $k=2$ (bottom). The intensity of the blue and orange colors increases with the increasing of the electric field and energy density, respectively.}
		\label{figEFfpsi2}
		\end{figure}

The conditions for $\sigma$ and $k$ are the same of the ones discussed in Sec.~\ref{twospatial}. The solutions $\phi(r)$ and $\chi(r)$ are displayed in Fig.~\ref{figdensityn2} for $\sigma=5$, $\alpha=2$ and some values of $k$. The energy density of these solutions, $\rho_1$, is given by Eq.~\eqref{rho1ele}, which leads us to
\be\begin{aligned}
    \label{energydensityd2fpsi2}
   \rho_{1}&=\frac{4k^{2}}{r^2}\bigg(\frac{r^{2\alpha}-1}{r^{2\alpha}+1}\bigg)^{2}\bigg({\sech}^{4}{(\xi)}\\
   &+{\sech}^{2}{(\xi)}{\tanh}^{2}{(\xi)}\bigg(\frac{\sigma}{k}-2\bigg)\bigg).
\end{aligned}
\ee
The electric field, which we write again in the form $\textbf{E}(r) = E_r \hat{r}$, is obtained from Eq.~\eqref{estd} with permittivity in Eq.~\eqref{permit2}. Its expression is
\be
\begin{aligned}
\label{EFD2fpsi2}
   E_r&=\bigg[4k^{2}{\sech}^{2}{(\xi)}\bigg(\frac{r^{2\alpha}-1}{r^{2\alpha}+1}\bigg)^{2}\bigg({\sech}^{2}{(\xi)}\\
    &+{\tanh}^{2}{(\xi)}\bigg(\frac{\sigma}{k}-2\bigg)\bigg)+\frac{16\alpha^{2}r^{4\alpha}}{(r^{2\alpha}+1)^4}\bigg]\frac{1}{re}.
\end{aligned}
\ee
The electric field and the energy density are displayed in Fig.~\ref{figEFfpsi222}. We see that the parameter $k$, which controls the strength of the coupling between $\phi$ and $\chi$, controls the location of the peaks in the energy density and electric field. As $k$ increases, the valley at $r=0$ becomes larger and the peaks tends to accumulate near the valley outside the origin. To better illustrate these results, we display the radial component of the electric field and the energy density in Fig.~\ref{figEFfpsi2} in the plane. We see that both engender a ringlike character which becomes sharper as $k$ increases. The total energy of the configuration is $E=E_{1}+E_{2}=8\pi(\sigma+\alpha)/3$ as we expected from Eq.~\eqref{ebogoelectric1}.

%%%%%%%%%%%%
%%%%%%%%%%%%%

\subsubsection{Second model}\label{model2db}
\begin{figure}[t!]
		\centering
		\includegraphics[width=6.2cm,trim={0cm 0cm 0 0},clip]{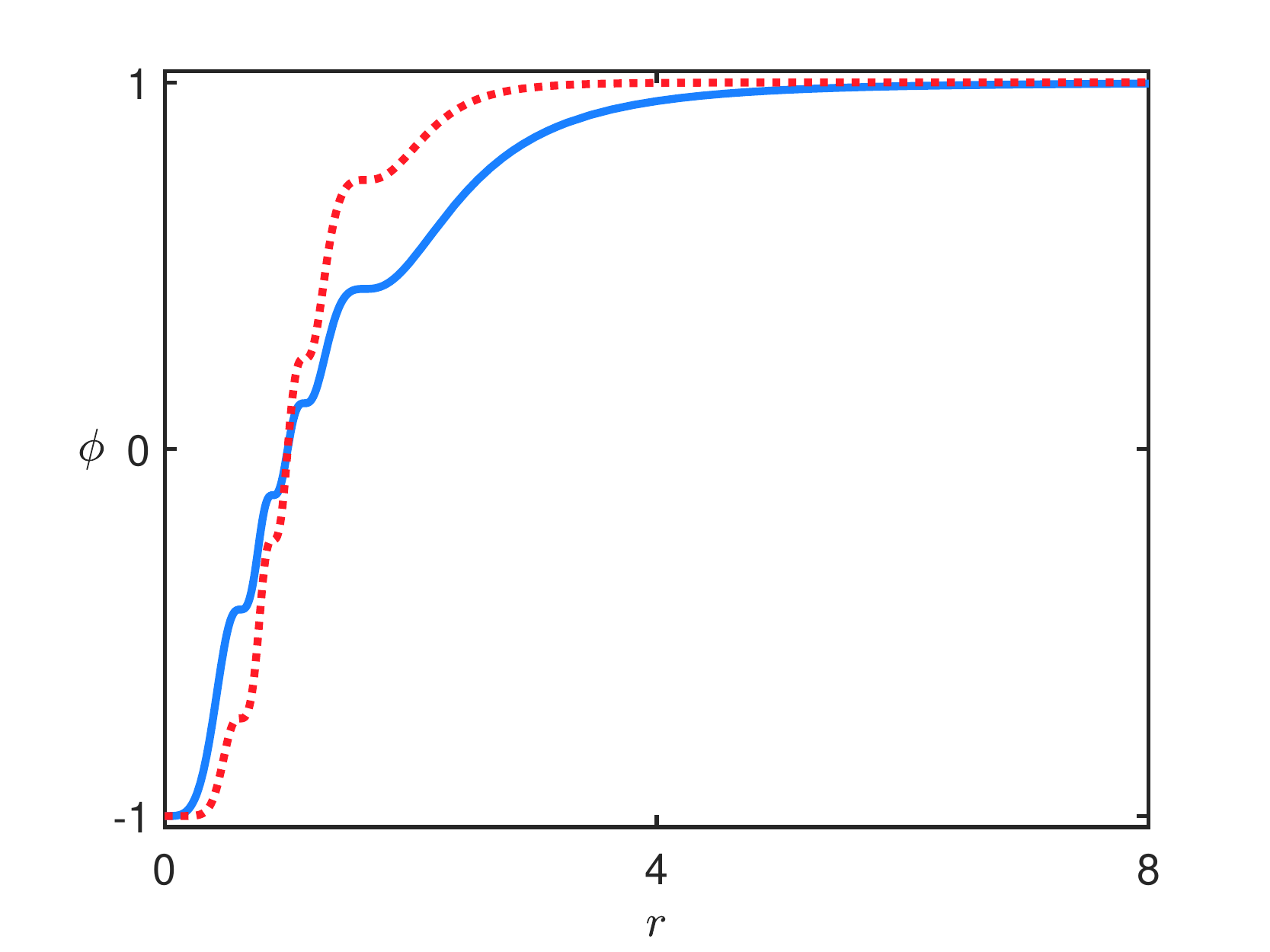}
		\includegraphics[width=6.2cm,trim={0cm 0cm 0 0},clip]{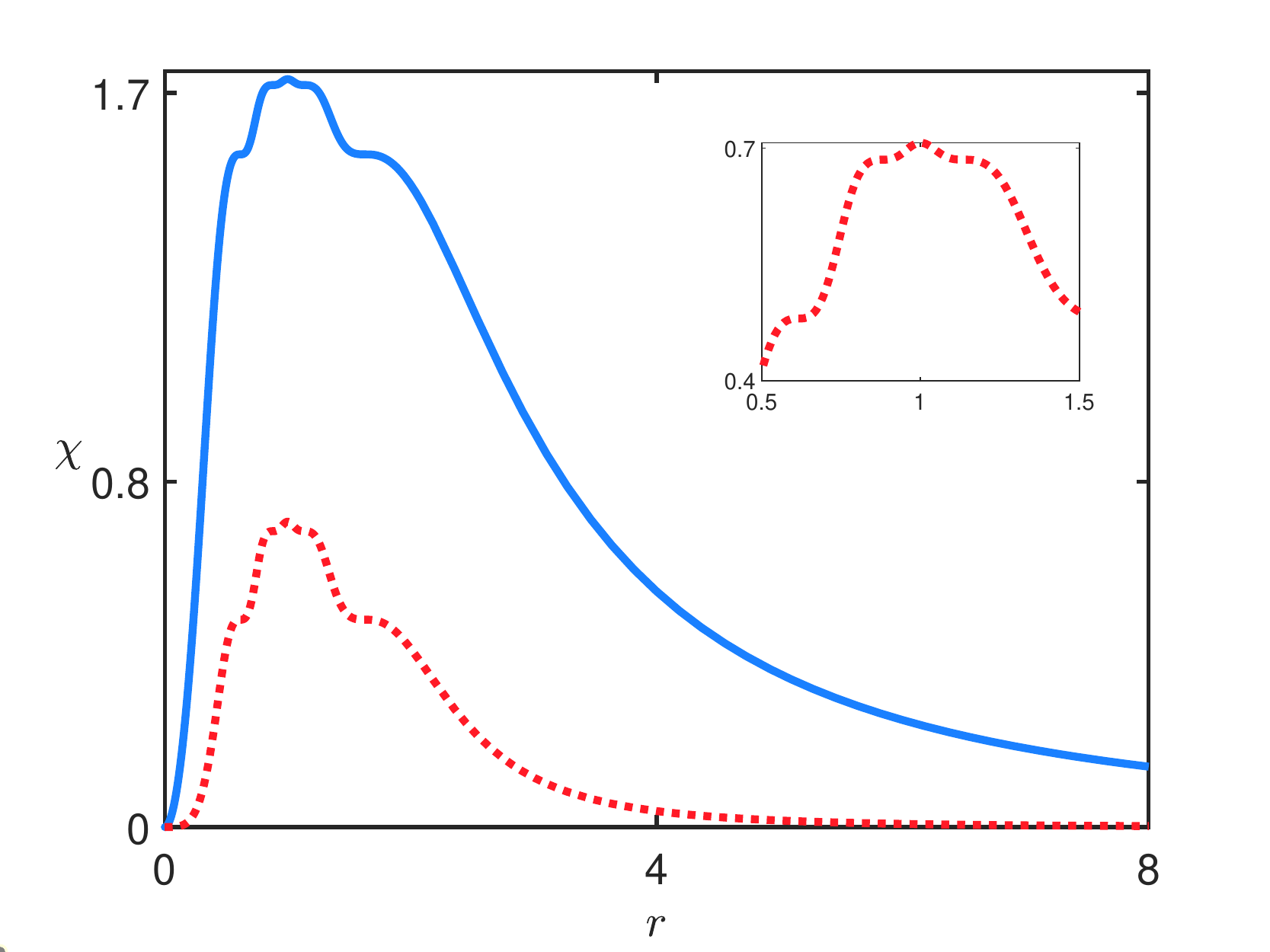}
		\caption{The solutions $\phi(r)$ (top) and $\chi(r)$ (bottom) associated to the model in Sec.~\ref{model2db} for $\sigma=5$, $\alpha=2$ and $n=2$, with $k=1$ (solid, blue line) and $2$ (dotted, red line). The inset in the bottom figure shows the interval $r\in[0.5,1.5]$ for $k=2$.}
		\label{figEFfpsi23}
		\end{figure}

\begin{figure}[t!]
		\centering
		\includegraphics[width=6.2cm,trim={0cm 0cm 0 0},clip]{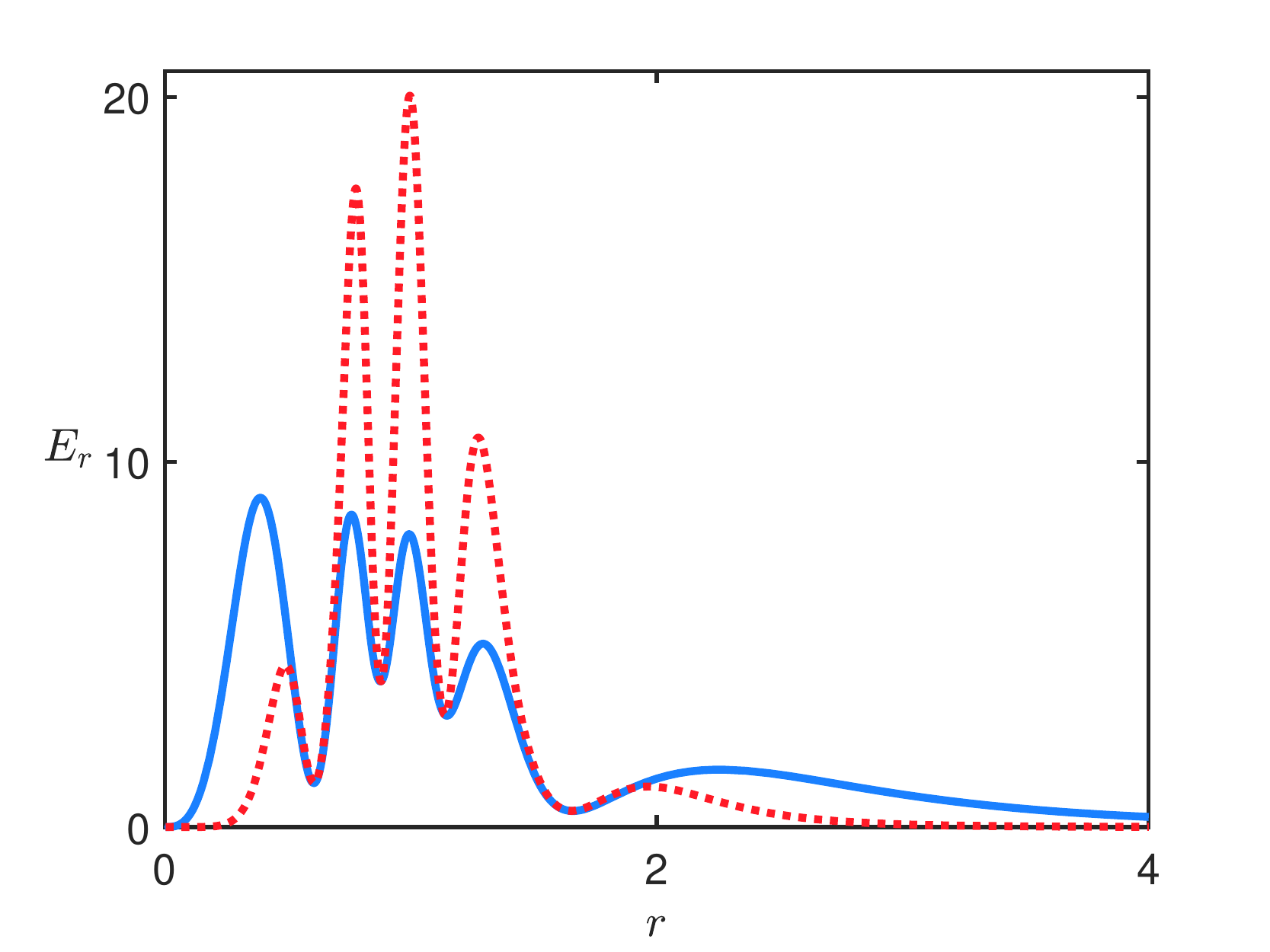}
		\includegraphics[width=6.2cm,trim={0cm 0cm 0 0},clip]{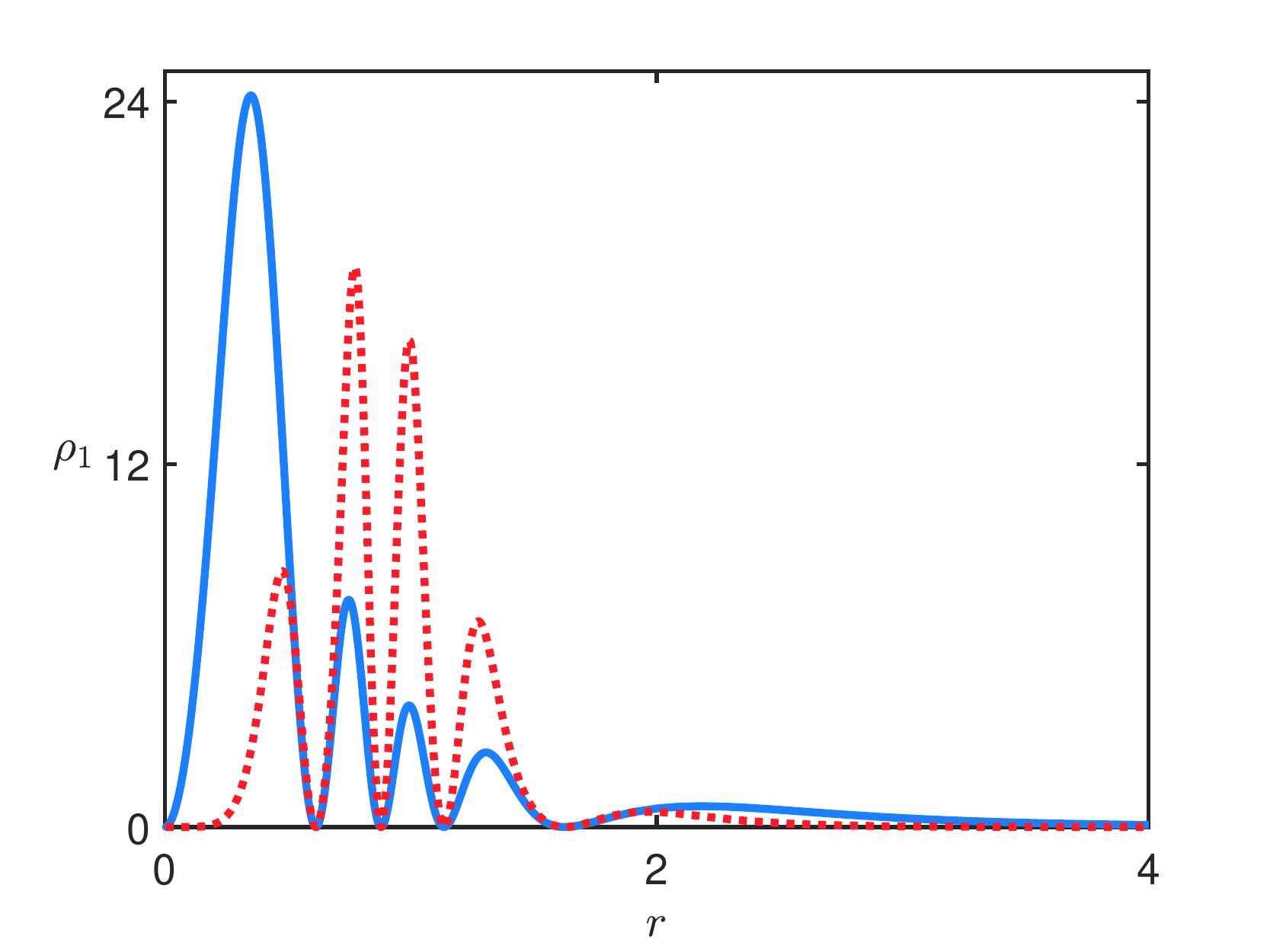}
		\caption{The radial component of the electric field (top) with $e=1$ and the energy density (bottom) associated to the model in Sec.~\ref{model2db} for $\sigma=5$, $\alpha=2$ and $n=2$, with $k=1$ (solid, blue line) and $k=2$ (dotted, red line).}
		\label{CEDEK}
		\end{figure}

\begin{figure}[t!]
		\centering
		\includegraphics[width=4.27cm,trim={0cm 0cm 0 0},clip]{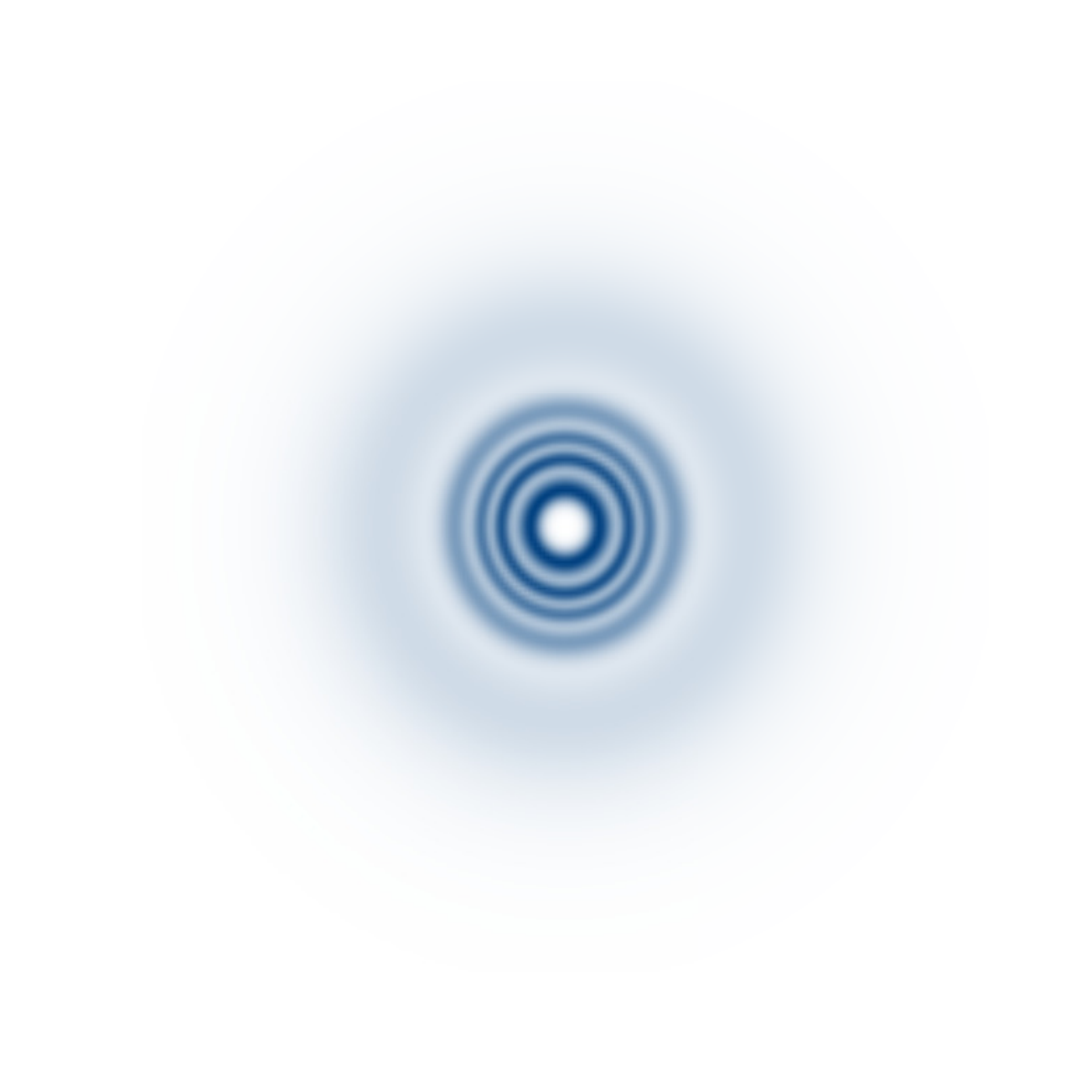}
		\includegraphics[width=4.27cm,trim={0cm 0cm 0 0},clip]{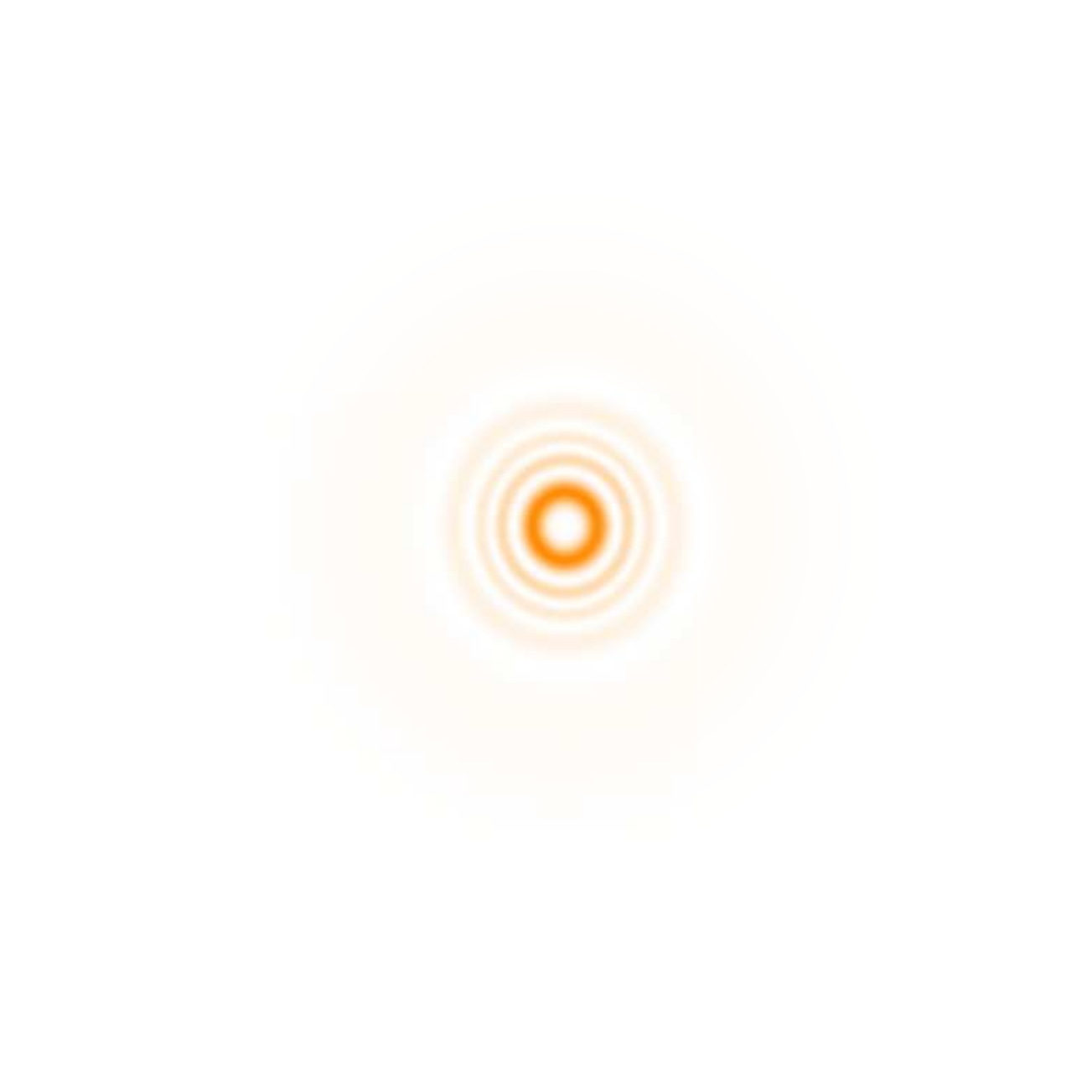}
		\includegraphics[width=4.27cm,trim={0cm 0cm 0 0},clip]{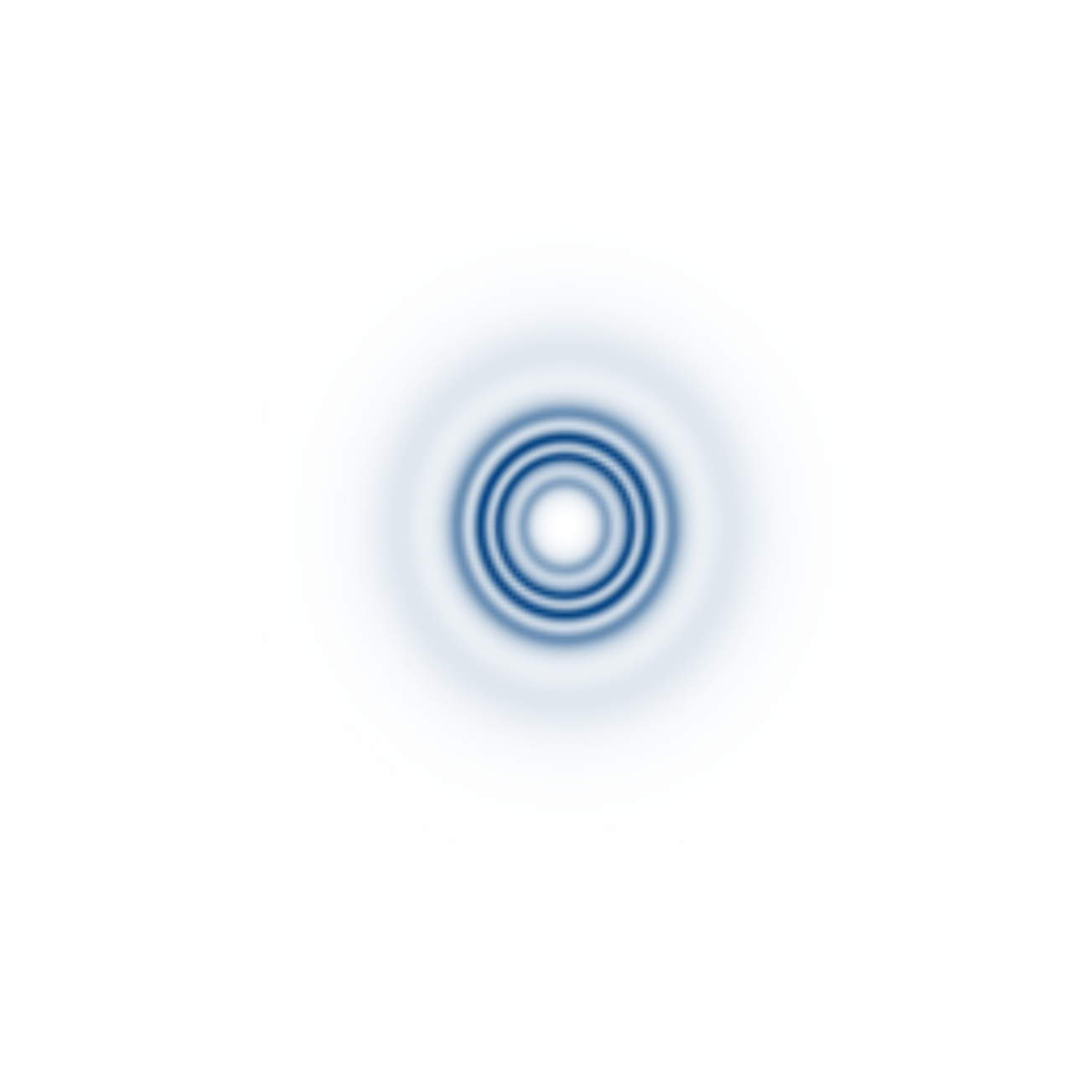}
		\includegraphics[width=4.27cm,trim={0cm 0cm 0 0},clip]{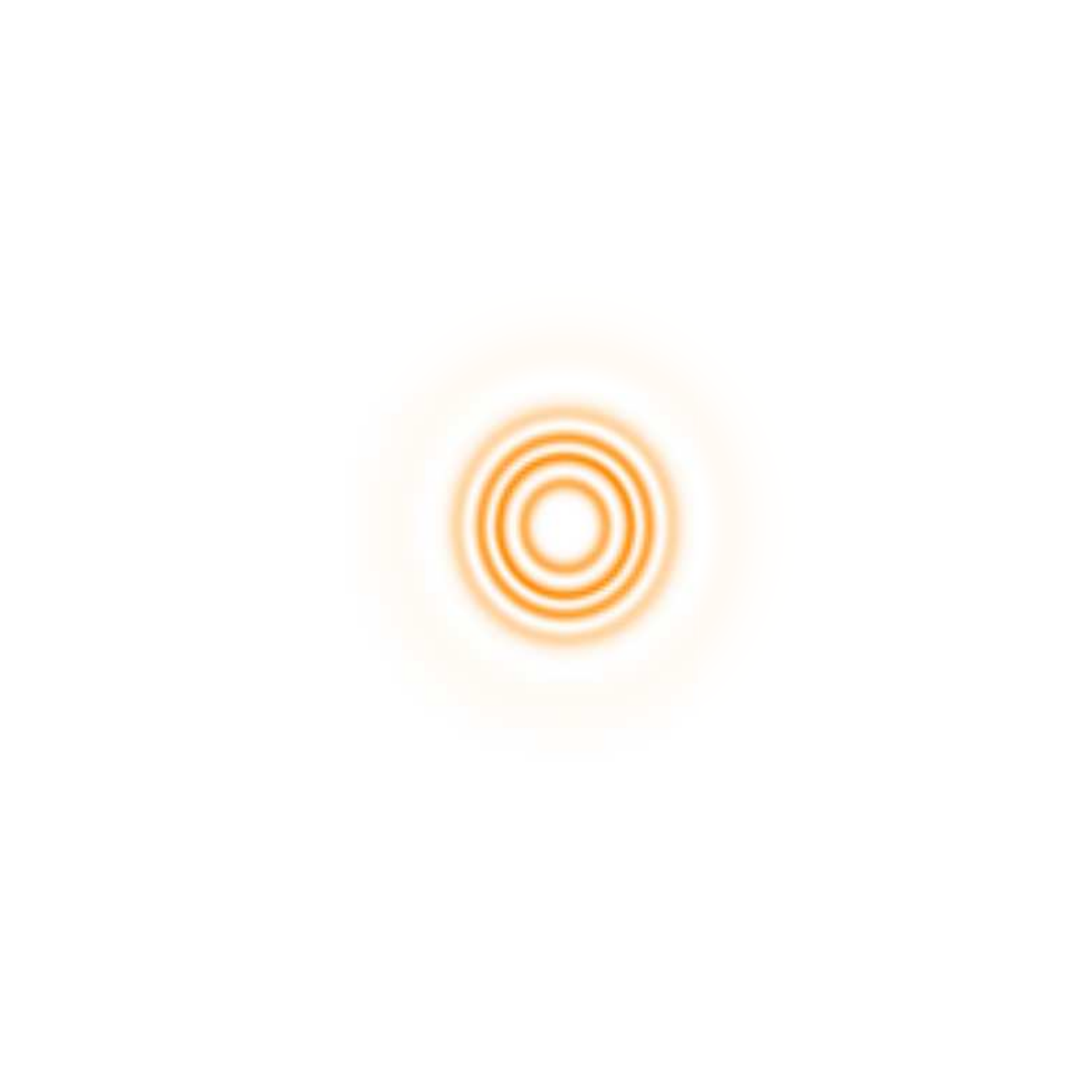}
		\caption{The electric field (left, blue) with $e=1$ and the energy density (right, orange) associated to the model in Sec.~\ref{model2db} depicted in the plane for $\sigma=5$, $\alpha=2$ and $n=2$, with $k=1$ (top) and $2$ (bottom). The intensity of the blue and orange colors increases with the increasing of the electric field and energy density, respectively.}
		\label{surfaced2}
		\end{figure}

The second model that we investigate is driven by $f={\sec}^{2}{(n\pi\psi)}$. This model supports solutions similar to that in Eqs.~\eqref{1phi}--\eqref{1chi} with a new geometrical coordinate, $\eta(r)$, replacing $\xi(r)$, given by
\begin{align}
    \eta(r)=k\ln(r)+\frac{k}{2\alpha}\big[\textrm{Ci}(\xi_{+}(r))-\textrm{Ci}(\xi_{-}(r))\big].
\end{align}
In the above expression, Ci stands for the cosine integral function and $\xi_{\pm}(r)=2n\pi[1\pm (r^{2\alpha}-1)/(r^{2\alpha}+1)]$. In Fig.~\ref{figEFfpsi23} we show the solutions $\phi(r)$ and $\chi(r)$ for $\sigma=5$, $\alpha=2$, $n=2$ and some values of $k$. The solutions now present $2n$ inflection points that appear due to the form of the function $f$. 
%%%%%%%%%%%%%
%%%%%%%%%%%%%
The energy density in Eq.~\eqref{rho1ele} reads
\be
\begin{aligned}
\label{densitycosn2}
    \rho_{1}&=\bigg[4k^{2}{\sech}^{2}{(\eta)}\cos^{2}{\bigg(n\pi\bigg(\frac{r^{2\alpha}-1}{r^{2\alpha}+1}\bigg)\bigg)}\bigg({\sech}^{2}{(\eta)}\\
    &+{\tanh}^{2}{(\eta)}\bigg(\frac{\sigma}{k}-2\bigg)\bigg)\bigg] \frac{1}{r^2},
\end{aligned}
\ee
and the electric field in Eq.~\eqref{estd} takes the form
\be\begin{aligned}
\label{EFd2fcos}
   E_r&=\bigg[4k^{2}{\sech}^{2}{(\eta)}\cos^{2}{\bigg(n\pi\bigg(\frac{r^{2\alpha}-1}{r^{2\alpha}+1}\bigg)\bigg)}\bigg({\sech}^{2}{(\eta)}\\
    &+{\tanh}^{2}{(\eta)}\bigg(\frac{\sigma}{k}-2\bigg)\bigg)+\frac{16\alpha^{2}r^{4\alpha}}{(r^{2\alpha}+1)^4}\bigg] \frac{1}{re}.
\end{aligned}
\ee
In Fig.~\ref{CEDEK} we depict the electric field and the energy density for $\sigma=5$, $\alpha=2$, $n=2$ and some values of $k$. Notice that the internal structure in these quantities are much richer than before and, as $k$ increases, the inner ring becomes less and less visible. To better illustrate these features, we display the electric field and the energy density in the plane in Fig.~\ref{surfaced2}. This model has the same energy of the previous one, that is, $E=8\pi(\sigma+\alpha)/3$ as we expected from Eq.~\eqref{ebogoelectric1}, since it does not depend on $f(\psi)$.
%%%%%%%%%%%%%
%%%%%%%%%%%%%
%%%%%%%%%%%%%
%%%%%%%%%%%%%

\subsection{Three spatial dimensions}
\begin{figure}[t!]
		\centering
		\includegraphics[width=6.2cm,trim={0cm 0cm 0 0},clip]{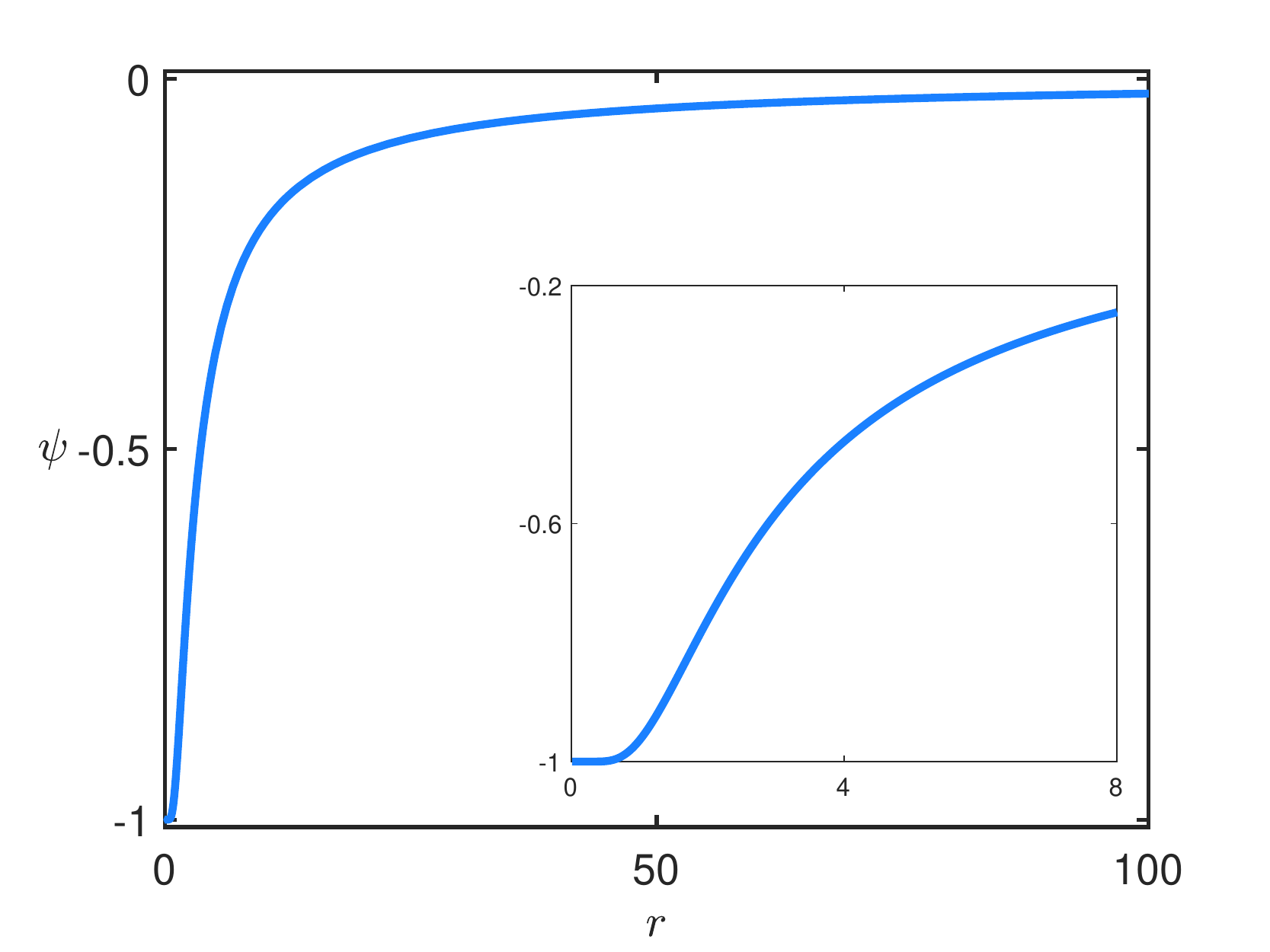}
		\includegraphics[width=6.2cm,trim={0cm 0cm 0 0},clip]{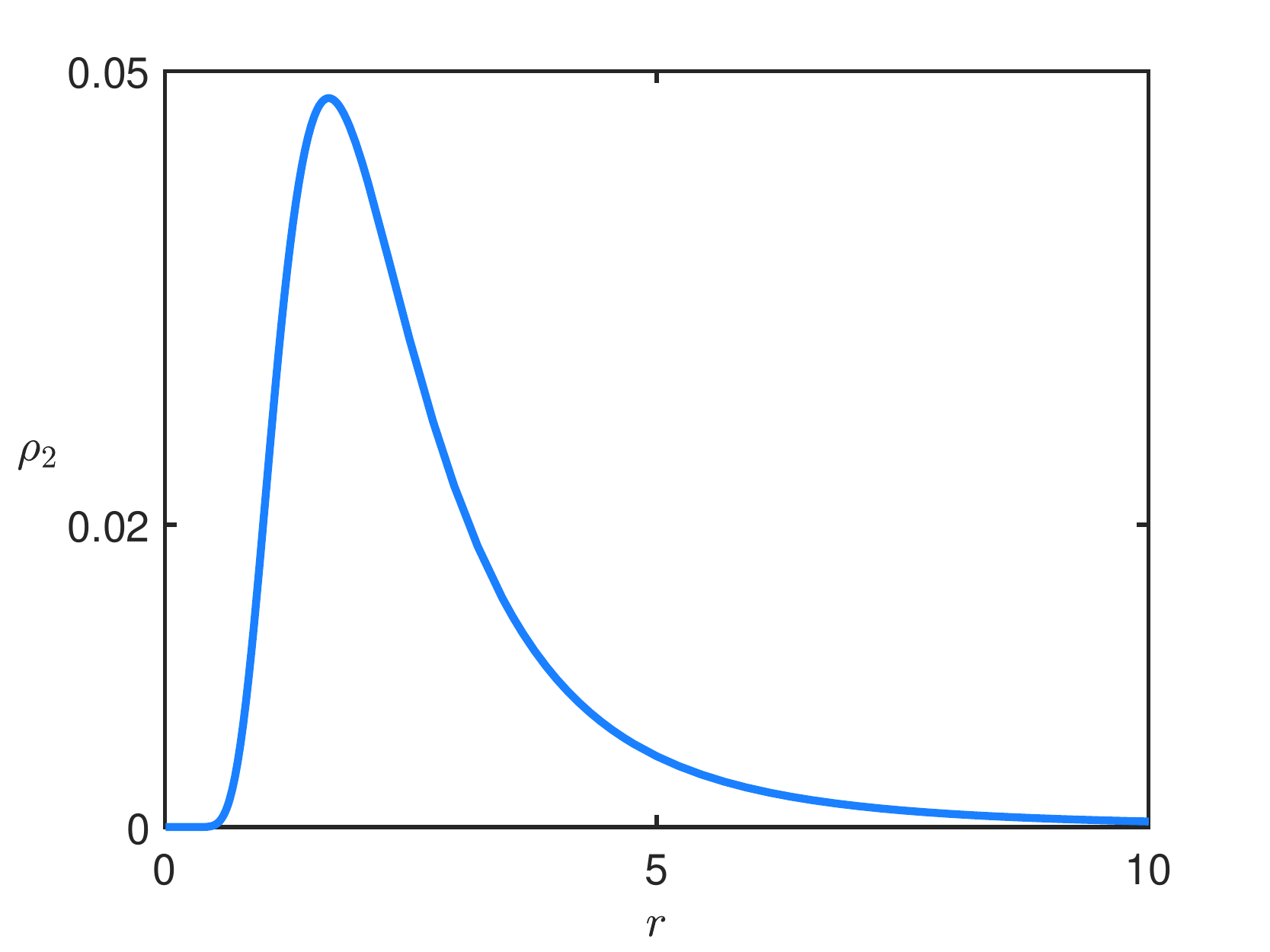}
		\caption{The solution $\psi(r)$ in Eq.~\eqref{psisolution} (top) and the energy density $\rho_{2}$ in Eq.~\eqref{energydensityd322} (bottom) for $\alpha=2$. The inset in the top figure shows the solution near the origin.}
		\label{psid311}
		\end{figure}

We continue our investigation of the model \eqref{lele}, studying it in three spatial dimensions. In this case, it supports the solution
\begin{align}
\label{psisolution}
    \psi=-\tanh\left(\frac{\alpha}{r}\right),
\end{align}
whose associated energy density \eqref{rho2ele} is 
\begin{equation}
     \label{energydensityd322}
   \rho_{2} =\frac{\alpha^2}{r^4}{\sech}^{4}\left(\frac{\alpha}{r}\right).
\end{equation}
In Fig.~\ref{psid311} we depict the solution \eqref{psisolution} and the energy density \eqref{energydensityd322} for $\alpha=2$. Notice that there is a hole at the origin in the above energy density.
%%%%%%%%%%%%%%%
%%%%%%%%%%%%%%
Next, we study the behavior of the electrical structure generated by the scalar fields associated to the Bloch wall for the same functions $f(\psi)$ considered in Sec.~\ref{secftwo}.
 
\subsubsection{First model}\label{model3d1}
\begin{figure}[t!]
		\centering
		\includegraphics[width=6.2cm,trim={0cm 0cm 0 0},clip]{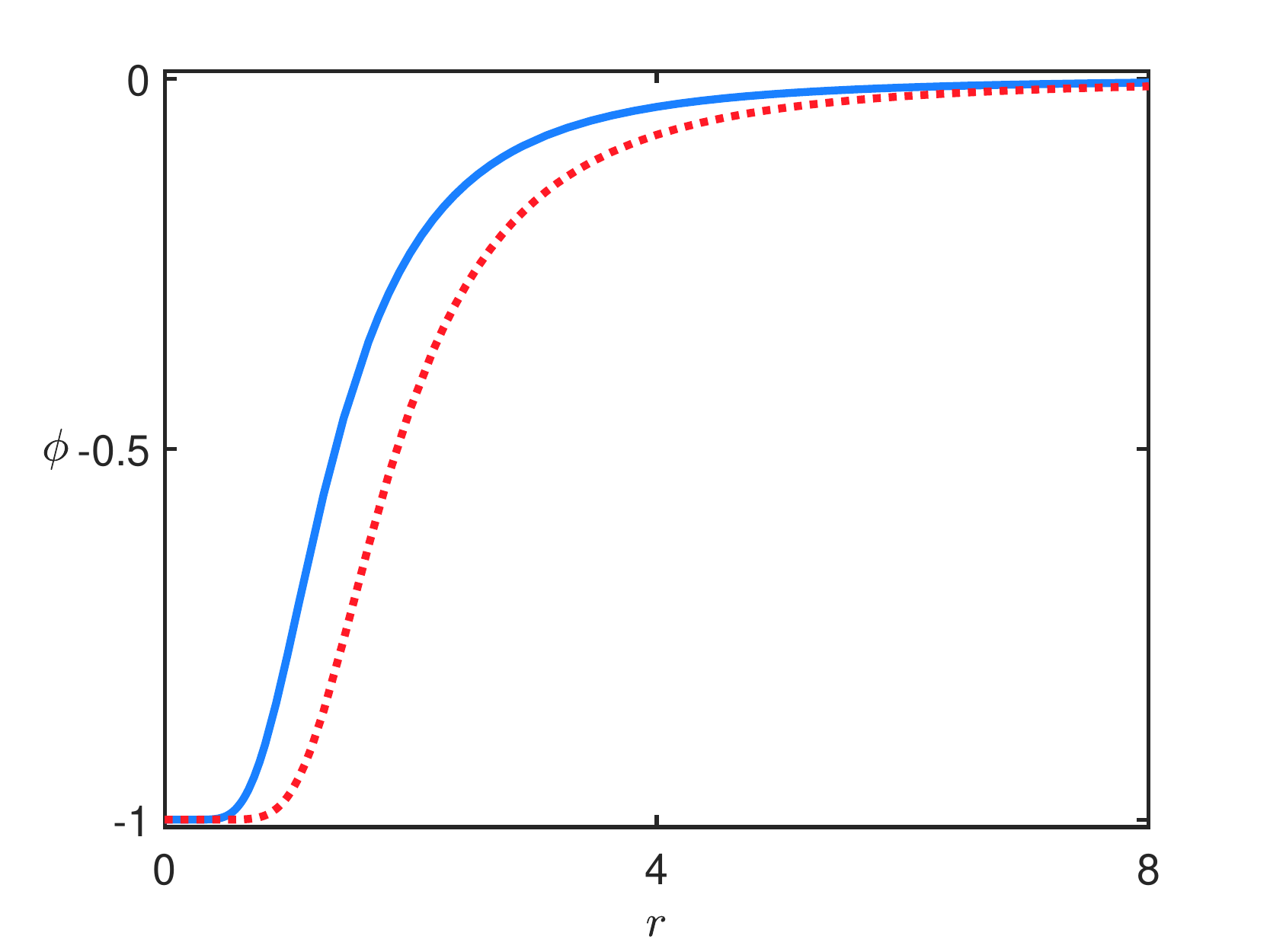}
		\includegraphics[width=6.2cm,trim={0cm 0cm 0 0},clip]{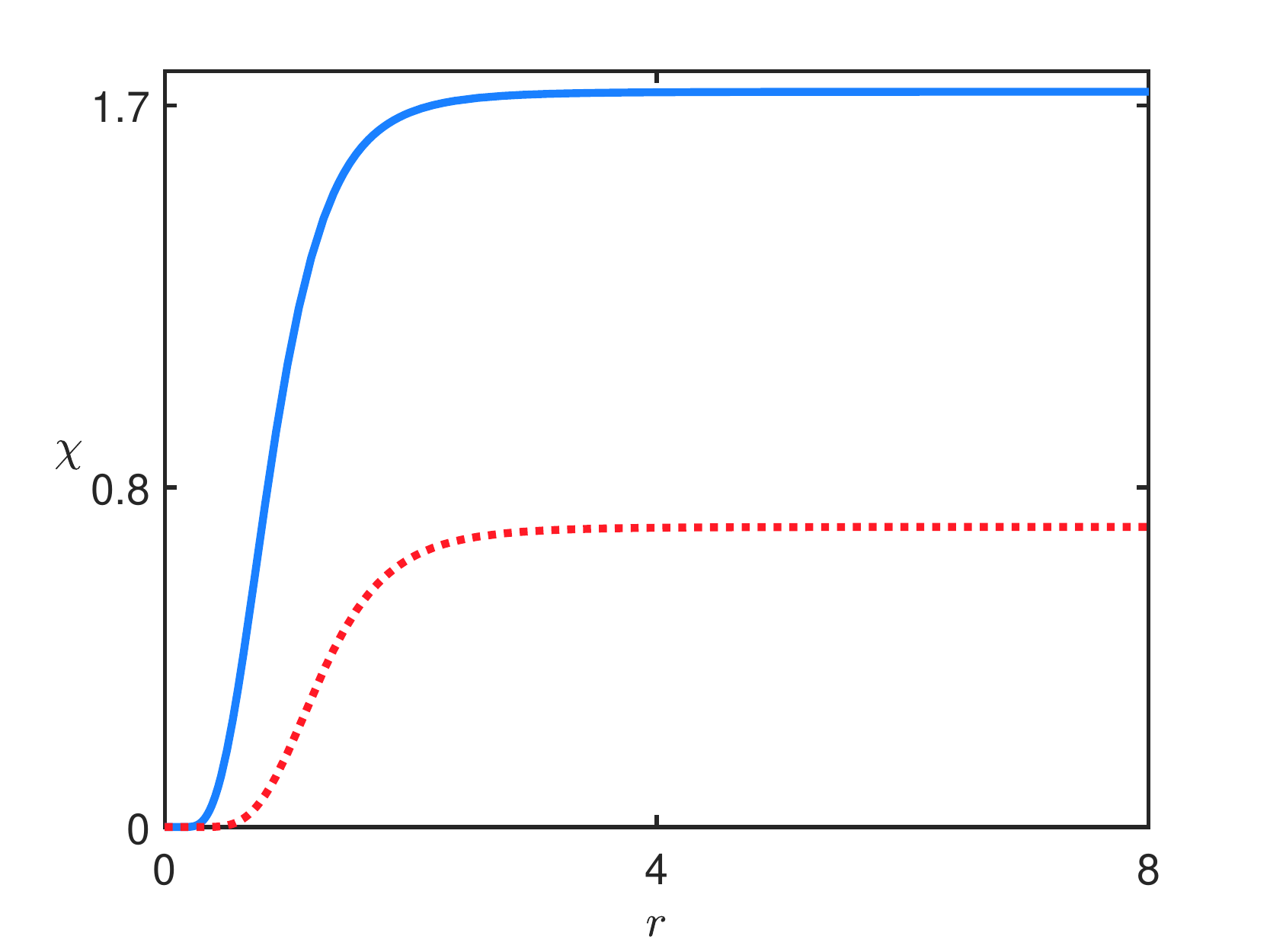}
		\caption{The solution $\phi(r)$ (top) and $\chi(r)$ (bottom) associated to the model in Sec.~\ref{model3d1} for $\sigma=5$, $\alpha=2$ and $n=2$ with $k=1$ (solid, blue line) and $2$ (dotted, red line).}
		\label{phichid31}
		\end{figure}

\begin{figure}[t!]
		\centering
		\includegraphics[width=6.2cm,trim={0cm 0cm 0 0},clip]{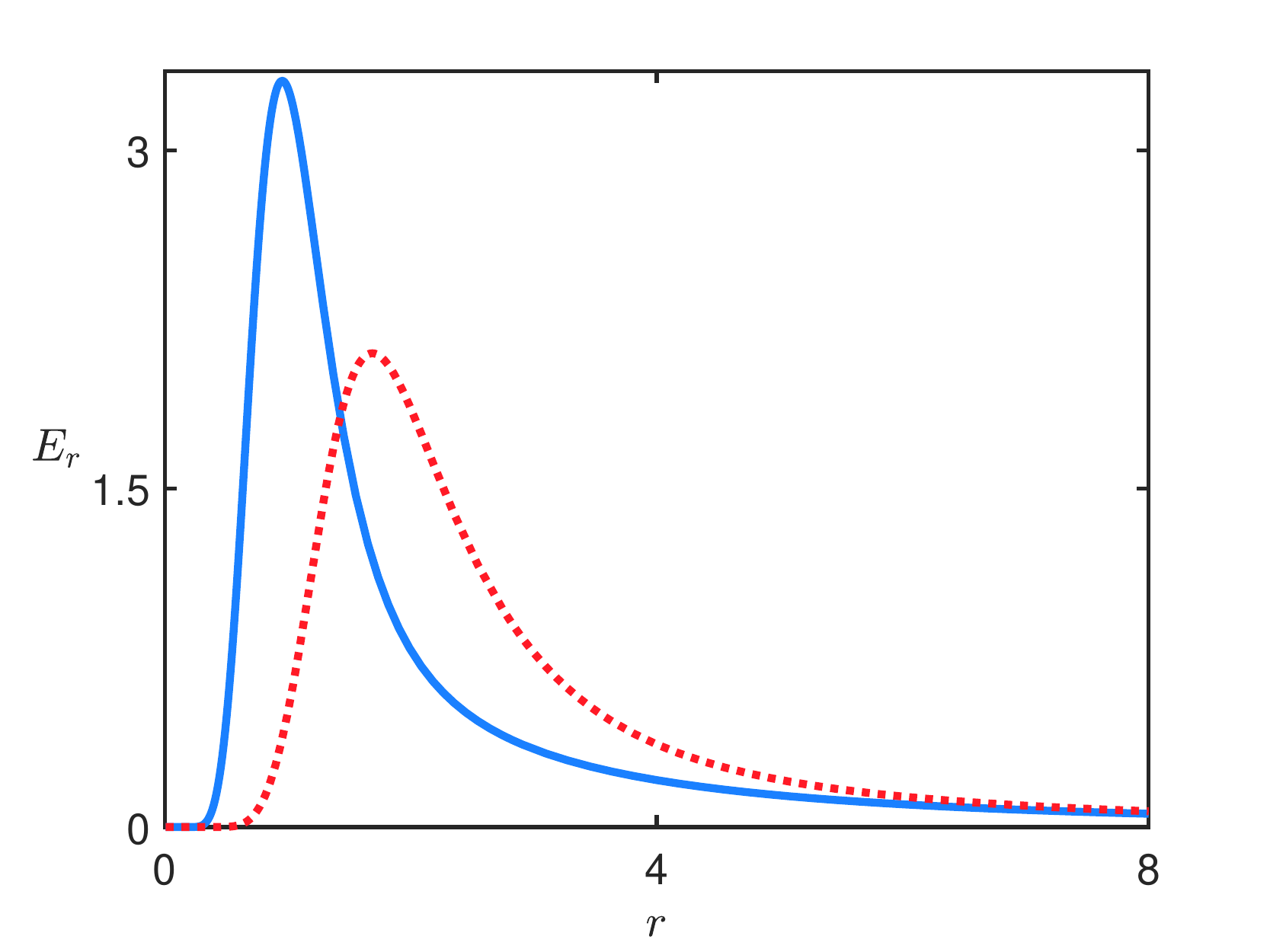}
		\includegraphics[width=6.2cm,trim={0cm 0cm 0 0},clip]{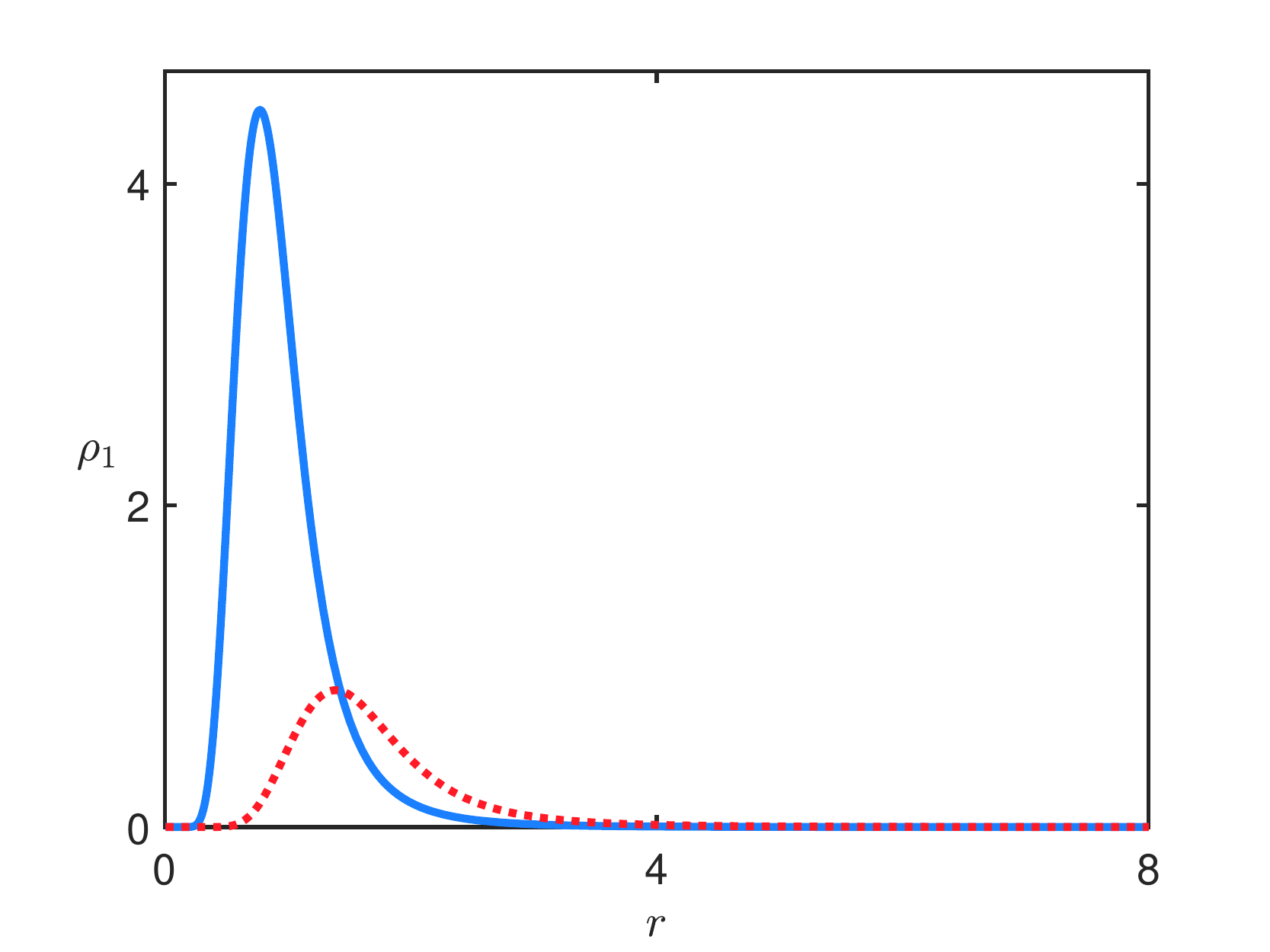}
		\caption{The radial component of the electric field (top) with $e=1$ and the energy density (bottom) associated to the model in Sec.~\ref{model3d1} for $\sigma=5$, $\alpha=2$, and $k=1$ (solid, blue line) and $2$ (dotted, red line).}
		\label{CED3fpsi22}
		\end{figure}

\begin{figure}[t!]
		\centering
		\includegraphics[width=4.27cm,trim={0cm 0cm 0 0},clip]{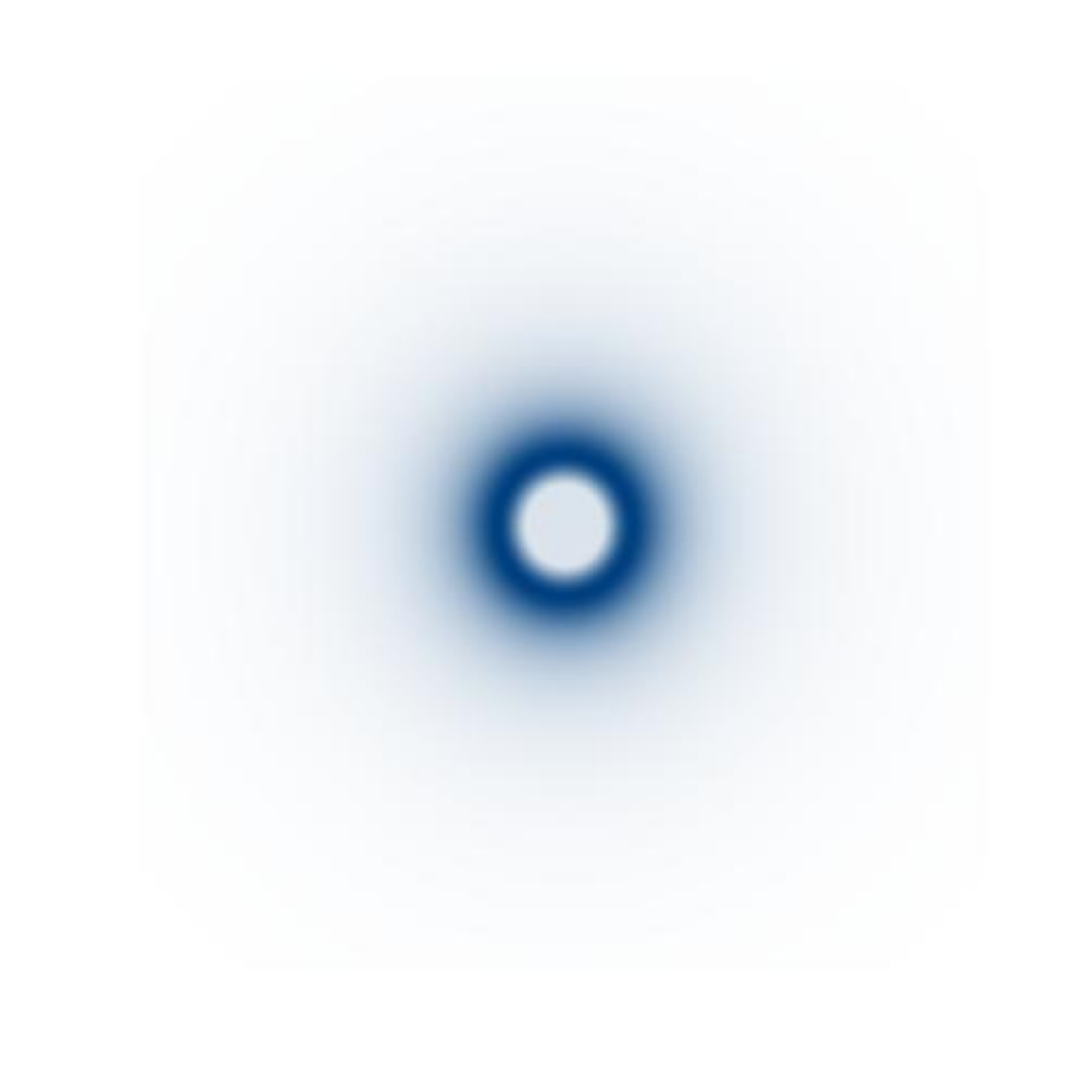}
		\includegraphics[width=4.27cm,trim={0cm 0cm 0 0},clip]{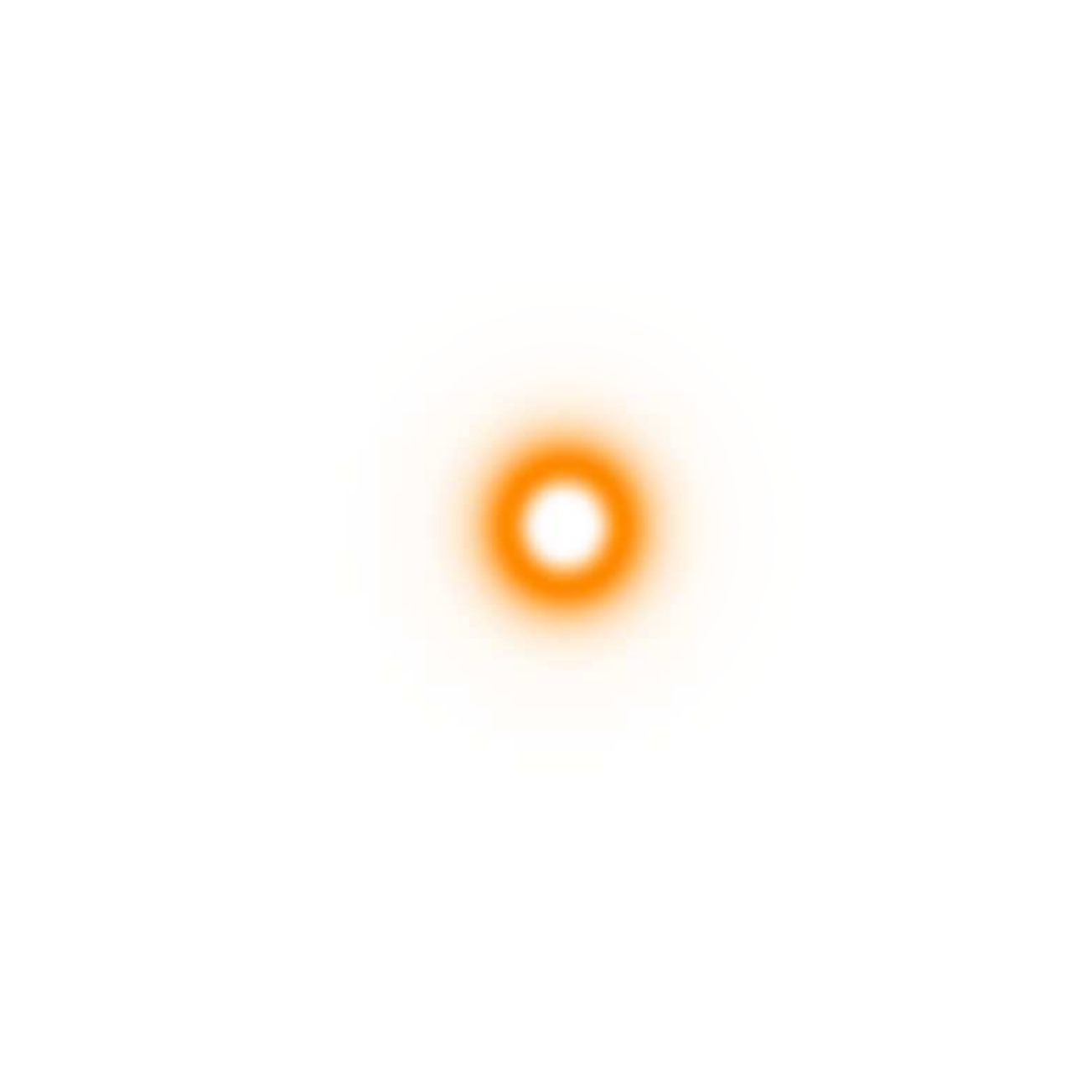}
		\caption{The section passing through the center of the structure, representing the electric field (left, blue) with $e=1$ and the energy density (right, orange) associated to the model in Sec.~\ref{model3d1} for $\sigma=5$, $\alpha=2$ and $k=2$. The intensity of the blue and orange colors increases with the increasing of the electric field and energy density, respectively.}
		\label{CED3fpsi2}
		\end{figure}

Considering $f=1/\psi^2$, the solutions in Eqs.~\eqref{1phi} and \eqref{1chi} are now described by the geometrical coordinate
\begin{equation}
    \xi(r)=\frac{2k}{\alpha}\bigg(\tanh\bigg(\frac{\alpha}{r}\bigg)-\frac{\alpha}{r}\bigg).
\end{equation}
The solution $\phi(r)$ connects the points $\phi=-1$ and $\phi=0$, and $\chi(r)$ goes from $\chi=0$ to $\chi=\sqrt{\sigma/k-2}$. They can be seen in Fig.~\ref{phichid31}.
%%%%%%%%%%%%%%
%%%%%%%%%%%%%%
The energy density in Eq.~\eqref{rho1ele} is
\begin{align}
\label{energydensityd3}
    \rho_{1}&=\bigg[4k^{2}{\sech}^{2}{(\xi)}\tanh^{2}{\bigg(\frac{\alpha}{r}\bigg)}\bigg({\sech}^{2}{(\xi)}\nonumber  \\
 &\quad+{\tanh}^{2}{(\xi)}\bigg(\frac{\sigma}{k}-2\bigg)\bigg)\bigg]\frac{1}{r^{4}}\;,
\end{align}
and the electric field \eqref{Efield} is given by
\be
\begin{aligned}
\label{EFD3f3}
    E_{r}&=\bigg[4k^{2}{\sech}^{2}{(\xi)}\tanh^{2}{\bigg(\frac{\alpha}{r}\bigg)}\bigg({\sech}^{2}{(\xi)}\\
    &\quad+{\tanh}^{2}{(\xi)}\bigg(\frac{\sigma}{k}-2\bigg)\bigg)+\alpha^2{\sech}^{4}{\bigg(\frac{\alpha}{r}\bigg)}\bigg]\frac{1}{er^{2}}.
\end{aligned}
\ee
In Fig.~\ref{CED3fpsi22} we display the electric field \eqref{EFD3f3} and the energy density \eqref{energydensityd3}. Notice that, as $k$ increases the hole around the origin gets wider and the height of both the electric field and energy density decreases. In Fig.~\ref{CED3fpsi2} we illustrate their behavior depicting the section passing through the center of the structure, from which one can see that the energy density has the form of a shell. In this situation, the energy is $E=8\pi(\sigma+\alpha)/3$.
%%%%%%%%%%%%%%%%%%%%%
%%%%%%%%%%%%%%%%%%%%%%%%
%%%%%%%%%%%%%%%%%%%%
%%%%%%%%%%%%%%%%%%%%

\subsubsection{Second model}\label{model3d2}
\begin{figure}[t!]
		\centering
		\includegraphics[width=6.2cm,trim={0cm 0cm 0 0},clip]{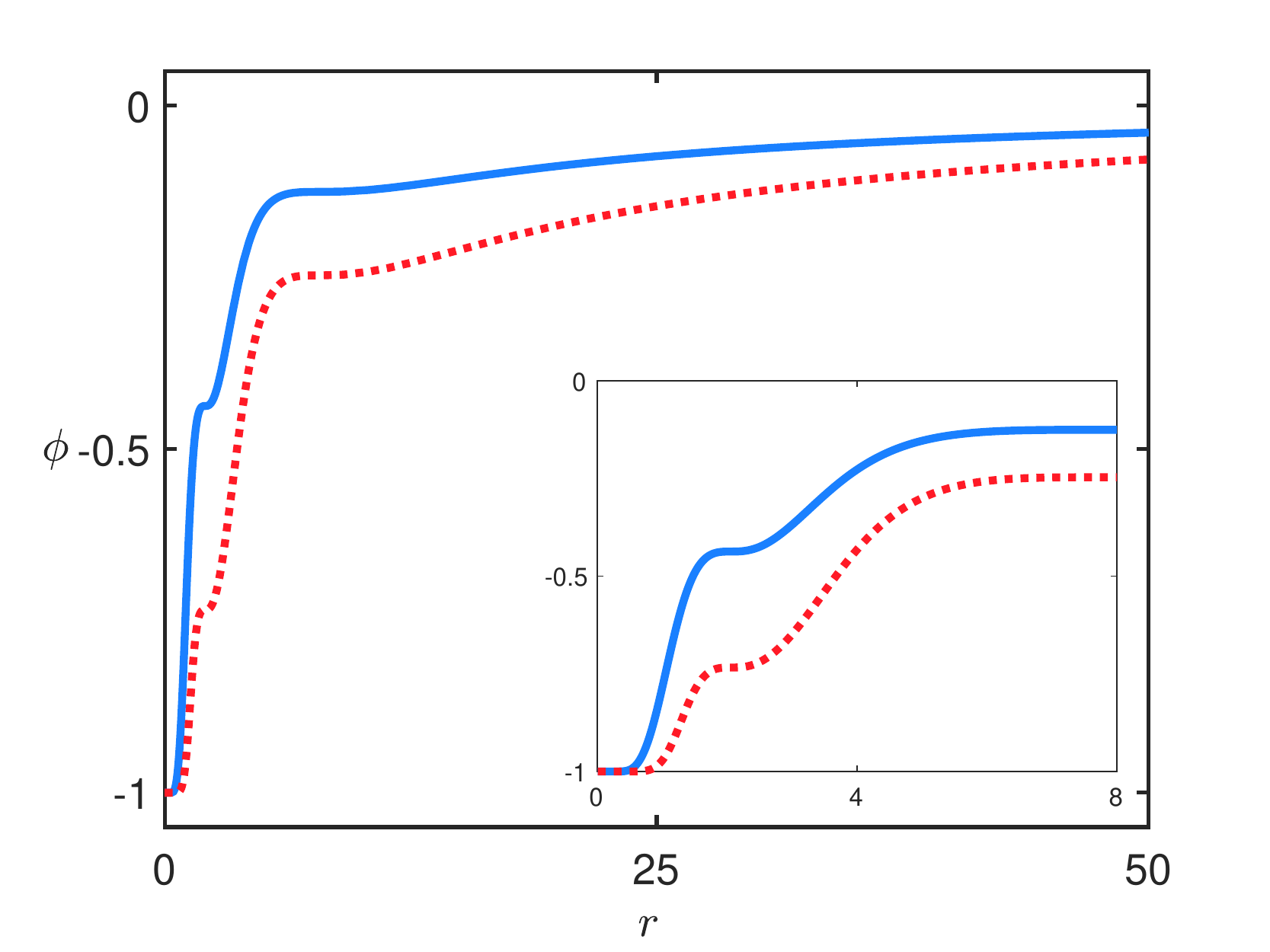}
		\includegraphics[width=6.2cm,trim={0cm 0cm 0 0},clip]{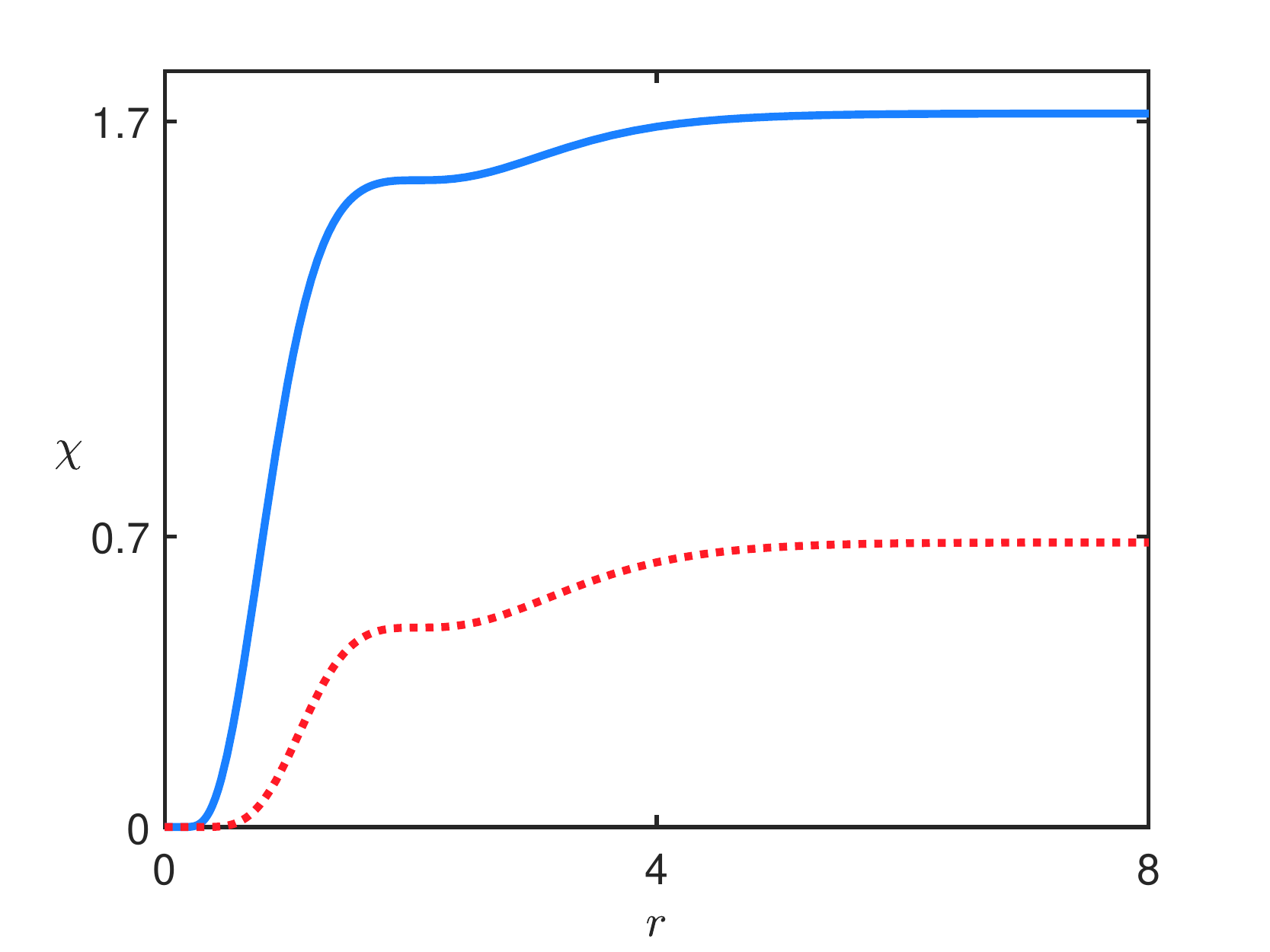}
		\caption{The solutions $\phi(r)$ (top) and $\chi(r)$ (bottom) associated to the model in Sec.~\ref{model3d2} for $\sigma=5$, $\alpha=2$ and $n=2$ with $k=1$ (solid, blue line) and $2$ (dotted, red line). The inset in the top figure shows the behavior for small $r$.}
		\label{figphichicosd3}
		\end{figure}

\begin{figure}[t!]
		\centering
		\includegraphics[width=6.2cm,trim={0cm 0cm 0 0},clip]{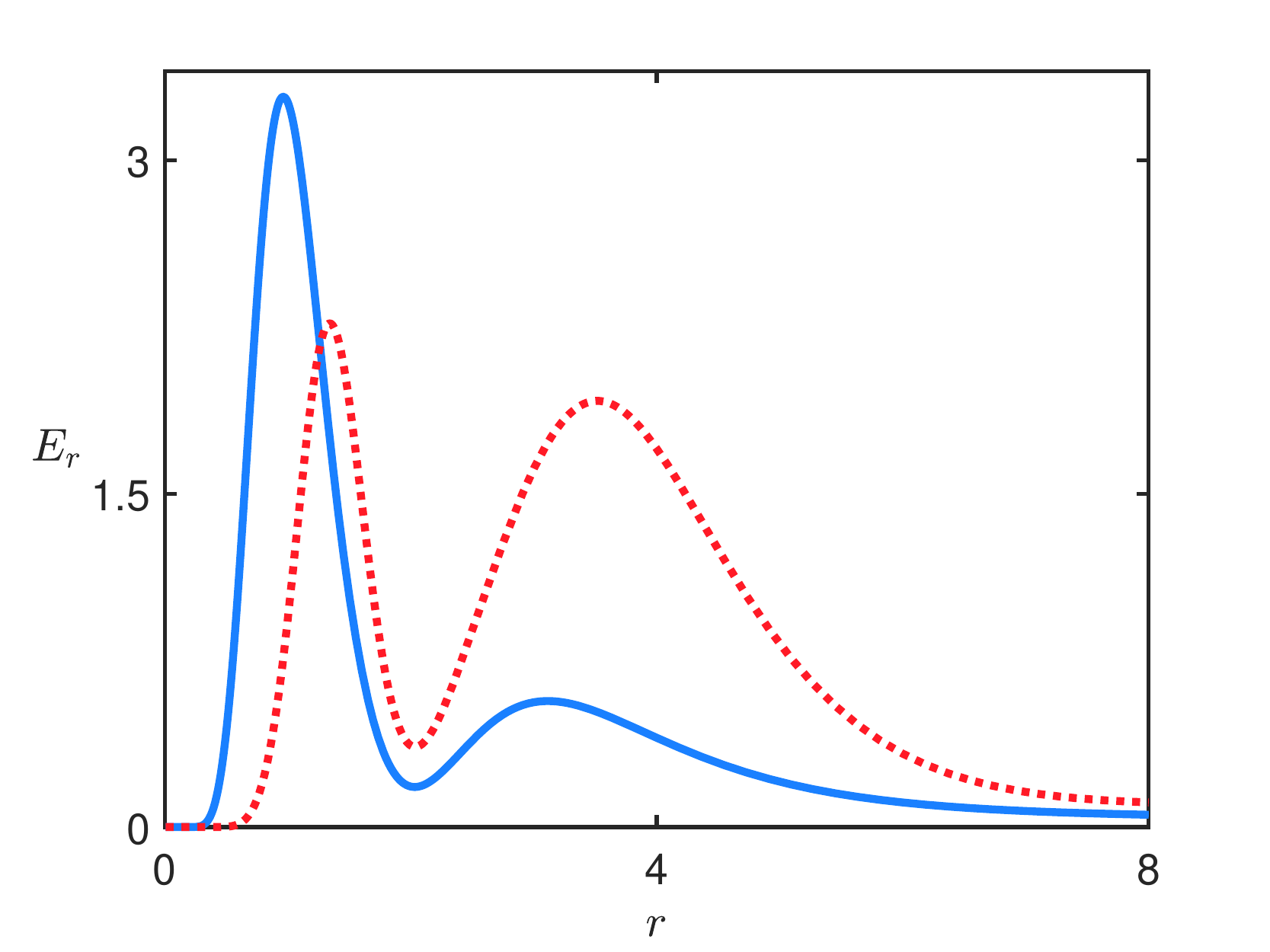}
		\includegraphics[width=6.2cm,trim={0cm 0cm 0 0},clip]{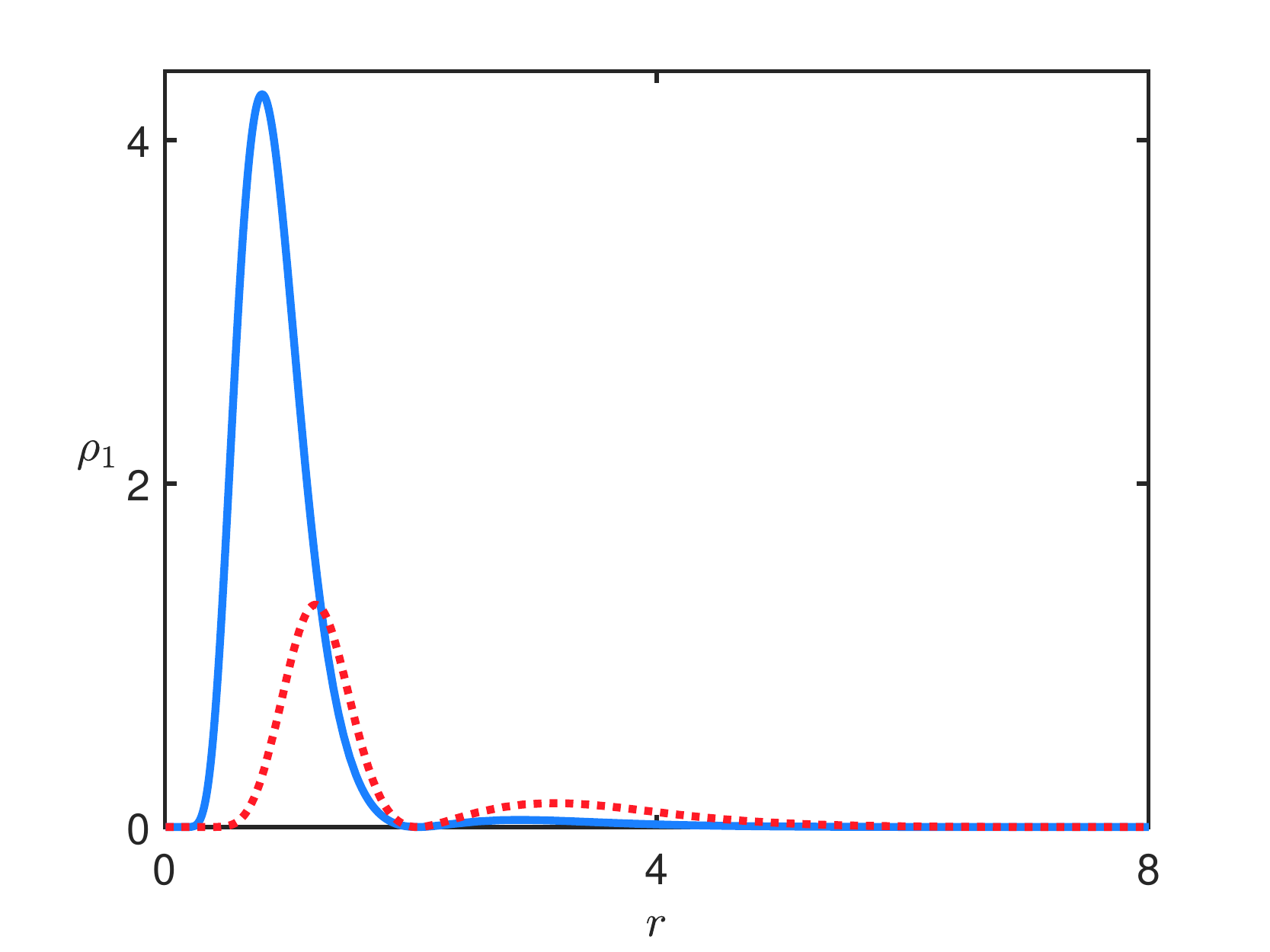}
		\caption{The radial component of the electric field (top) with $e=1$ and energy density (bottom) associated to the model in Sec.~\ref{model3d2} for $\sigma=5$, $\alpha=2$ and $n=2$, with $k=1$ (solid, blue line) and $2$ (dotted, red line). For $k=2$, we have depicted $2E_r$ and $2\rho_1$ to better illustrate the behavior.}
		\label{cedensityd3cos}
		\end{figure}

\begin{figure}[t!]
		\centering
		\includegraphics[width=4.27cm,trim={0cm 0cm 0 0},clip]{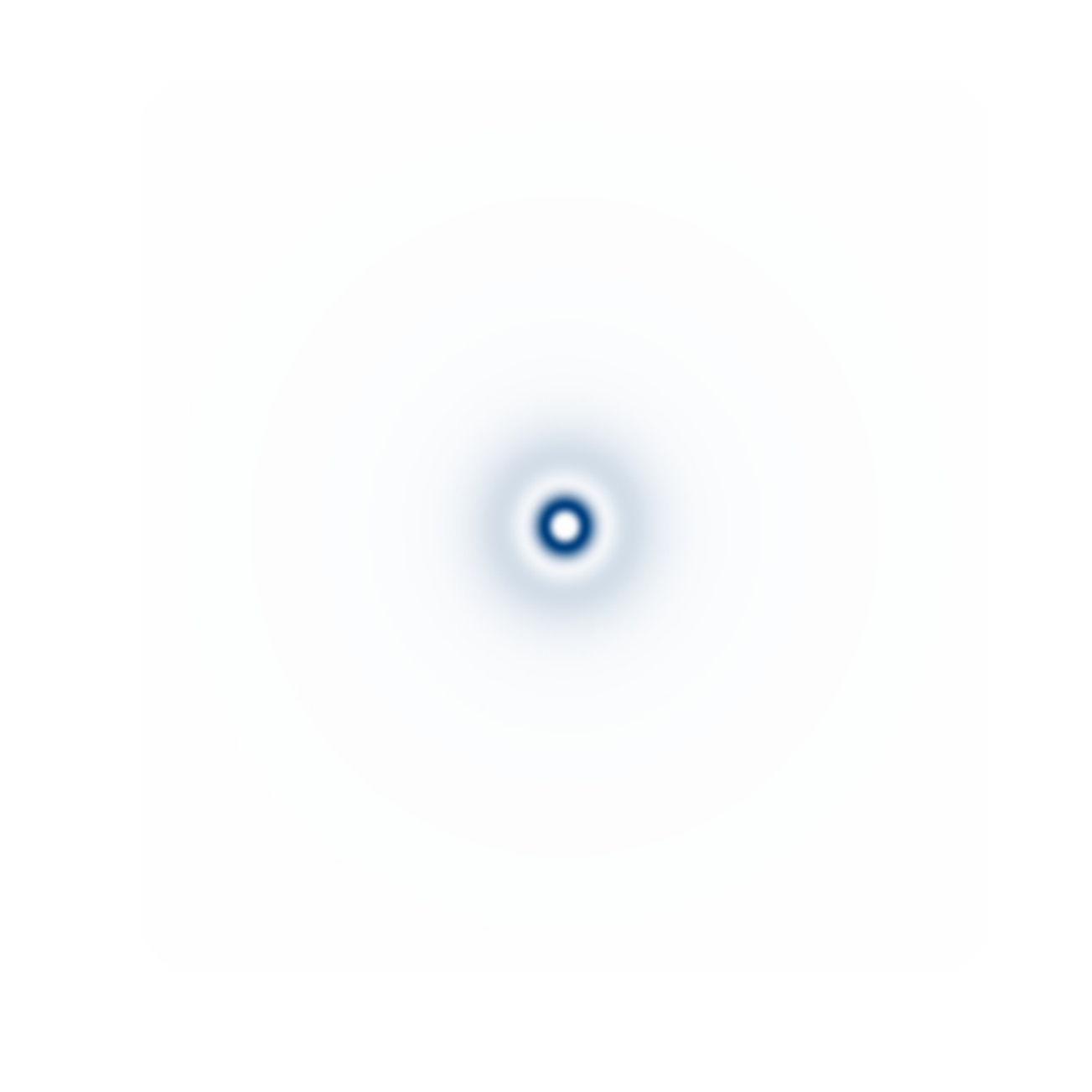}
		\includegraphics[width=4.27cm,trim={0cm 0cm 0 0},clip]{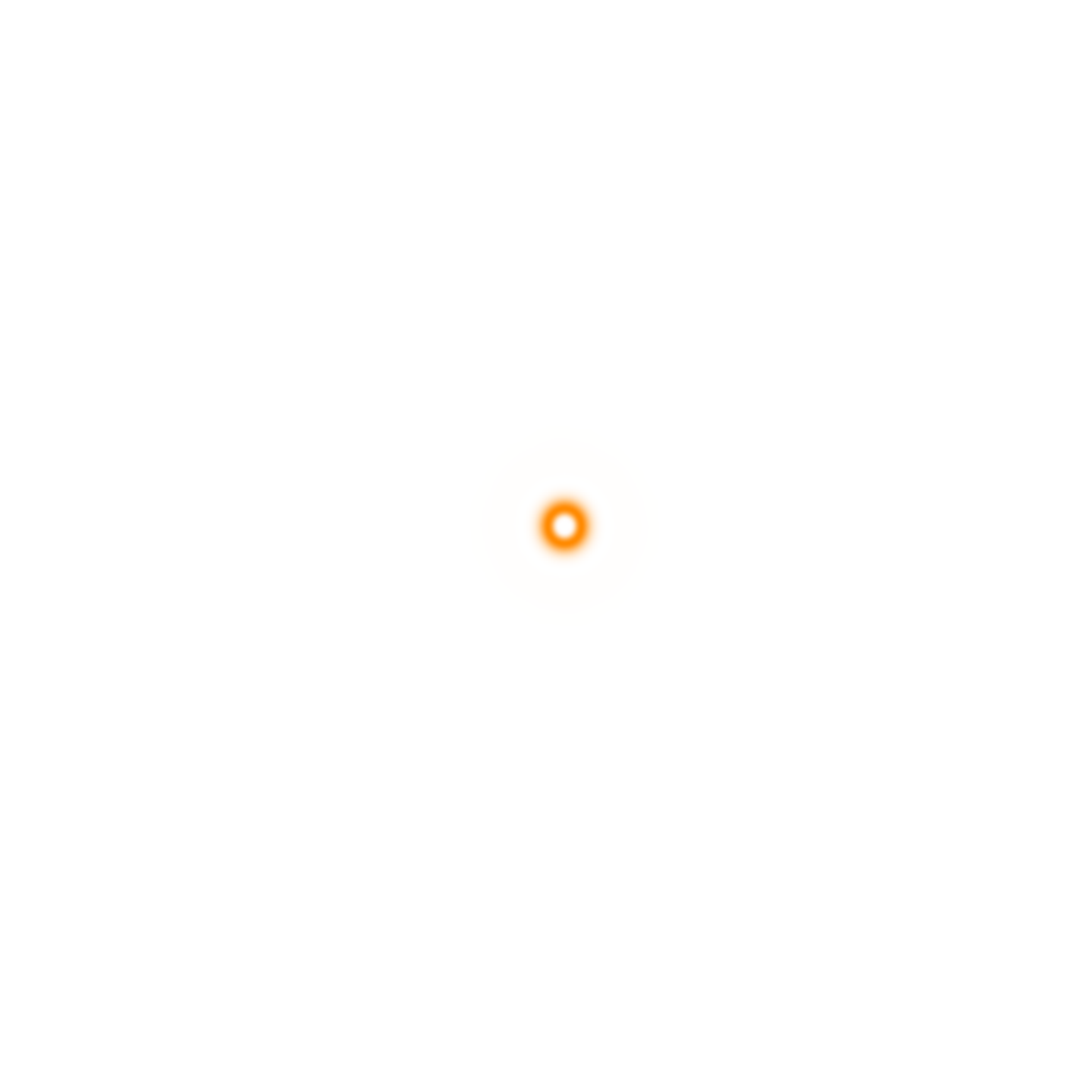}
		\includegraphics[width=4.27cm,trim={0cm 0cm 0 0},clip]{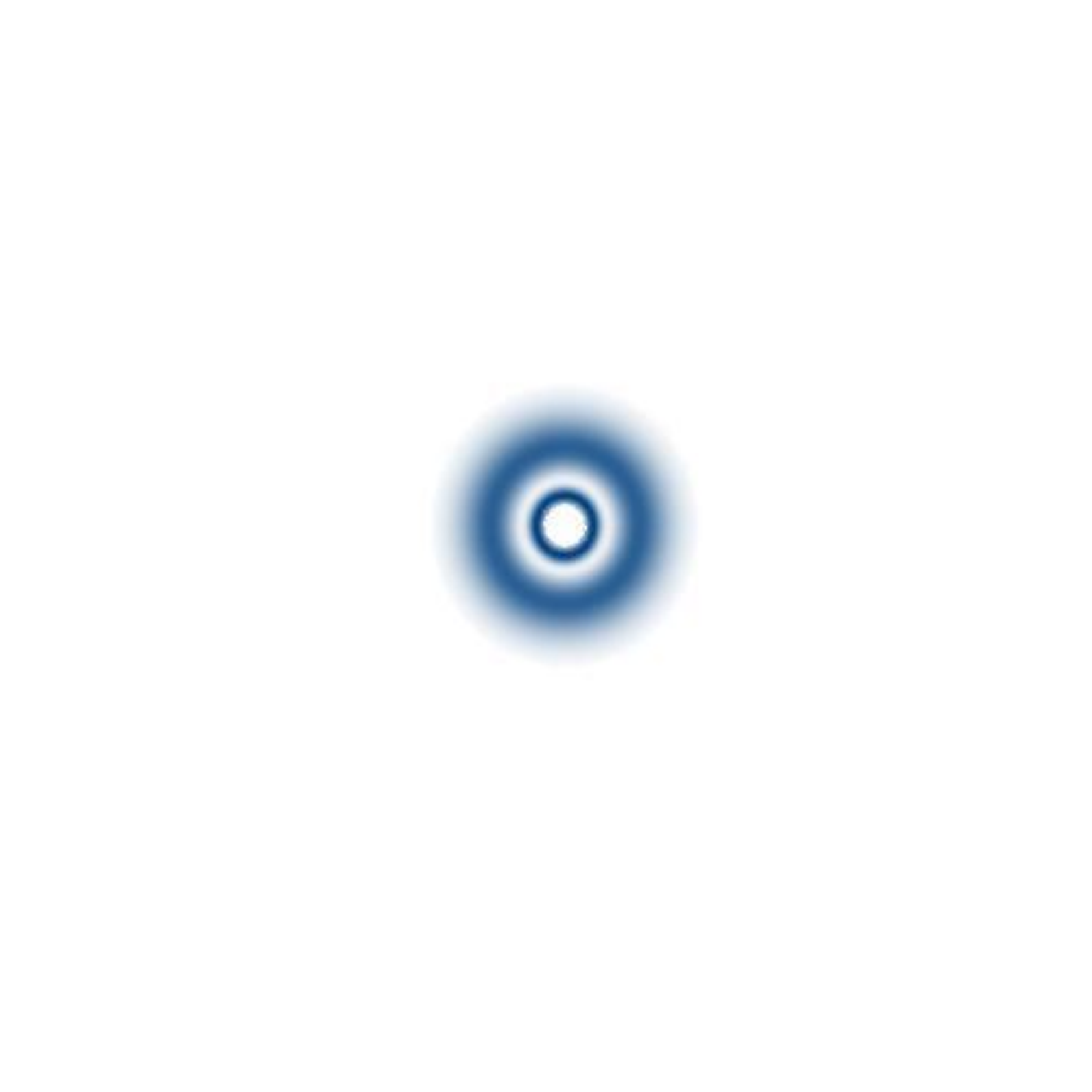}
		\includegraphics[width=4.27cm,trim={0cm 0cm 0 0},clip]{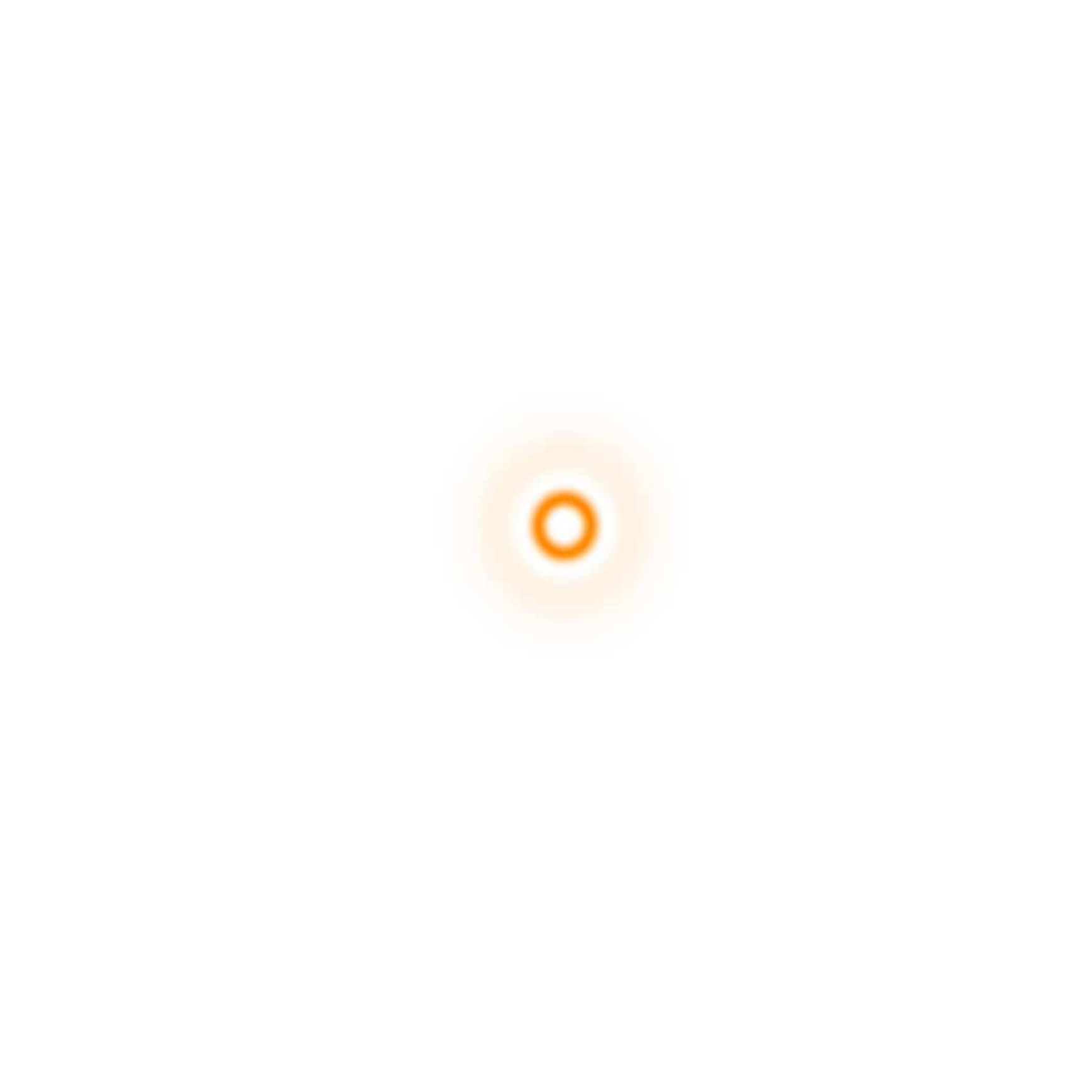}
		\caption{The section passing through the center of the structure, representing the electric field (left, blue) with $e=1$ and the energy density (right, orange) associated to the model in Sec.~\ref{model3d2} for $\sigma=5$, $\alpha=2$, with $k=1$ (top) and $2$ (bottom). The intensity of the blue and orange colors increases with the increasing of the electric field and energy density, respectively.}
		\label{CED3fcoseps}
		\end{figure}

To find a richer internal structure, we take $f={\sec}^{2}(n\pi \psi)$. It leads us to the solutions in Eqs.~\eqref{1phi} and \eqref{1chi} with geometrical coordinate $\xi$ replaced by
\begin{align}
    \eta(r)=-\frac{k}{r}+\frac{k}{2\alpha}\big[\textrm{Ci}(\xi_{+}(r))-\textrm{Ci}(\xi_{-}(r))\big],
\end{align}
where $\xi_{\pm}(r)=2n\pi(1\mp\tanh(\alpha/r))$. We depict the solutions in Fig.~\ref{figphichicosd3}.
%%%%%%%%%%%%%%%%%
%%%%%%%%%%%%%%%%
In this case the energy density is given by
\begin{align}
\label{EFD3fcosdensityd3}
   \rho_{1}&=\bigg[4k^{2}{\sech}^{2}{(\eta)}\cos^{2}{(n\pi\tanh(\alpha/r))}\bigg({\sech}^{2}{(\eta)}\nonumber  \\
    &\quad+{\tanh}^{2}{(\eta)}\bigg(\frac{\sigma}{k}-2\bigg)\bigg)\bigg] \frac{1}{r^{4}},
\end{align}
and the electric field is
\be
\begin{aligned}
\label{EFD3fcosd3}
    E_{r}&=\bigg[4k^{2}{\sech}^{2}{(\eta)}\cos^{2}{(n\pi\tanh(\alpha/r))}\bigg({\sech}^{2}{(\eta)}\\
    &\quad+{\tanh}^{2}{(\eta)}\bigg(\frac{\sigma}{k}-2\bigg)\bigg)+\alpha^2{\sech}^{4}{(\alpha/r)}\bigg] \frac{1}{er^{2}}.
\end{aligned}
\ee
In Fig.~\ref{cedensityd3cos} we depict the electric field \eqref{EFD3fcosd3} and the energy density \eqref{EFD3fcosdensityd3} for $k=1$ and $k=2$. Since the object is spherically symmetric, we see that the structure has a multi-shell form; the number of shells are controlled by $n$. As $k$ increases, the electric field and energy tend to become less and less concentrated at the center, as it spreads more and more to the external rings. We also display the planar section passing through the center of the structure in Fig.~\ref{CED3fcoseps}. The energy of this configuration is the very same of the one in Sec.~\ref{model3d1}.\\
%%%%%%%%%%%%%%%%%%%
%%%%%%%%%%%%%%%%%%%%%%%
%%%%%%%%%%%%%%%%%%%%%
%%%%%%%%%%%%%%%%%%%%

\section{Conclusion}\label{sec5}
In this work, we investigated the behavior of a single point charge in a medium with electric permittivity controlled by two scalar fields that can be associated to the so called Bloch domain wall. We have shown that the equations of motion that govern the scalar fields are independent and, based on the energy minimization, we have developed a first order formalism for our model. In this situation, it was found that an elliptic orbit allows the scalar fields to be decoupled, leading to analytical solutions. This procedure was illustrated with two models, which lead us to conclude that the presence of the scalar fields regularizes the behavior of the electric field and energy density at the origin and localizes the structure in a region outside the origin, with a ring- or shell-like profile. The results show that the strength of the coupling between the scalar fields controls the location of the ring or shell. We have also investigated how the inclusion of a third scalar field that simulates geometrical constrictions modifies the solutions. In this case, it was possible to obtain richer configurations, exhibiting several rings or shells whose location is also controlled by the parameter that couples the fields of the Bloch wall.

The results obtained in the present study motivate us to suggest distinct perspectives of future work. In particular, we are now investigating other extensions, as in the recent study on electric dipole \cite{ED}, to propose new devices of the electric type. Despite the intrinsic interest in high energy physics, we also think the present study may find applications to the investigation of plasmonic materials that exhibit high refractive index at the nanometric scale. It is perhaps possible to adapt the Maxwell-scalar model to describe properties of thermoplasmonics interest, such as the generation of nano-sources of heat through the use of plasmonic nanoparticles \cite{AA,BB,CC}. The Maxwell-scalar model can also be studied in the presence of curvature, which may respond to other related effects, as explored in the case of boson stars, for instance, which have many interesting application \cite{BS}. In this context, we can also think of five-dimensional holographic model that includes spacetime metric, an Abelian gauge field and a real scalar field giving rise to Einstein-Maxwell-scalar model \cite{EMs}, which can be used to calculate thermodynamics properties of a hot and dense quark-gluon plasma; see, e.g., Ref. \cite{Jorge}. Other possibilities concern the presence of radially symmetric, stable solutions in four dimensional static, radially symmetric spacetime, as studied in \cite{He,Mo1,Mo2} and in references therein. We hope to report on some of these issues in the near future.

\acknowledgements{The work is supported by the Brazilian agencies Coordena\c{c}\~ao de Aperfei\c{c}oamento de Pessoal de N\'ivel Superior (CAPES), grant No.~88882.440276/2019-01 (MP), Conselho Nacional de Desenvolvimento Cient\'ifico e Tecnol\'ogico (CNPq), grant No. 303469/2019-6 (DB), Paraiba State Research Foundation (FAPESQ-PB), grant No. 0015/2019, and Federal University of Para\'\i ba (UFPB/PROPESQ/PRPG) project code PII13363-2020.}

\end{document}